\documentclass[twocolumn]{aastex62}
\usepackage{appendix}
\usepackage{natbib}
\usepackage{amsmath}
\usepackage{mathrsfs}

\usepackage{lineno}
\usepackage{tabularx}
\usepackage[normalem]{ulem}
\usepackage{graphicx}
\usepackage{booktabs}
\usepackage{url}
\usepackage{color,subfigure,lineno}
\usepackage[T1]{fontenc}
\usepackage{academicons}
\usepackage{xcolor}
\usepackage{hyperref}

\newcommand{\mbh}{$M_{\rm BH}$}

\newcommand{\RNum}[1]{\uppercase\expandafter{\romannumeral #1\relax}}
\newcommand{\dotm}{\ifmmode {\dot{\mathscr{M}}} \else $\dot{\mathscr{M}}$\fi}

\newcommand{\hbeta}{H{$\beta$}}
\newcommand{\hb}{H{$\beta$}}

\newcommand{\FeII}{\ion{Fe}{2}}

\def\HeII{He\,{\sc ii}}
\def\HeIIopt{He\,{\sc ii}\,$\lambda$4687}

\def \OIII {[O\,{\sc iii}]}
\newcommand{\OIIIa}{[O{\sevenrm\,III}]\,$\lambda$4959}
\newcommand{\OIIIb}{[O{\sevenrm\,III}]\,$\lambda$5007}

   \font\sevenrm=cmr7 scaled 1000
\newcommand{\comments}[1]{}

\newcommand*{\rom}[1]
{\expandafter\@slowromancap\romannumeral #1@}

\newcommand{\snu}{\affil{Department of Physics \& Astronomy, Seoul National University, Seoul 08826, Republic of Korea, jhwoo@snu.ac.kr, wangshu100002@gmail.com}}
\newcommand{\kasi}{\affil{Korea Astronomy and Space Science Institute, Daejeon 34055, Republic of Korea}}
\newcommand{\umich}{\affil{Department of Astronomy, University of Michigan, Ann Arbor, MI 48109, USA}}
\newcommand{\nasa}{\affil{NASA/GSFC, Code 662, Greenbelt, MD 20771, USA}}
\newcommand{\nysc}{\affil{National Youth Space Center, Goheung 59567, Republic of Korea}}
\newcommand{\ucla}{\affil{Department of Physics and Astronomy, University of California, Los Angeles, CA 90095-1547, USA}}
\newcommand{\calpoly}{\affil{Physics Department, California Polytechnic State University, San Luis Obispo, CA 93407, USA}}

\newcommand{\knu}{\affil{Major in Astronomy and Atmospheric Sciences, Kyungpook National University, Daegu 41566, Republic of Korea}}
\newcommand{\uci}{\affil{Department of Physics and Astronomy, 4129 Frederick Reines Hall, University of California, Irvine, CA, 92697-4575, USA}}
\newcommand{\cbu}{\affil{Department of Astronomy and Space Science, Chungbuk National University, Cheongju 28644, Republic of Korea}}

\begin{document}

\title{The Seoul National University AGN Monitoring Project III: \hbeta\ lag measurements of 32 luminous AGNs and the high-luminosity end of the size-luminosity relation}

\author[0000-0002-8055-5465]{Jong-hak Woo} \snu
\author[0000-0002-2052-6400]{Shu Wang} \snu
\author[0000-0002-8377-9667]{Suvendu Rakshit} \snu \affiliation{Aryabhatta Research Institute of Observational Sciences, Manora Peak, Nainital-263001, Uttarakhand, India}
\author[0000-0003-2010-8521]{Hojin Cho} \snu
\author{Donghoon Son} \snu

\author[0000-0003-2064-0518]{Vardha N. Bennert} \calpoly
\author[0000-0001-5802-6041]{Elena Gallo} \umich
\author[0000-0002-2397-206X]{Edmund Hodges-Kluck} \nasa \umich
\author[0000-0002-8460-0390]{Tommaso Treu} \ucla

\author[0000-0002-3026-0562]{Aaron J.\ Barth} \uci
\author[0000-0002-4896-770X]{Wanjin Cho} \snu
\author{Adi Foord} \affil{Kavli Institute of Particle Astrophysics and Cosmology, Stanford University, Stanford, CA 94305, USA}
\author{Jaehyuk Geum} \knu
\author[0000-0001-8416-7059]{Hengxiao Guo} \affiliation{Key Laboratory for Research in Galaxies and Cosmology, Shanghai Astronomical Observatory, Chinese Academy of Sciences, 80 Nandan Road, Shanghai 200030, People’s Republic of China} \uci
\author{Yashashree Jadhav} \snu
\author{Yiseul Jeon} \snu
\author[0000-0003-2632-8875]{Kyle M. Kabasares} \uci
\author{Won-Suk Kang} \nysc
\author{Changseok Kim} \snu 
\author[0000-0002-3560-0781]{Minjin Kim} \knu
\author{Tae-Woo Kim} \nysc \cbu
\author[0000-0003-1270-9802]{Huynh Anh N. Le} \snu \affiliation{CAS Key Laboratory for Research in Galaxies and Cosmology, Department of Astronomy, University of Science and Technology of China, Hefei 230026, China}
\author[0000-0001-6919-1237]{Matthew A. Malkan} \ucla
\author{Amit Kumar Mandal} \snu
\author{Daeseong Park} \kasi \knu
\author{Chance Spencer} \calpoly
\author[0000-0001-6363-8069]{Jaejin Shin} \snu \knu \kasi
\author[0000-0001-9515-3584]{Hyun-il Sung} \kasi
\author[0000-0002-1912-0024]{Vivian U} \uci
\author[0000-0002-4645-6578]{Peter R. Williams} \ucla
\author{Nick Yee} \calpoly

\begin{abstract}

We present the main results from a long-term reverberation mapping campaign carried out for the Seoul National University Active Galactic Nuclei (AGN) Monitoring Project. 
High-quality data were obtained during 2015-2021 for 32 luminous AGNs (i.e., continuum luminosity in the range of $10^{44-46}$ erg s$^{-1}$) at a regular cadence, of 20-30 days for spectroscopy and 3-5 days for photometry. 
We obtain time lag measurements between the variability in the \hb\ emission and the continuum for 32 AGNs; twenty-five of those have the best lag measurements based on our quality assessment,
examining correlation strength, 
and the posterior lag distribution.
Our study significantly increases the current sample of reverberation-mapped AGNs, particularly at the moderate to high luminosity end. 
Combining our results with literature measurements, 
we derive a H$\beta$ broad line region size--luminosity relation with a shallower slope than reported in the literature. 
For a given luminosity, most of our measured lags are shorter than the expectation,
implying that single-epoch black hole mass estimators based on previous calibrations could suffer large systematic uncertainties.

\end{abstract}


\section{Introduction}

Black hole mass (\mbh) is a key parameter in understanding the physics of active galactic nuclei (AGN) and the connection of black hole growth with galaxy evolution.
\mbh\ can be determined based on the spatially resolved data by measuring the kinematics of stars, gas, and masers near the sphere of influence of supermassive black holes  \citep[e.g.,][]{Gultekin09, Barth01,Marconi03, Davies06, Scharwachter13, DenBrok15, Boizelle21,Kabasares22}, or by imaging black hole shadows along with theoretical approach \citep{EHTM87I,EHTSgA-I}. However, these methods are limited to relatively nearby objects due to the limited spatial resolution of current facilities.

In contrast, reverberation mapping \citep[RM;][]{Blandford82,Peterson93}  based on the variability of AGNs can be applied to mass-accreting black holes beyond the local universe. Currently, the RM technique and related indirect mass estimators are the primary methods for determining \mbh\ over a large range of cosmic time. The main idea of RM is to measure the time delay ($\tau$) between the flux variations of the continuum and broad emission lines, which represents the light travel time from the central photo-ionizing source to the photo-ionized gas, providing the size (or radius) of the broad-line region (BLR). Based on the virial assumption that the dynamics of gas in the BLR is governed by the gravitational potential of the central black hole, \mbh\ is determined by combining the measured size ($R_{\rm BLR}$) with the velocity measure ($V$) from the width of broad emission lines as:
\begin{equation}
    M_{\rm BH}=f\frac{(c\tau) V^2}{G},
\end{equation}
where $c$ is the speed of light, $G$ is the gravitational constant and $f$ is a factor representing the unknown geometry of the BLR. While $f$ can be different for each AGN, an average $f$ factor is calibrated based on the black hole mass - stellar velocity dispersion relation of the local galaxies \citep[e.g.,][]{Onken04,Woo10,Park12b,Woo15}. Note that the $f$ factor is the main source of the uncertainty of \mbh, up to 0.4 dex in the case of H$\beta$ reverberation based \mbh\ \citep{Park12b,Woo15}. The $f$ factor has been constrained for a small number of individual AGNs based on the dynamical modeling of the BLR combined with the velocity-resolved reverberation-mapping data \citep[e.g.,][]{Pancoast11, Pancoast14a, Pancoast14b, Li14, Williams18,Villafana22}. 

Early studies of RM  reported a correlation between the measured \hb\ BLR size and the monochromatic continuum luminosity at 5100\AA\ (L$_{\rm 5100}$) \citep{Wandel1999, Kaspi00}, opening an indirect way of estimating BLR size and \mbh\ from single-epoch spectra since monitoring data for RM is not required \citep[e.g.,][]{Woo&Urry02, Vestergaard06,Shen11,Shen12,DallaBonta20}. 
While the best-fit slope was initially reported as 0.7 \citep{Kaspi00}, following studies based on HST images calibrated the slope as $\sim$0.5 
as expected from photoionization 
after correcting for the host galaxy contribution to the observed L$_{5100}$, particularly for low-luminosity AGNs with relatively strong stellar continuum \citep[e.g.,][]{Bentz09b, Bentz13}.

It is of importance to investigate the photoionization and the BLR size-luminosity relation for a general population of AGNs over a large dynamic range of \mbh\ and AGN luminosity. In the past, however, RM studies were limited to low-to-moderate luminosity AGNs due to the observational challenges.
The main difficulty is that long-term spectroscopic monitoring with good cadence is required to obtain an accurate time lag measurement between AGN continuum and emission line flux variations. Over the last decades, a number of intensive programs were dedicated to RM studies, dramatically increasing the sample size and the dynamic range of the reverberation-mapped AGNs \citep{Barth15, Du15, Grier17b,Zhang19,U22,Malik22ArXiv}. However, it is still important to extend the RM study to more luminous AGNs, particularly in the high-luminosity regime (i.e., L$_{5100}$ = $ 10^{45\sim46}$ erg s$^{-1}$), which are the representative luminosity of high-z AGNs.
For example, an AGN with L$_{5100}$ = 10$^{46}$ at z=1 is expected to have a H$\beta$ time lag of 500-600 days in the observed-frame, which then has to be determined based on a monitoring campaign of 5-10 years. Such a long timeline explains why there is a relative lack of very luminous AGNs among the reverberation-mapped AGNs (see Figure 1).

Currently, the H$\beta$ time lag has been reported for more than 200 AGNs \citep[e.g.,][]{Wandel1999, Kaspi00, Peterson04,Bentz09c,Bentz13,Barth15,Grier17b,Park17, Du_Wang19,Zhang19,Martinez-Aldama19,DallaBonta20,Hu21,Li-SS21,U22, Malik22ArXiv}. These AGNs show a larger scatter in the H$\beta$ BLR size-luminosity relation compared to the previously reported relations \citep[e.g.,][]{Bentz13}. While it is possible that the intrinsic scatter may not be larger than previously thought, a consistent study of cross-correlations with uniform measurements of H$\beta$ lag and uncertainty is required to properly constrain the intrinsic scatter as well as the slope of the size-luminosity relation. Note that various studies performed by different groups adopted different criteria to select reliable lag measurements and inconsistent methods to derive the uncertainty of the lag.

In particular, AGNs with a super-Eddington ratio seem to deviate from the size-luminosity relation according to the results from the recent project, the Super Eddington Accreting Massive BHs (SEAMBHs) \citep{Wang14,Du14,Zhang19,Hu21,Li-SS21}. 
It is claimed that the deviation from the size-luminosity relation correlates with the accretion rate or the flux ratios between \ion{Fe}{2} and \hbeta\ ($R_{\rm Fe}$) or between \OIII\ and \hbeta\ lines. This systematic trend is crucial to verify since it will introduce strong bias in \mbh\ determination for high Eddington ratio AGNs from single-epoch spectra \citep[see the discussion by][]{Li-SS21}.

To extend the RM study to moderate-to-high luminosity AGNs and investigate the H$\beta$ BLR size-luminosity relation at the high-luminosity end, we performed an intensive long-term campaign, the Seoul National University AGN Monitoring Project (SAMP). 
Using a 100 AGNs with moderate to high luminosity (L$_{5100}$ $>$ $10^{44.0}$ erg s$^{-1}$) out to z$<$0.4, we started test photometry in 2015. Then, we carried out photometric and spectroscopic monitoring with the Lick 3-m, MDM 2.4-m and other 1-m class telescopes until the middle of 2021. The project strategy and sample selection were presented by \citet[][hereafter Paper \RNum{1}]{Woo19}, and initial measurements of the H$\beta$\ lag for two targets based on the first three-year data were reported by \citet{Rakshit19}. In this work, we present the final sample of 32 AGNs from the six-year spectroscopic campaign and the H$\beta$\ lag analysis. The measurements of H$\alpha$ lag are presented by \citet{Cho+23}, and the photometry monitoring results of 72 AGNs will be presented by Son et al. (in preparation). 
We briefly describe the sample selection in Section 2 and data analysis in Section 3. Results and discussion are presented in Sections 4 and 5, respectively, followed by Discussion in Section 5 and Summary in Section 6. Throughout this paper, we use the $\Lambda$CDM cosmology, with $H_0=72.0$ km s$^{-1}$ Mpc$^{-1}$, and $\Omega_{m}=0.3$.

\section{Observations and data reduction}

\subsection{Sample}

\begin{figure}[!h]
    \centering
    \includegraphics[width=0.48\textwidth]{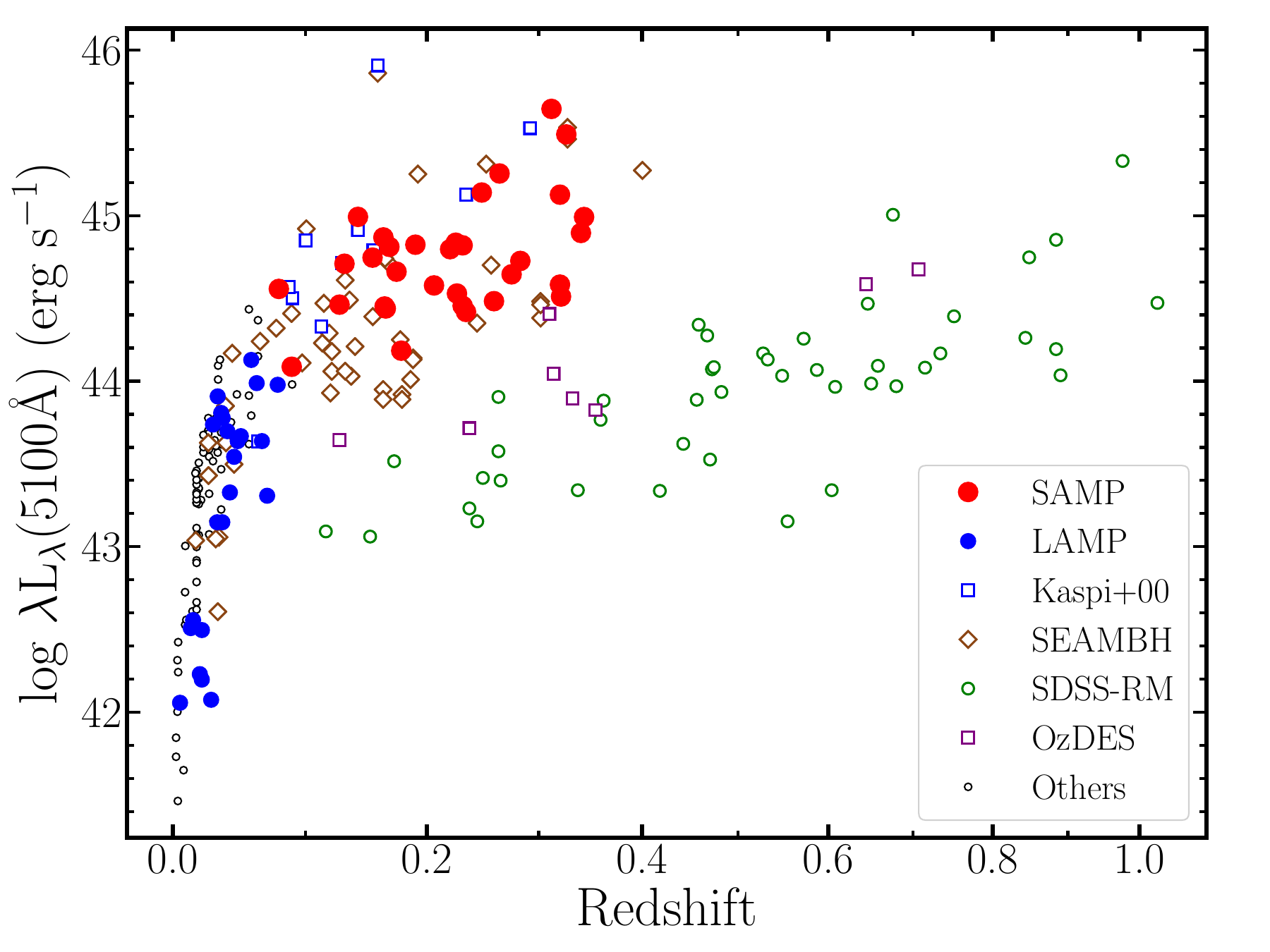}
    \caption{Luminosity and redshift distribution of the SAMP final sample (red filled circles) compared to the \hbeta\ lag measured AGNs in other RM campaigns, including the LAMP 2008 and 2016 \citep{Bentz09c,U22}, the monitoring campaign of PG quasars by \citet{Kaspi00}, the SEAMBHs \citep{Du14, Wang14, Du15, Du16, Du18b, Zhang19, Hu21, Li-SS21},  OzDES \citep{Malik22ArXiv}, the SDSS-RM \citep{Grier17b}, and other AGNs in the collection by \citet{DallaBonta20}. 
 }
    \label{fig:L-z-plane}
\end{figure}

We selected the best available type 1 AGNs for a multi-year monitoring program for our facilities, by considering the expected lag, observability, and feasibility of H$\beta$ lag measurements along with the simulation of light curves and spectral decomposition. The details of the project strategy and sample selection of the SAMP were presented by \citet{Woo19}, and here we briefly describe the sample. 

We selected relatively high-luminosity AGNs using the MILLIQUAS catalog \citep[Milliquas v4.5 (2015) update,][]{Flesch2015, Boroson_Green_92}, in order to test the high-end of the \hbeta\ BLR size-luminosity relation. As summarized by \citet{Woo19},
we initially selected 100  AGNs with the $V$-band  magnitude $<17.0$ at $z<0.5$. The \hbeta\ lag is expected to range from $\sim$40 to $\sim$250 days in the observed-frame, which is estimated based on the monochromatic continuum luminosity at 5100\AA\ (L$_{5100}$) and the size-luminosity relation of \citet{Bentz13}. In this process, we used SDSS spectra or the spectra provided by  \citet{Boroson_Green_92} to measure $\lambda L_{\lambda}(5100$\AA). $B$-band photometry is used instead if there is no available spectrum. We identified 48 AGNs with an expected lag longer than 70 days in the initial sample as
the first priority targets (i.e., SAMP ID started with Pr1 in Table 1) and 37 AGNs with an expected lag shorter than 70 days as the second priority targets (i.e., SAMP ID started with Pr2). We also included 15 PG QSOs \citep{Boroson_Green_92} for filling up the seasonal gaps and increasing the sample size (i.e., SAMP ID started with P). These PG QSOs are also medium-to-high luminosity objects with an expected lag of $\sim$50-300 days.

Using the initial sample of 100 AGNs, we performed the variability test of continuum and \hbeta\ line emission based on photometry and spectroscopy during the first few years, and excluded weakly-varying objects. Based on these variability check processes, we continued 6-year monitoring for a final sample of 32 AGNs. This final AGNs are out to z$\sim0.4$ with luminosity L$_{\rm 5100}$ $>$ 10$^{44}$ erg s$^{-1}$ as presented in Figure \ref{fig:L-z-plane}. We emphasize that the SAMP final sample covers relatively higher luminosity ranges compared to previous \hbeta\ monitoring campaigns \citep[e.g.,][]{Kaspi00,Bentz09c, Barth15, Wang14, Du14, Du15, Du16, Grier17b, U22, Malik22ArXiv}. The properties of the sample are summarized in Table \ref{tab:sample-information}.

{\begin{table*}[!h]
    \centering
    \small
    \caption{Sample properties and observation parameters} 
    \begin{tabular}{@{\extracolsep{8pt}}r l c  c c  c c c c c l }
    \hline \hline
    & Name  & SDSS Identifier  & $z$ & $m_{V}$  &${A_{V}}$ & PA & Exp. & S/N & Band  & SAMPID \\ 
  & &   &  & (mag)  & (mag) & ($^{\circ}$) & (s) \\ 
  &  (1) & (2) & (3) & (4) & (5) & (6) & (7) & (8) & (9) & (10)  \\ \hline
1  & Mrk~1501        & J001031.00$+$105829.4   &  0.089  &  15.8  & 0.273 &  Para  &  800   &  41  &   $B$, $V$  & P02        \\ 
2  & PG~0026$+$129   & J002913.70$+$131603.9   &  0.142  &  15.2    & 0.195 &  Para  &  600   &  73  &   $B$, $V$  & P03        \\ 
3  & PG~0052$+$251   & J005452.11$+$252539.0   &  0.154  &  15.7    & 0.129 &  Para  &  800   &  46  &   $B$, $V$  & P06        \\ 
4  & J0101$+$422     & J010131.17$+$422935.5    &  0.190  &  16.3    & 0.244 &  96.4  &  1200  &  46  &   $B$, $V$  & Pr1\_ID01  \\ 
5  & J0140$+$234     & J014035.01$+$234451.1   &  0.320  &  17.0    & 0.362 &  67.1  &  1200  &  31  &   $B$, $V$  & Pr1\_ID03  \\ 
6  & Mrk~1014        & J015950.25$+$002340.8   &  0.163  &  15.5    & 0.079 &  Para  &  800   &  39  &   $B$, $V$  & P07        \\ 
7  & J0801$+$512     & J080112.02$+$512812.0   &  0.321  &  18.4    & 0.136 &  90.5  &  3600  &  14  &   $B$, $V$  & Pr1\_ID11  \\ 
8  & J0939$+$375     & J093939.69$+$375705.8   &  0.231  &  17.2    & 0.049 &  100.8 &  2400  &  27  &   $B$, $V$  & Pr1\_ID14  \\ 
9  & PG~0947$+$396   & J095048.39$+$392650.4   &  0.206  &  16.7    & 0.053 &  161.8 &  1200  &  31  &   $B$, $V$  & Pr1\_ID15  \\ 
10 & J1026$+$523     & J102613.90$+$523751.2    &  0.259  &  17.9    & 0.037 &  94.8  &  2400  &  16  &   $B$, $V$  & Pr1\_ID17  \\ 
11 & J1059$+$665     & J105935.50$+$665757.9   &  0.340  &  17.4    & 0.042 &  67.9  &  2400  &  24  &   $B$, $V$  & Pr1\_ID18  \\ 
12 & PG~1100$+$772   & J110413.87$+$765858.1   &  0.312  &  15.4    & 0.100 &  66.3  &  600   &  41  &   $B$, $V$  & Pr1\_ID19  \\ 
13 & J1105$+$671     & J110527.25$+$671636.4    &  0.320  &  17.7    & 0.050 &  112.5 &  2400  &  23  &   $V$, $R$  & Pr1\_ID20  \\ 
14 & J1120$+$423     & J112007.43$+$423551.3   &  0.226  &  17.2    & 0.051 &  72.7  &  2400  &  30  &   $B$, $V$  & Pr1\_ID23  \\ 
15 & PG~1121$+$422   & J112439.18$+$420145.0    &  0.225  &  16.5    & 0.062 &  133.0 &  1200  &  33  &   $B$, $V$  & Pr1\_ID24  \\ 
16 & J1203$+$455     & J120348.08$+$455951.1    &  0.343  &  17.2    & 0.045 &  158.2 &  2400  &  25  &   $B$, $V$  & Pr1\_ID26  \\ 
17 & PG~1202$+$281   & J120442.10$+$275411.8   &  0.165  &  16.7    & 0.057 &  168.7 &  1800  &  30  &   $B$, $V$  & Pr1\_ID27  \\ 
18 & J1217$+$333     & J121752.16$+$333447.2   &  0.178  &  17.3    & 0.037 &  169.9 &  2400  &  26  &   $B$, $V$  & Pr1\_ID29  \\ 
19 & VIII~Zw~218     & J125337.71$+$212618.2  &  0.127  &  15.9    & 0.139 &  126.2 &  1800  &  38  &   $B$, $V$  & Pr1\_ID30  \\ 
20 & PG~1322$+$659   & J132349.52$+$654148.1   &  0.168  &  15.9    & 0.053 &  97.4  &  1200  &  39  &   $B$, $V$  & Pr2\_ID18  \\ 
21 & J1415$+$483     & J141535.94$+$483543.6   &  0.275  &  17.6    & 0.040 &  192.9 &  3600  &  28  &   $B$, $V$  & Pr1\_ID36  \\ 
22 & PG~1427$+$480   & J142943.07$+$474726.2   &  0.220  &  16.5    & 0.047 &  90.8  &  1800  &  38  &   $B$, $V$  & Pr2\_ID24  \\ 
23 & PG~1440$+$356   & J144207.47$+$352622.9   &  0.079  &  14.6    & 0.039 &  157.1 &  600   &  70  &   $B$, $V$  & Pr2\_ID26  \\ 
24 & J1456$+$380     & J145608.63$+$380038.5    &  0.283  &  17.1    & 0.030 &  55.4  &  1800  &  25  &   $B$, $V$  & Pr1\_ID38  \\ 
25 & J1526$+$275     & J152624.02$+$275452.1    &  0.231  &  16.8    & 0.113 &  137.4 &  1800  &  43  &   $B$, $V$  & Pr1\_ID40  \\ 
26 & J1540$+$355     & J154004.25$+$355050.1   &  0.164  &  16.7    & 0.071 &  8.10  &  1200  &  37  &   $B$, $V$  & Pr1\_ID41  \\ 
27 & PG~1545$+$210   & J154743.53$+$205216.6   &  0.264  &  16.0    & 0.117 &  59.10 &  1200  &  41  &   $B$, $V$  & Pr1\_ID42  \\ 
28 & PG~1612$+$261   & J161413.20$+$260416.2   &  0.131  &  15.4    & 0.151 &  17.2  &  600   &  56  &   $B$, $V$  & Pr2\_ID35  \\ 
29 & J1619$+$501     & J161911.24$+$501109.2   &  0.234  &  16.0    & 0.055 &  46.70 &  1800  &  25  &   $B$, $V$  & Pr1\_ID43  \\ 
30 & J1935$+$531 & J19352118$+$5314119$^{\dagger}$  &  0.248  &  16.4    & 0.312 &  169.5 &  1200  &  34  &   $B$, $V$  & Pr1\_ID47  \\ 
31 & PG~2251$+$113   & J225410.42$+$113638.8   &  0.326  &  15.8    & 0.236 &  Para  &  720   &  59  &   $V$, $R$  & P13        \\ 
32 & PG~2349$-$014   & J235156.12$-$010913.3  &  0.174  &  16.1    & 0.075 &  Para  &  1200  &  37  &   $B$, $V$  & P15        \\ 
    \hline
     \multicolumn{11}{p{0.99\textwidth}}{{\bf Notes.} Column (1): Object name in the sequence of R.A. with the sequence No. shown in the front. Column (2): SDSS identifier. Column (3): Redshift. Column (4):  the median apparent $V$-band magnitudes measured from the SAMP light curves. Column (5): Extinction in $V$-band extracted from NED based on \citet{SF11} dust map. Column (6): Slit position angle. The label Para means that the PA is set to be parallactic. Column (7): Typical exposure time of single-night spectroscopic observation. Column (8): Average signal-to-noise ratio per pixel of all spectra of this object. Column (9): Primary and secondary band in photometric observation. The primary band light curves are used as the continuum light curves. Column (10): Object ID defined for the project as used in Paper \RNum{1}. 

     $^{\dagger}$ This object J1935$+$531 is not in the SDSS footprint. Its main identifier is 2MASS J19352118$+$5314119. } 
    \end{tabular}

\end{table*}
}\label{tab:sample-information}

\subsection{Photometry} 

We performed photometric monitoring using several telescopes: the MDM 1.3-m and 2.4-m telescopes at the Kitt Peak, Tucson, Arizona, USA, the Lemmonsan Optical Astronomy Observatory (LOAO) 1-m telescope located on Mt. Lemmon, Tucson, Arizona, USA, the Lick observatory 1-m telescope located at Mt. Hamilton, California, USA, the Las Cumbres Observatory Global Telescope (LCOGT) network, and the Deokheung Optical Astronomy Observatory (DOAO) 1-m telescope. We used the $B$ and $V$, or $V$ and $R$ band filters to monitor continuum variability, depending on the redshift of each object (see Table \ref{tab:sample-information}).

Details of the photometric data reduction and variability analysis including the final photometric light curves will be presented in a forthcoming paper (Son et al. in preparation). Here we briefly describe the basic information for completeness. We followed the standard reduction procedure for bias subtraction and flat-fielding using {\tt IRAF} package \citep{Tody86,Tody93}. We used the {\tt LA-Cosmic}\footnote{http://www.astro.yale.edu/dokkum/lacosmic/} task \citep{vanDokkum01} to remove cosmic rays and adopted the {\tt Astrometry.net}\footnote{http://astrometry.net} \citep{lang+10} for astrometric calibration. Sub-exposure images were combined using the {\tt SWarp}\footnote{https://www.astromatic.net/software/swarp/} software \citep{Bertin02} by matching the position of all stars in the Field-of-View (FoV) of each image.
We performed aperture photometry using the {\tt SExtractor}\footnote{https://www.astromatic.net/software/ extractor/} software \citep{BA96} with an initial aperture size to be 3 times of the seeing size. By generating the magnitude growth curve as a function of the aperture size, we tested whether the magnitude based on the seeing was fainter than the brightest magnitude in the growth curve, and enlarged the aperture size accordingly in order to avoid any aperture loss. A small fraction of images with bad quality due to the full moon, gust wind, or thick clouds were excluded from the final photometric light curves based on visual inspection of each image.  
We performed flux-calibration using non-variable stars in the FoV of each AGN, using the star catalogs of SDSS and APASS\footnote{www.aavso.org/apass}. 
Finally, we obtained the photometric light curves of the sample in each band at each telescope. 

We found systematic offset among the light curves obtained with different telescopes, presumably caused by various weather conditions and difference in the filter properties as often reported by previous studies \citep[e.g.,][]{Peterson95, Pancoast19}. Thus, it is critical to inter-calibrate these light curves in the merging process. We performed the inter-calibration by adopting the python software {\tt PyCALI}\footnote{\url{https://github.com/LiyrAstroph/PyCALI}} \citep{Li14} which used damped random walk (DRW) model to describe the AGN variability and determine the best scaling factors based on Bayesian statistics. For each AGN, we adopted the light curve from the MDM 1.3-m telescope as a reference, and aligned all the other light curves. Note that the MDM 1.3-m light curves have the largest number of epochs, which are also most evenly distributed over the monitoring time baseline. Systematic uncertainties were added to each telescope's light curve during the inter-calibration process. 

Visual inspection of the merged light curves showed that for most targets DOAO light curves were still relatively scattered and deviated from the general trend of the light curves obtained at other telescopes, presumably due to flux calibration issues, i.e., high humidity and quickly changing seeing condition at the DOAO observatory. In the case of the light curves obtained at the Lick 1-m and LCOGT telescopes, we sometimes found a similar problem for some objects. Under these circumstances, we decided to exclude the light curve from Lick 1-m and LCOGT as well. 

We utilized the $B$-band light curve as the continuum light curves for all objects except for two higher redshift objects at z $>$0.3, namely, J1105$+$671 and PG~2251$+$113, for which we instead adopted the $V$-band  light curve. For seven AGNs, we found that the $g$-band light curves from the Zwicky Transient Facility (ZTF) were available, and combined them to the SAMP continuum light curves, in order to improve the temporal coverage and cadence. Note that ZTF is a time-domain survey, which started in 2018 and overlapped with the SAMP monitoring baseline.We used the ZTF DR8 to include the photometric data from March 2018 to September 2021. 
Following the previous studies based on the ZTF data \citep[e.g.,][]{Sanchez-Saez21}, we cleaned the ZTF light curves by requiring the {\tt catflags} $=0$, to avoid the effect of bad weather conditions (e.g., clouds, moon contamination, and large seeing).
We assumed that the time lag between the $B$-band and ZTF $g$-band continuum fluxes is much smaller than that of \hbeta\ emission line as is the case according to the previous continuum reverberation studies \citep[e.g.,][]{Jha21,Netzer22}. For example, \citet{Wang23} reported that a typical size of the continuum emitting region is a factor of $\sim$8 smaller than that of \hbeta. 
By directly testing the lag between $B$ and $g$-bands of our sample, we found that the lags between the two broadband light curves were much smaller than the \hbeta\ time lags for the SAMP AGNs (Mandal et al. in preparation).
 By combining the SAMP and ZTF light curves, we obtained the improved light curves with better temporal coverage with a less than $3$-day cadence, increasing the constraints of the continuum variability and the reliability of the lag measurements. 

\subsection{Spectroscopy}

We carried out the spectroscopic monitoring using two telescopes, the Shane 3-m telescope at the Lick observatory and the 2.4-m telescope at the MDM observatory. For the Lick observations, we utilized the Kast double spectrograph\footnote{\url{https://mthamilton.ucolick.org/techdocs/instruments/kast/}}, which consists of two spectrographs, optimized for red and blue wavelength ranges, respectively. In this study, we focused on the \hbeta\ emission line and used only the red side of the observed spectra. We used the 600-line mm$^{-1}$ grating, covering the spectral range of 4300--7100$\,$\AA\ and a dispersion of $2.33\,$\AA\ pixel$^{-1}$ until September 2016. After the upgrade of the CCD in September 2016, the spectral range was changed to be 4450--7280$\,$\AA, and a dispersion of $1.27\,$\AA\ pixel$^{-1}$. We used a 4$^{\prime\prime}$ slit width to minimize slit loss. Combined with the grating, Lick spectral setup provided a spectral resolution $R$ of $\sim$624. We measured the instrumental resolution (FWHM) of 481 km s$^{-1}$ by utilizing unblended skylines. 
Calibration frames, i.e., bias, dome flats, and arc lamps (He, Ne, Ar, and Hg-Cd), were obtained at each night. For most objects, we used a fixed slit position angle (PA) for observations at airmass less than 1.3, while we adopted a parallactic angle for observations at higher airmass. The on-source exposure time of Lick observations was set between 360 and 1800 seconds depending on the magnitude of each target, in order to obtain a signal-to-noise ratio (S/N) per pixel $>$ 15-20 calculated over the entire spectral range.

For the MDM observations, we utilized the VPH blue grism with a spectral coverage of $3970-6870\,$\AA\ and a dispersion of $0.715\,$\AA\ pixel$^{-1}$. We used a 3$^{\prime\prime}$ slit width before February 2017, after which we ordered and replaced it with a customized 4$^{\prime\prime}$ slit, in order to make a consistent setup compared to the Lick spectroscopy. The corresponding instrumental resolution is $R=617$. Calibration frames, including bias, dome flats, and Ar/Xe arc lamps were obtained at each night. The PA of slits was set to be the same as the Lick observations. The on-source exposure time of the MDM observations was set between 600 and 2400 seconds depending on the magnitude of target AGNs. 

The sample of SAMP covers a large range of R.A., and individual targets show various levels of flux variability. To optimize the monitoring efficiency, we continuously checked the variability and the feasibility of the lag measurements based on the updated light curves in 1-2 month time scales. Note that we can predict strong variability of \hbeta\ line emission if we see a strong variability pattern in the photometric continuum light curves in advance. Thus, we reduced the time allocation of relatively non-varying targets, while we provided more spectroscopic time to promising targets, for which strong variability was detected in the photometric light curves. Consequently, cadence and the total observed number of epochs varied for each target. 
In Table 2 we summarize the observations of each object in the final sample.

We performed standard spectroscopic data reduction including overscan subtraction, bias, and flat-fielding using the standard IRAF package. The cosmic-ray rejection was done using the {\tt LA-Cosmic} task. For the MDM spectra, we used a single sensitivity function, which was averaged over monitoring seasons, because the difference among various epochs was sufficiently small. In the case of the Lick calibration, we tested the consistency using individual sensitivity functions obtained at each night, which were constructed by fitting the spectra of spectrophotometric standard stars observed at each night \citep{Oke90} with polynomial functions using a script provided by {\tt PypeIt} v1.4 \citep{Prochaska20,Prochaska20b}. We found that the results were almost consistent with those from the analysis with the IRAF, but provided better consistency at the edge of the spectral range, especially for 2017-2021 observations. Thus, we adopted the {\tt PypeIt}-reduced spectra for the data obtained in 2017-2021. For the 2016 data, the individual sensitivity function could not be accurately constrained at the blue edge of the spectral range. Thus, we decided to use the reduced spectra based on the IRAF analysis for 2016 Lick observations.

\section{Data Analysis}

\subsection{Flux re-calibration}\label{sec:mapspec}

To reliably measure the spectral variability of the \hbeta\ line emission, we first perform flux re-calibration using the non-varying narrow \OIII\ line emission. This re-scaling approach described by \citet{VGW92} is often adopted by various reverberation-mapping studies since the non-intrinsic variation caused by slightly inconsistent spectral resolution and slit loss due to the nightly change of the focus,  seeing, and centering of AGN in the slit, etc \citep[e.g.,][]{Peterson95}. 

We utilize the python package {\tt mapspec\footnote{\url{https://github.com/mmfausnaugh/mapspec/}}} \citep{Fausnaugh17a}, which follows the same approach proposed by \citet{VGW92}, but works in a Bayesian framework, enabling assessment of the uncertainties of the calibration. As {\tt mapspec} subtracts a linear continuum to extract the \OIII\ line flux within a user-defined window, we first transfer each spectrum into the rest-frame in order to use essentially the same \OIII\ extraction window for all objects. The extraction window of \OIII\ is defined as [4968, 5055]\AA\, and the two adjacent continuum windows are selected to be [4963, 4968] and [5055,5060]\AA. After extracting \OIII\ of each epoch, {\tt mapspec} matches the \OIII\ line profile of each epoch with that of the reference epoch using three parameters: a wavelength shift factor, a flux scaling factor, and a line broadening factor (i.e., the width of a Gauss-Hermite broadening kernel \citep{Fausnaugh17a}). We test various choices of the \OIII\ window and the broadening kernel, finding that the results are essentially the same. 

As a reference epoch, we choose the broadest \OIII\ profile with S/N $>$ 20 and degrade the \OIII\ line profiles of the other epochs to align them with the reference.The selected reference epoch is typically one of the epochs with bad seeing conditions and suffers from slit loss of the flux. One of the advantages of {\tt mapspec} is that the uncertainty of each parameter can be estimated using the half amplitude of the 16th-84th quadrature interval of the distribution of MCMC samples. For each individual epoch, the derived uncertainty of the multiplicative factor is added in quadrature to the \hbeta\ flux uncertainty. Thus, the epochs with more uncertain \OIII-based flux calibration have larger uncertainties in the \hbeta\ light curves. 

We further calibrate the absolute flux level of the spectroscopy by matching the synthetic $V$-band  light curves with the photometric $V$-band  light curves (or $R$-band for two objects, J1105$+$671 and PG~2251$+$113 at z$>$0.3) using {\tt PyCALI}. In this process we obtain an average scale factor and rescale the H$\beta$ light curve for each object. The synthetic $V$($R$)-band flux is calculated by performing synthetic photometry on the mapspec-calibrated spectra.

To verify the quality of the flux calibration, we calculate the normalized excess variance $\sigma_{\rm nx}^2$ of the \OIII\ line flux \citep{Barth15} using the calibrated spectra of each target:
\begin{equation}
    \sigma_{\rm nx}^2 = \frac{1}{N\langle F\rangle ^2} \sum^{N}_{i=1}[(F_{i}-\langle F\rangle )^2 - \delta_{i}^2],
\end{equation}
where $N$ is the number of spectra, $F_{i}$ and $\delta_{i}$ is the \OIII\ flux and flux uncertainty of epoch $i$, and $\langle F\rangle$ is the average of the \OIII\ flux. The normalized excess variance is the fractional residual scatter of the \OIII\ line flux, representing the systematic uncertainty of the flux calibration. While several previous studies uniformly added a single value of $\sigma_{\rm nx}^2$  to each epoch's flux uncertainty in quadrature \citep[e.g.,][]{Barth15, U22}, we add the uncertainty of the \OIII\ scaling factor derived by {\tt mapspec} to \OIII\ flux uncertainty in quadrature for each epoch.
 For completeness, we present $\sigma_{\rm nx}$ of each target in Table  \ref{tab:varibility_lcstats} to demostratre the quality of the flux scaling.

We note that for AGNs with a weak \OIII\ line, this procedure introduces a large systematic error since it is difficult to define the \OIII\ line profile because of the blending of strong  \ion{Fe}{2} and \OIII\ lines. In our sample, five objects, namely, J0939$+$375, PG~1121$+$422, PG~1322$+$659, PG~1440$+$356, and J1526$+$275, show  a relatively small \OIII\ equivalent width. As a different approach, we obtain a flux scale factor for each epoch by matching the synthetic $V$ band magnitude with the $V$ band magnitude from the interpolated photometry light curve based on the DRW model provided by {\tt PyCALI}.  By comparing the results from two different calibration approaches, we find that only one target J1526$+$275 with the weakest (almost no narrow) \OIII\ line shows noticeably better cross-correlation results from the photometry-based scaling. Therefore, we adopt the photometry-based scaling result for J1526$+$275, while we use the \OIII-based scaling results for the rest of the targets. 

\begin{figure*}
    \centering
    \includegraphics[width=0.95\textwidth]{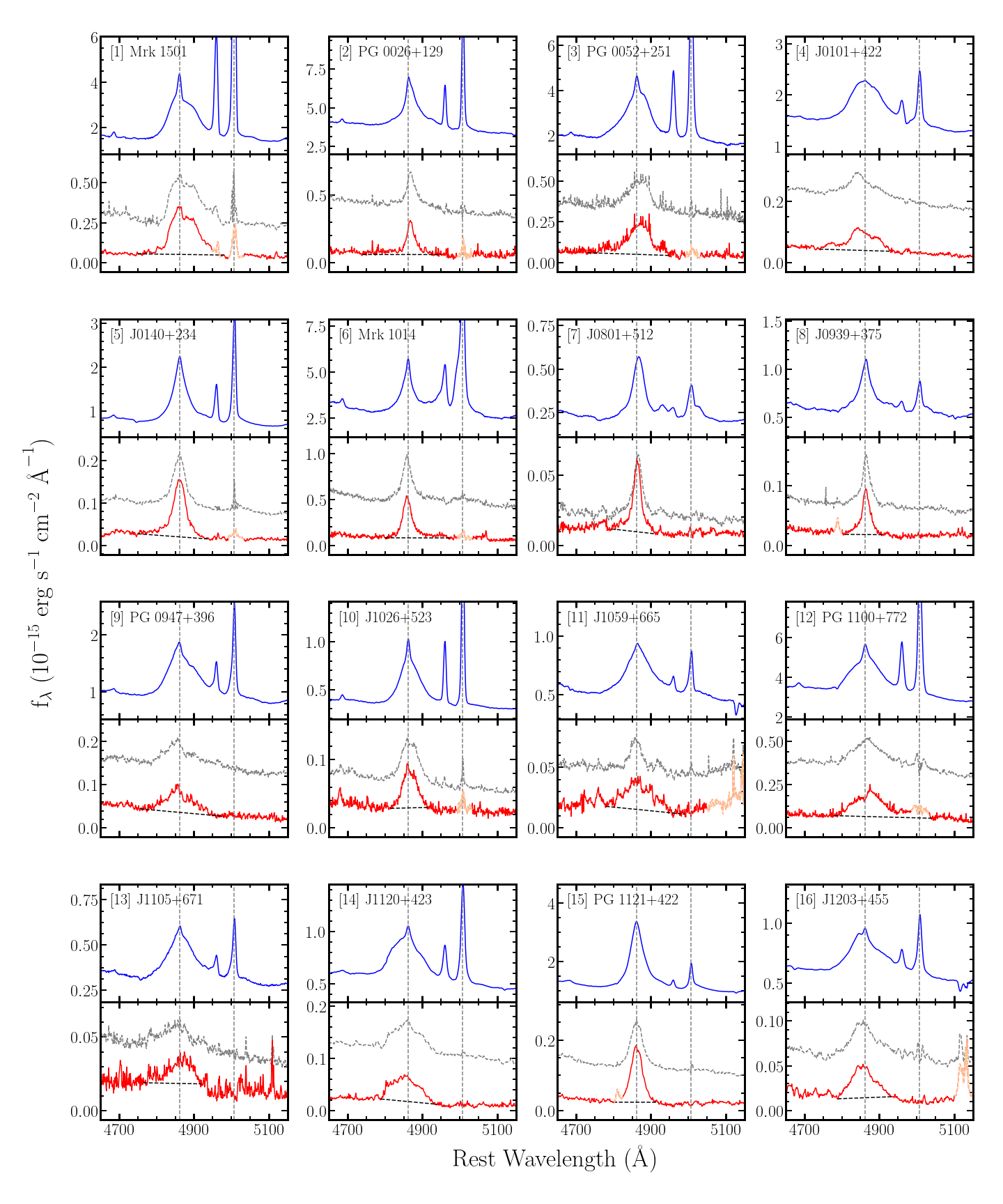}
    \caption{The \hbeta\ emission line profiles in the mean (upper panel) and rms spectra (lower panel) of each AGN.
    The red and grey lines represent the rms spectra, calculated from the individual spectra with and without subtracting continuum and narrow lines, respectively. The two vertical dashed lines indicate the peak of \hbeta\ and \OIIIb. Some features caused by skylines or \OIII\ residuals are masked (orange dashed lines).
    }
    \label{fig:mean-rms-0}
\end{figure*}

\addtocounter{figure}{-1}
\begin{figure*}
    \centering
    \includegraphics[width=0.95\textwidth]{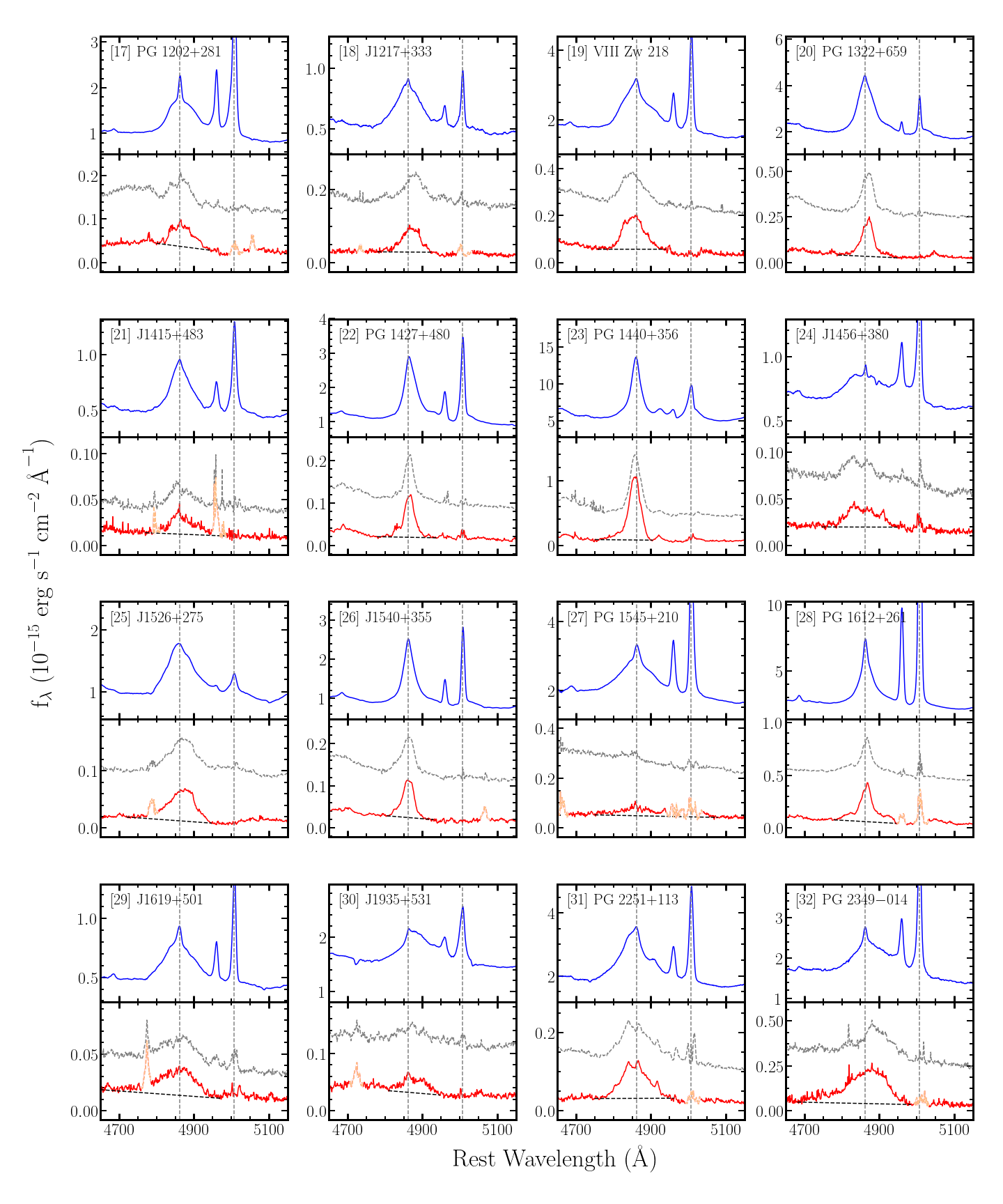}
    \caption{Continued.}
    \label{fig:mean-rms-1}
\end{figure*}

\subsection{Mean and rms spectra}

We generate the mean and rms spectra using the flux re-calibrated spectra of each target, as done by previous studies \citep[e.g.,][]{Peterson04, Barth15, Rakshit19}. The mean spectra are calculated by averaging the flux of all epochs, while the rms spectra are generated using the definition: 
\begin{equation}
    S_{\lambda} = \left(\frac{1}{N-1}\sum_{i=1}^{N}(f_{\lambda, i}-\overline{f}_{\lambda})^2 \right)^{1/2},
\end{equation}
where N is the number of spectra, $f_{\lambda,i}$ and $\overline{f}_{\lambda}$ is the flux density of the spectrum from the ith epoch and the mean spectrum, respectively. 

We provide two sets of rms spectra, using the individual epoch's spectra with and without subtracting continuum and narrow emission lines (see \S 3.3 for the decomposition process). 
It has been shown that the rms spectra generated without subtracting continuum and narrow lines can suffer from a bias in the line width measurement
because of the different variations between the continuum and emission line fluxes \citep{Barth15, Wang19}. As presented in Figure \ref{fig:mean-rms-0}, we find a small residual at the location of \OIIIb, indicating the high quality of the flux calibration. The noticeable residual of \OIII\ in the rms spectra of some targets, together with other small features caused by the telluric lines or the CCD cosmetics and cosmic rays are masked out (as indicated by orange color in Figure \ref{fig:mean-rms-0}) and interpolated before the line width calculation. Note that because we have different wavelength coverages owing to three instrumental setups of Lick spectroscopy, there are small rises of the continuum at the blue side of \hbeta\ (e.g., PG~1440$+$356  and Mrk~1501) due to the CCD effect.

\subsection{Spectral decomposition and measurements}

\begin{figure}
    \centering
    \includegraphics[width=0.47\textwidth]{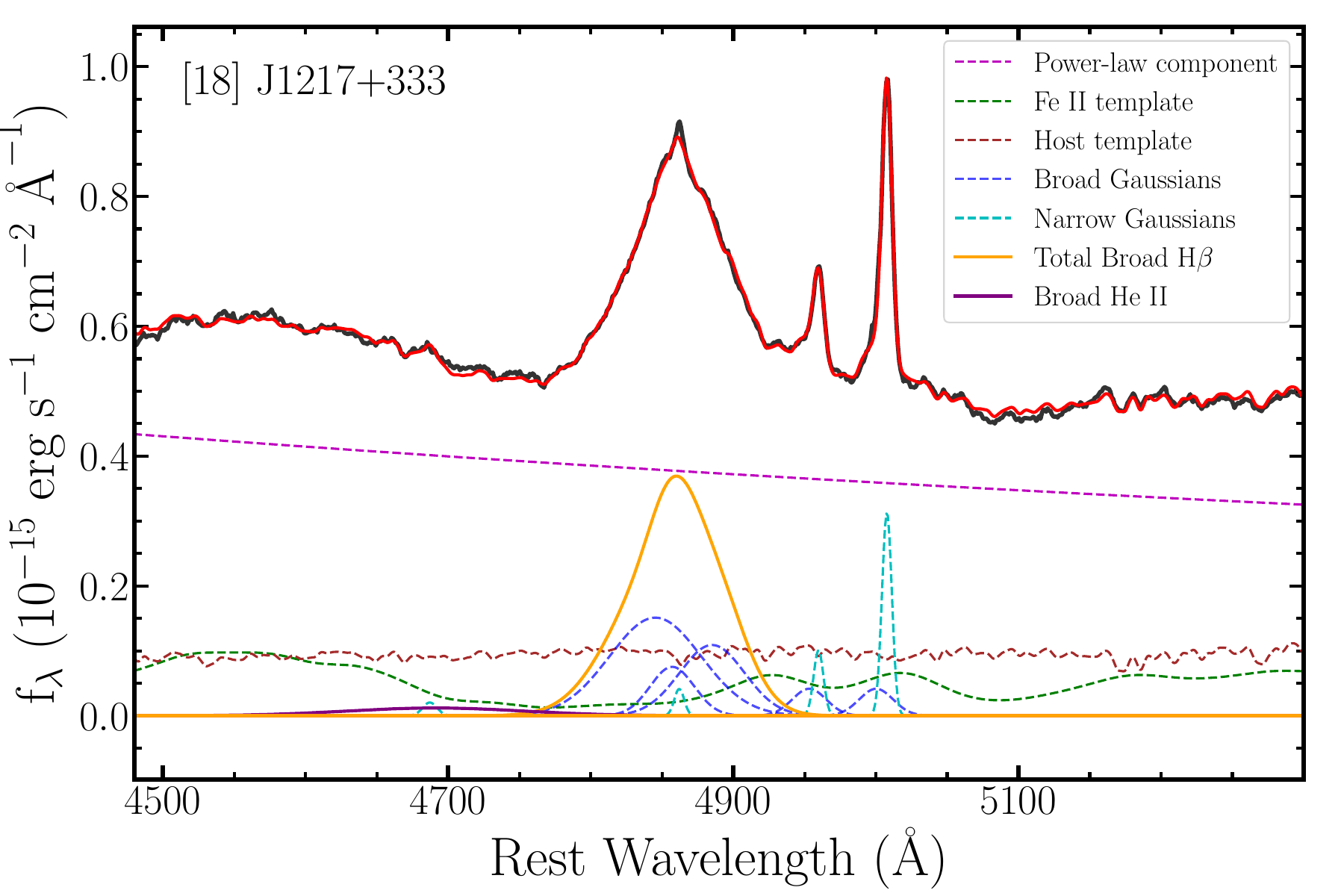}
    \caption{Decomposition of mean spectrum for J1217$+$333 as an example. The black and red solid lines represent the mean spectrum and best-fit model. The continuum model consists of a power-law (magenta dashed) component, a \FeII\ template (green dashed), and a host galaxy template (brown dashed). Broad and narrow Gaussians are displayed using blue and cyan dashed lines, while the total broad \hbeta\ and \HeII\ profiles are shown by orange and purple solid lines.}
    \label{fig:decomposition}
\end{figure}

We perform spectral decomposition on the mean spectrum to remove the host contamination, if any, and measure the line flux and width of the broad \hbeta. The same procedure is adopted to fit the spectrum of each epoch to derive the light curve of the \hbeta\ line flux. Compared to the linear fit of the continuum under the \hbeta\ line profile, the spectral decomposition approach can better isolate the blended components, e.g., \hbeta\ and \FeII. Thus, the decomposition approach has been adopted in a number of recent studies, particularly based on high-quality spectra \citep[e.g.,][]{Park12a, Barth15, Hu15, Hu21, U22}.

We adopt the spectral decomposition procedure of the previous studies by \citet{Shen11, Shen19} with slight modifications. Here, we summarize the basics of the fitting procedure. First, we generated a pseudo-continuum model by combining a power law, an \FeII\ component, and a host galaxy stellar population model, in order to fit the observed spectra within two line-free windows, i.e., [4450,4600] and [5050, 5550] \AA\ in the rest-frame. We apply velocity shift and Gaussian velocity broadening to the iron and host templates, respectively. For the \FeII\ model, we test two commonly used templates provided by \citet{Boroson_Green_92} and \citet{Kovacevic10}. 
In the case of the template of \citet{Kovacevic10}, it is difficult to use consistent flux ratios of different emission line groups for individual epochs, because of the variation of the wavelength coverage due to the different spectroscopy setups of the MDM and Lick observations. 
Thus, we decide to use the \FeII\ template of \citet{Boroson_Green_92}. However, we find that the \FeII\ template of \citet{Kovacevic10} provides a more accurate fit for several objects
\citep[e.g., Mrk~1501 and Mrk~1014, see][]{Barth15, U22}. Thus, we adopt the fits using \citet{Kovacevic10} iron template for these objects. In the case of the host galaxy stellar population model, we used a single-burst 10 Gyr old stellar population model with solar metallicity \citep{Bruzual_Charlot03}. As our targets are moderate to high luminosity AGNs, the host galaxy stellar absorption line features are negligible in the mean spectra of most targets. We present one example of the spectral decomposition using the lowest luminosity AGN, J1217$+$333, in our sample in Figure \ref{fig:decomposition}. As expected from its low AGN luminosity, the host component, i.e., the strong of Mg\,{Ib} $\lambda\lambda$ 5167,5173,5184 triplet absorption, is clearly detected. In contrast, most of the other objects show no clear sign of host galaxy absorption lines in their mean spectrum.

After subtracting the best-fit pseudo continuum model, we fit the residual emission lines with Gaussian models by accounting for various kinematical features of individual emission lines. Specifically, we use three Gaussians for the broad H$\beta$ line and one Gaussian for the narrow H$\beta$ component. For \OIIIa\ and \OIIIb, we use two Gaussian models for the central and wing components. In addition, one narrow and one broad Gaussian models were used for fitting the narrow and broad components of \HeIIopt. All the narrow line centers and widths are tied together.

\begin{table*}[htbp]
    \centering
    \caption{Light curve statistics}
    
    \begin{tabular}{@{\extracolsep{8pt}} r l c c c c c  c c  c c c }
    \hline \hline
    & &  &  \multicolumn{4}{c}{Continuum light curve} & \multicolumn{4}{c}{\hbeta\ light curve} \\
    \cmidrule(r){4-7}  \cmidrule(r){8-11} 
    & Name   &  MJD range & Band &  ${\rm N}_{\rm obs}$  & $\overline{\Delta(T)}$ & $F_{\rm var}$ &  ${\rm N}_{\rm obs}$  & $\overline{\Delta(T)}$ &  $F_{\rm var}$  & $\sigma_{\rm nx}$\\
    & (1)    &  (2) & (3) & (4) &  (5) & (6) & (7) & (8) & (9) & (10) \\
    \hline

1 & Mrk~1501 & 57689-59430 & $B+g$ &375   &  1   &  0.20$\pm$0.01 & 38   &  15   &  0.17$\pm$0.02 & 0.002 \\ 
2 & PG~0026$+$129 & 57689-59395 & $B+g$ &381   &  1   &  0.10$\pm$0.00 & 30   &  19   &  0.06$\pm$0.01 & 0.003 \\ 
3 & PG~0052$+$251 & 57689-59391 & $B+g$ &378   &  2   &  0.20$\pm$0.01 & 37   &  16   &  0.09$\pm$0.01 & 0.002 \\ 
4 & J0101$+$422 & 57327-59430 & $B$ & 256   &  4   &  0.15$\pm$0.01 & 84   &  13   &  0.10$\pm$0.01 & 0.006 \\ 
5 & J0140$+$234 & 57327-59391 & $B$ & 232   &  4   &  0.15$\pm$0.01 & 87   &  14   &  0.10$\pm$0.01 & 0.002 \\ 
6 & Mrk~1014 & 57689-59261 & $B+g$ &330   &  2   &  0.14$\pm$0.01 & 31   &  20   &  0.20$\pm$0.03 & 0.006 \\ 
7 & J0801$+$512 & 57327-59342 & $B$ & 213   &  6   &  0.15$\pm$0.01 & 31   &  25   &  0.11$\pm$0.01 & 0.005 \\ 
8 & J0939$+$375 & 57327-59377 & $B$ & 241   &  5   &  0.16$\pm$0.01 & 57   &  16   &  0.12$\pm$0.01 & 0.008 \\ 
9 & PG~0947$+$396 & 57327-59377 & $B$ & 268   &  4   &  0.15$\pm$0.01 & 72   &  14   &  0.08$\pm$0.01 & 0.004 \\ 
10 & J1026$+$523 & 57327-59386 & $B$ & 272   &  5   &  0.21$\pm$0.01 & 73   &  15   &  0.14$\pm$0.01 & 0.003 \\ 
11 & J1059$+$665 & 57359-59377 & $B$ & 214   &  6   &  0.10$\pm$0.01 & 36   &  28   &  0.08$\pm$0.01 & 0.011 \\ 
12 & PG~1100$+$772 & 57327-59386 & $B$ & 241   &  5   &  0.08$\pm$0.00 & 40   &  28   &  0.11$\pm$0.01 & 0.002 \\ 
13 & J1105$+$671 & 57351-59366 & $V$ & 204   &  6   &  0.17$\pm$0.01 & 29   &  29   &  0.13$\pm$0.02 & 0.009 \\ 
14 & J1120$+$423 & 57328-59387 & $B$ & 260   &  4   &  0.18$\pm$0.01 & 48   &  16   &  0.11$\pm$0.01 & 0.003 \\ 
15 & PG~1121$+$422 & 57328-59402 & $B$ & 234   &  4   &  0.13$\pm$0.01 & 61   &  17   &  0.05$\pm$0.01 & 0.004 \\ 
16 & J1203$+$455 & 57345-59386 & $B$ & 232   &  4   &  0.12$\pm$0.01 & 49   &  18   &  0.11$\pm$0.01 & 0.005 \\ 
17 & PG~1202$+$281 & 57345-59402 & $B$ & 250   &  4   &  0.16$\pm$0.01 & 84   &  14   &  0.08$\pm$0.01 & 0.002 \\ 
18 & J1217$+$333 & 57334-59386 & $B$ & 247   &  4   &  0.20$\pm$0.01 & 60   &  15   &  0.14$\pm$0.02 & 0.014 \\ 
19 & VIII~Zw~218 & 57346-59402 & $B$ & 240   &  4   &  0.15$\pm$0.01 & 79   &  14   &  0.11$\pm$0.01 & 0.003 \\ 
20 & PG~1322$+$659 & 57361-59404 & $B$ & 249   &  4   &  0.14$\pm$0.01 & 65   &  17   &  0.06$\pm$0.01 & 0.022 \\ 
21 & J1415$+$483 & 57384-59358 & $B$ & 203   &  5   &  0.11$\pm$0.01 & 30   &  21   &  0.07$\pm$0.01 & 0.003 \\ 
22 & PG~1427$+$480 & 57388-59395 & $B$ & 227   &  5   &  0.11$\pm$0.01 & 57   &  14   &  0.05$\pm$0.01 & 0.001 \\ 
23 & PG~1440$+$356 & 57389-59404 & $B$ & 236   &  4   &  0.11$\pm$0.01 & 62   &  15   &  0.12$\pm$0.01 & 0.002 \\ 
24 & J1456$+$380 & 57384-59404 & $B$ & 271   &  4   &  0.12$\pm$0.01 & 79   &  15   &  0.12$\pm$0.01 & 0.004 \\ 
25 & J1526$+$275 & 57384-59404 & $B$ & 237   &  5   &  0.10$\pm$0.01 & 51   &  18   &  0.05$\pm$0.01 & 0.006 \\ 
26 & J1540$+$355 & 57385-59404 & $B$ & 237   &  4   &  0.15$\pm$0.01 & 81   &  16   &  0.06$\pm$0.01 & 0.003 \\ 
27 & PG~1545$+$210 & 57385-59431 & $B$ & 239   &  4   &  0.14$\pm$0.01 & 69   &  18   &  0.06$\pm$0.01 & 0.001 \\ 
28 & PG~1612$+$261 & 57389-59431 & $B$ & 210   &  4   &  0.19$\pm$0.01 & 32   &  17   &  0.10$\pm$0.01 & 0.001 \\ 
29 & J1619$+$501 & 57385-59431 & $B$ & 277   &  4   &  0.11$\pm$0.01 & 84   &  15   &  0.08$\pm$0.01 & 0.004 \\ 
30 & J1935$+$531 & 57333-59430 & $B+g$ &778   &  1   &  0.07$\pm$0.00 & 81   &  14   &  0.07$\pm$0.01 & 0.006 \\ 
31 & PG~2251$+$113 & 57711-59404 & $V+g$ &295   &  2   &  0.07$\pm$0.00 & 28   &  17   &  0.06$\pm$0.01 & 0.002 \\ 
32 & PG~2349$-$014 & 57689-59392 & $B+g$ &233   &  3   &  0.18$\pm$0.01 & 34   &  16   &  0.18$\pm$0.02 & 0.005 \\ 
\hline
\multicolumn{11}{p{0.99\textwidth}}{
{\bf Notes.} 
Column (1): object name. Column (2): monitoring time baseline in MJD. Column (3): Continuum band. Column (4): number of epochs in photometry. 
Column (5): Median cadence in days.
Column (6): Noise-corrected fractional variability $F_{\rm var}$ and its uncertainty as defined in equation \ref{equ:fvar} and \ref{equ:e_fvar}. 
Column (7): number of epochs in spectroscopy.
Column (8): Median cadence in days.
Column (9): Noise-corrected fractional variability $F_{\rm var}$ and its uncertainty.
Column (10): normalized excess standard deviation of \OIII\ flux after spectral calibration by {\tt mapspec} (see \S \ref{sec:mapspec}).
}

    \end{tabular}
    \label{tab:varibility_lcstats}
\end{table*}

\subsection{\hbeta\ Light Curves}

\begin{table}[]
    \centering
    \caption{\hbeta\ line flux extraction window and average flux}   \label{tab:flux_window}

    \begin{tabular}{@{\extracolsep{5pt}}r l c c}
    \hline\hline
    & (1) & (2) & (3) \\
     & Name   &  Window  &  $\overline{F_{\rm H\beta}}$  \\ \hline
 1 & Mrk~1501 & 4810-4929 & 156$\pm$26  \\ 
2 & PG~0026$+$129 & 4783-4941 & 168$\pm$11  \\ 
3 & PG~0052$+$251 & 4790-4935 & 186$\pm$17  \\ 
4 & J0101$+$422 & 4778-4945 & 94$\pm$11  \\ 
5 & J0140$+$234 & 4810-4915 & 67$\pm$7 \\ 
6 & Mrk~1014 & 4806-4918 & 103$\pm$21  \\ 
7 & J0801$+$512 & 4814-4910 & 16$\pm$2  \\ 
8 & J0939$+$375 & 4808-4916 & 23$\pm$3  \\ 
9 & PG~0947$+$396 & 4787-4946 & 74$\pm$6  \\ 
10 & J1026$+$523 & 4799-4925 & 34$\pm$5  \\ 
11 & J1059$+$665 & 4790-4934 & 37$\pm$3  \\ 
12 & PG~1100$+$772 & 4779-5050 & 159$\pm$18  \\ 
13 & J1105$+$671 & 4790-4935 & 25$\pm$3  \\ 
14 & J1120$+$423 & 4779-4945 & 45$\pm$5  \\ 
15 & PG~1121$+$422 & 4810-4914 & 108$\pm$7  \\ 
16 & J1203$+$455 & 4769-4956 & 32$\pm$4  \\ 
17 & PG~1202$+$281 & 4779-4945 & 76$\pm$7  \\ 
18 & J1217$+$333 & 4795-4930 & 34$\pm$6  \\ 
19 & VIII~Zw~218 & 4778-4946 & 122$\pm$14  \\ 
20 & PG~1322$+$659 & 4811-4913 & 129$\pm$9  \\ 
21 & J1415$+$483 & 4805-4919 & 29$\pm$2  \\ 
22 & PG~1427$+$480 & 4812-4912 & 79$\pm$4  \\ 
23 & PG~1440$+$356 & 4821-4903 & 238$\pm$29  \\ 
24 & J1456$+$380 & 4764-4960 & 27$\pm$3  \\ 
25 & J1526$+$275 & 4796-4928 & 55$\pm$4 \\ 
26 & J1540$+$355 & 4814-4911 & 65$\pm$5  \\ 
27 & PG~1545$+$210 & 4769-4996 & 132$\pm$8  \\ 
28 & PG~1612$+$261 & 4795-4929 & 197$\pm$21  \\ 
29 & J1619$+$501 & 4783-4941 & 37$\pm$3  \\ 
30 & J1935$+$531 & 4775-4920 & 51$\pm$4  \\ 
31 & PG~2251$+$113 & 4772-4952 & 205$\pm$14  \\ 
32 & PG~2349$-$014 & 4748-4994 & 102$\pm$18  \\

\hline
\multicolumn{4}{p{0.44\textwidth}}{{\bf Notes.} 
Column (1): object name. 
Column (2): Extraction window for the line flux measurements in the rest-frame in the unit of \AA.
Column (3): average flux ($\overline{F_{\rm H\beta}}$) of \hbeta\ light curves as well as its standard deviation in the unit of 10$^{-15}$ erg s$^{-1}$ cm$^{-2}$.
}
\end{tabular}

\end{table}

We measure the \hbeta\ line flux based on the decomposition and generated the light curves over the 6-year monitoring period. Note that we apply a window to each epoch's spectrum, within which the \hbeta\ line flux was summed, as listed in Table \ref{tab:flux_window}. These additional windows are added to avoid any over-fitting to the noise at the edge of the line profile.  

Note that while both Lick and MDM spectra were aligned to the same reference spectrum during the flux re-calibration process using the non-varying \OIII\ emission line, there still can be a small offset between two sets of the light curves from two different telescopes, Lick and MDM, due to slightly different aperture effects \citep{Peterson95}. We further match the two light curves from Lick and MDM telescopes by assuming the spectra obtained at similar time should have the same flux level. By searching for pairs of close epochs within one day between the Lick and MDM observations, we calculated the median scaling ratio between the two spectra using these pairs. Finally, the MDM light curves are scaled with these ratios to match with the Lick light curves.

For each object, we calculate the noise-corrected fractional variability $F_{\rm var}$ of both continuum and \hbeta\ light curves as this quantity is frequently adopted to represent the level of variability \citep[e.g.,][]{Peterson04, Bentz09c}. The $F_{\rm var}$ and its uncertainty is defined by \citep{Rodriguez-Pascual97, Edelson02} as:
\begin{equation}
    F_{\rm var}=\frac{1}{\langle f\rangle}\sqrt{\sigma^2-\delta^2} \label{equ:fvar}
\end{equation}
\begin{equation}
    \sigma_{F_{\rm var}}=\frac{1}{\sqrt{2N}F_{\rm var}} \frac{\sigma^2}{f^2},
    \label{equ:e_fvar}
\end{equation}
where $N$ is the number of epochs, the $\langle f\rangle$ and $\sigma$ are the mean flux and standard deviation of the light curve, and the $\delta$ is the rms of individual flux uncertainties. We list the $F_{\rm var}$ of continuum and \hbeta\ in Table \ref{tab:varibility_lcstats}. For our sample, $F_{\rm var}$ of continuum ranges from 0.07 to 0.21, while the $F_{\rm var}$ of \hbeta\ ranges from 0.05 to 0.20. Note that 30 out of 32 objects in the sample show continuum $F_{\rm var} \geq 0.1$, indicating clear detection of the variability of the majority of the targets.

\begin{figure*}
\centering
\includegraphics[width=0.95\textwidth]{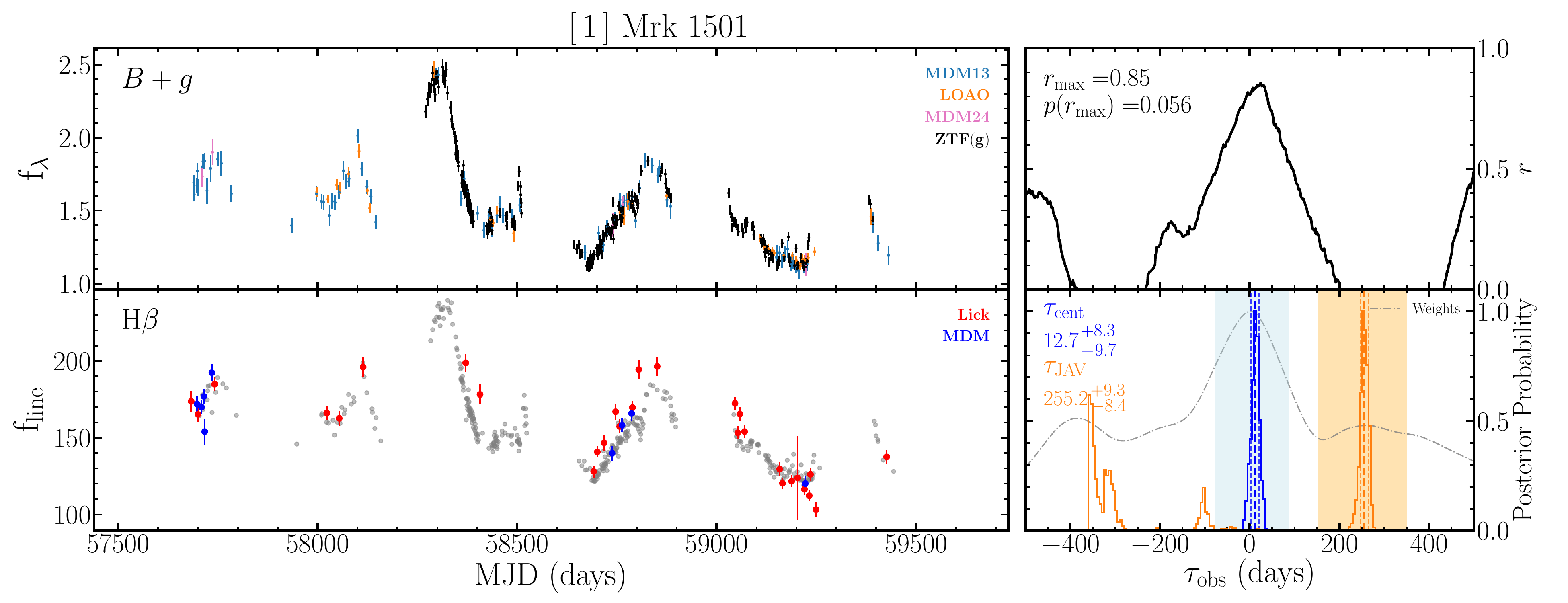}
\includegraphics[width=0.95\textwidth]{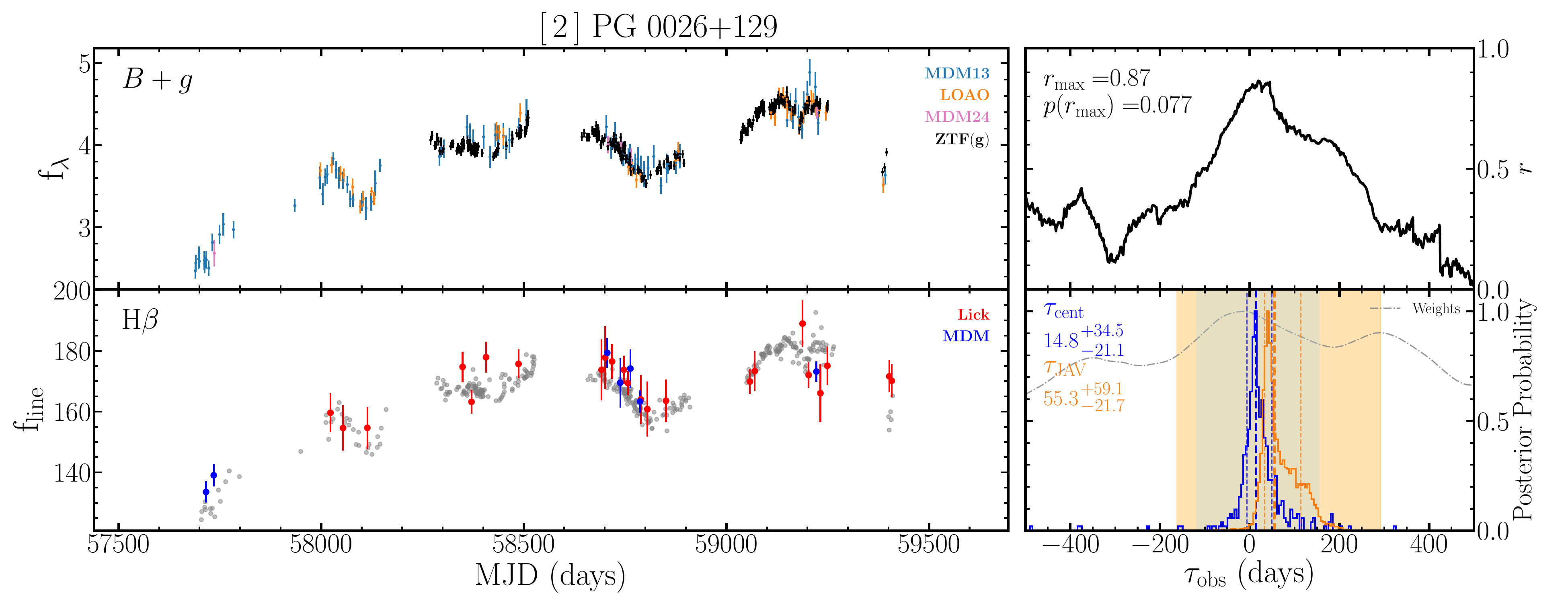}
\includegraphics[width=0.95\textwidth]{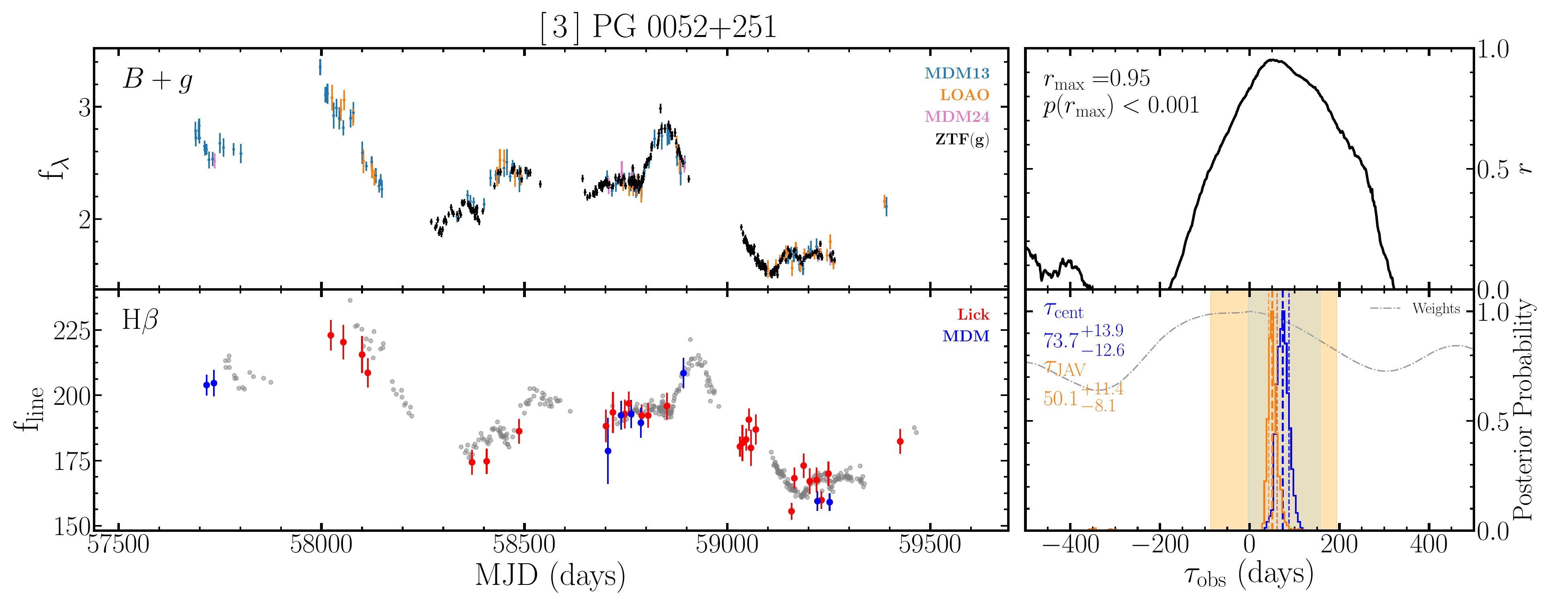}
\caption{Light curves of the continuum (upper left) and \hbeta\ (lower left)  in units of $10^{-15}\,$erg s$^{-1}\,$cm$^{-2}\,$\AA$^{-1}$ and $10^{-15}\,$erg s$^{-1}\,$cm$^{-2}$, respectively, along with the ICCF (upper right) and the posterior distribution in the observed-frame (lower right) for each AGN. The unweighted posterior probability distribution of ICCF $\tau_{\rm cent}$ (blue) and {\tt JAVELIN} $\tau_{\rm JAV}$ (orange) as well as the applied weight (dotted-dashed line) for searching the primary peak (see \S \ref{sec:lag-measurements}) are presented in the lower right panel. The blue and orange shadowed areas indicate the range of the primary peak for $\tau_{\rm cent}$ and $\tau_{\rm JAV}$, respectively. The vertical solid and dashed lines represent the location of the lag and its upper and lower limit calculated as the median, 16th, and 84th percentile within the primary peak range, respectively. $\tau_{\rm cent}$ is adopted as the final lag measurement, by which the continuum light curve is shifted and matched with the \hbeta\ light curves as visualized by grey points in the lower left panel. The two lag reliability indicators, i.e., the $r_{\rm max}$ and $p(r_{\rm max})$ (see \S \ref{sec:lag-reliability} for details), are displayed in the upper right panel.   }\label{fig:LC_1}
\end{figure*}

\begin{figure*}[htbp]
\centering
\includegraphics[width=0.95\textwidth]{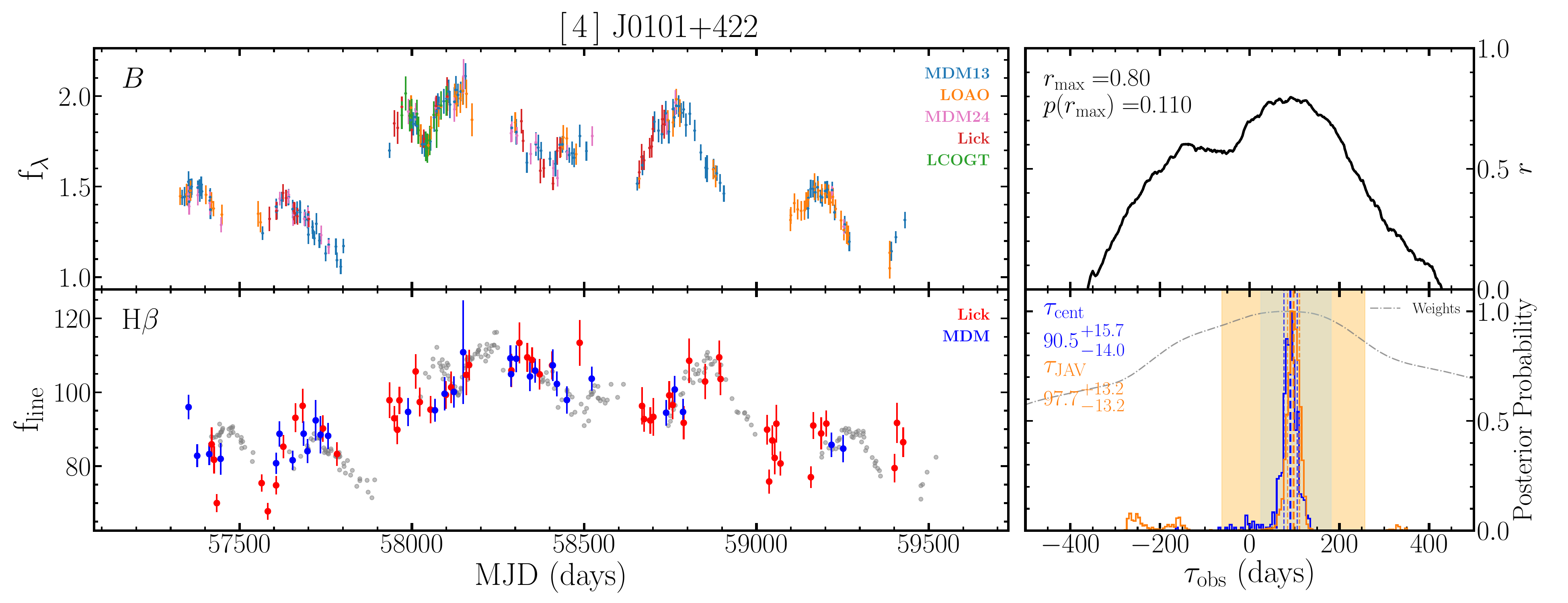}
\includegraphics[width=0.95\textwidth]{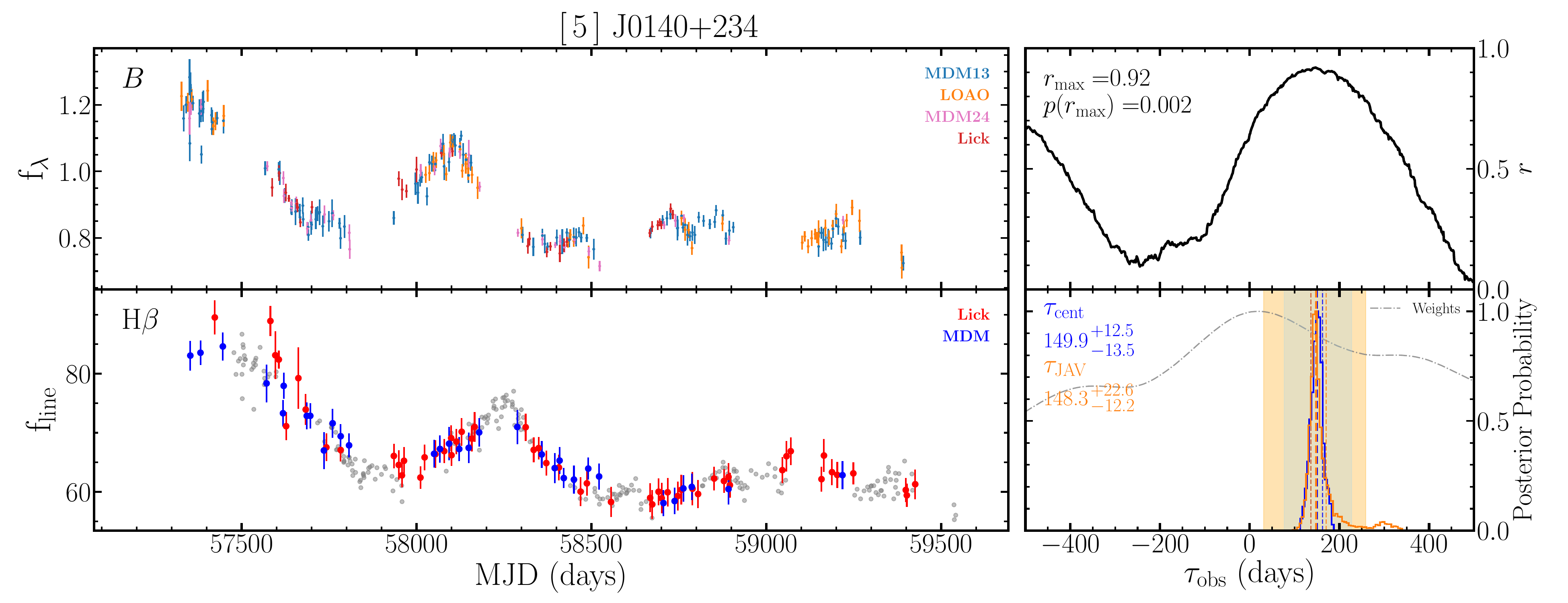}
\includegraphics[width=0.95\textwidth]{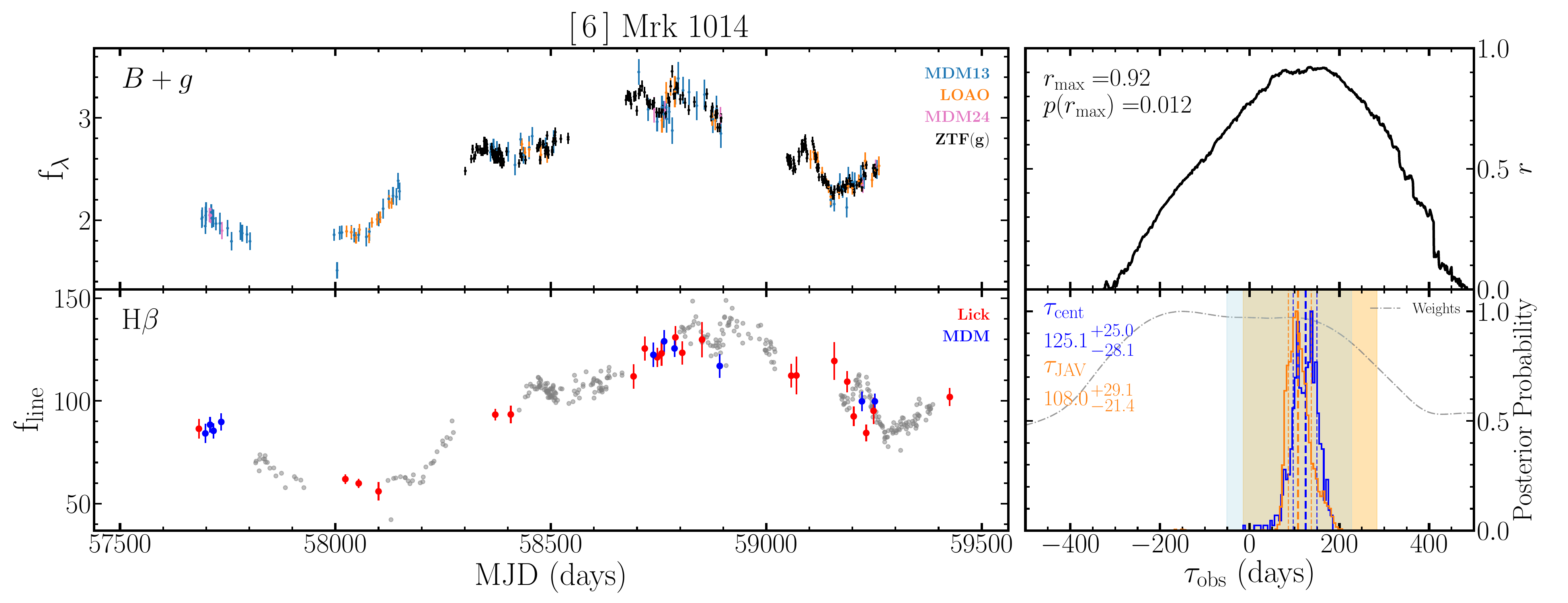}

\caption{Same as Figure \ref{fig:LC_1} but for J0101$+$422, Mrk~1014, and J0801$+$512.}
\label{fig:LC_3}
\end{figure*}

\begin{figure*}[htbp]
\centering
\includegraphics[width=0.95\textwidth]{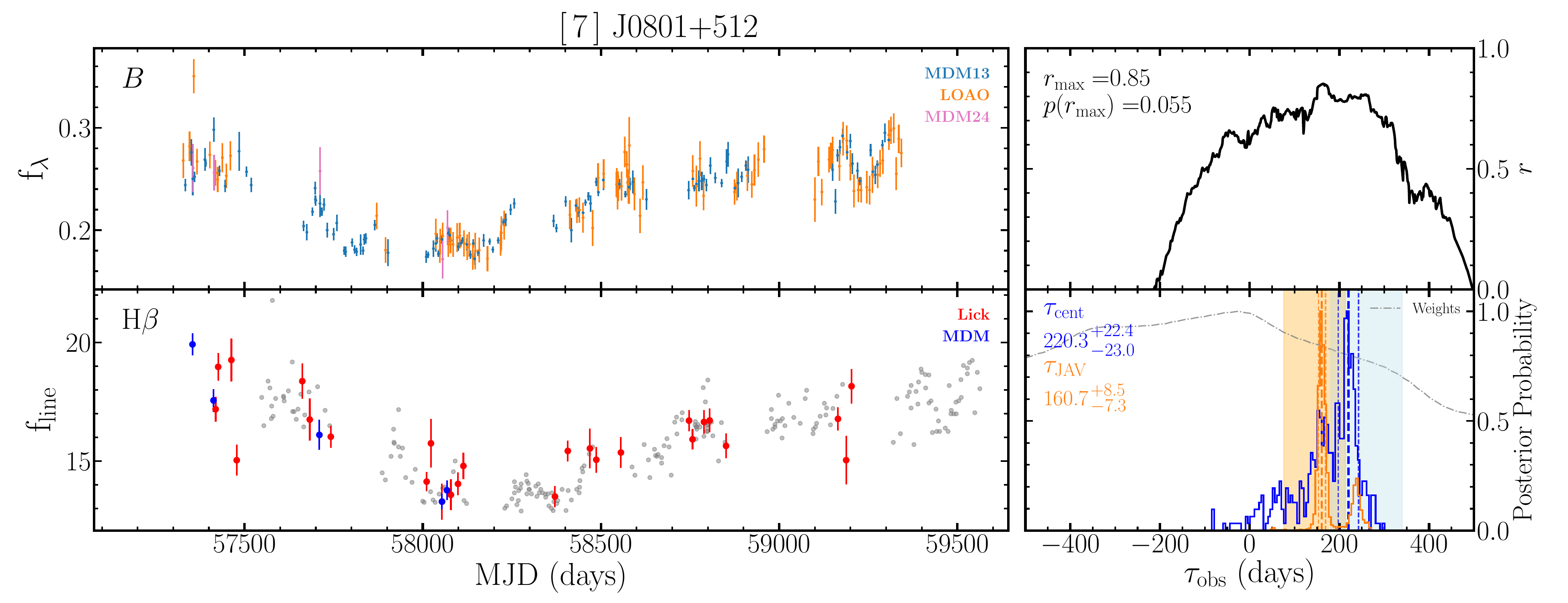}
\includegraphics[width=0.95\textwidth]{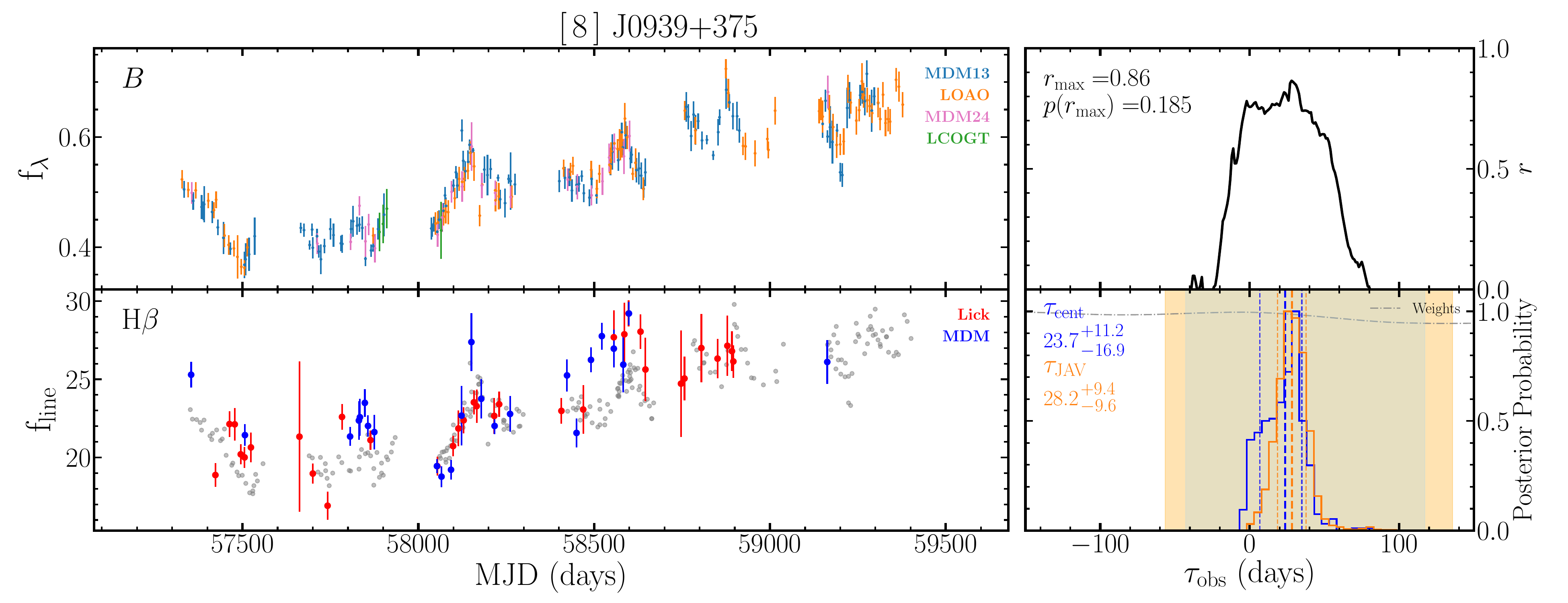}
\includegraphics[width=0.95\textwidth]{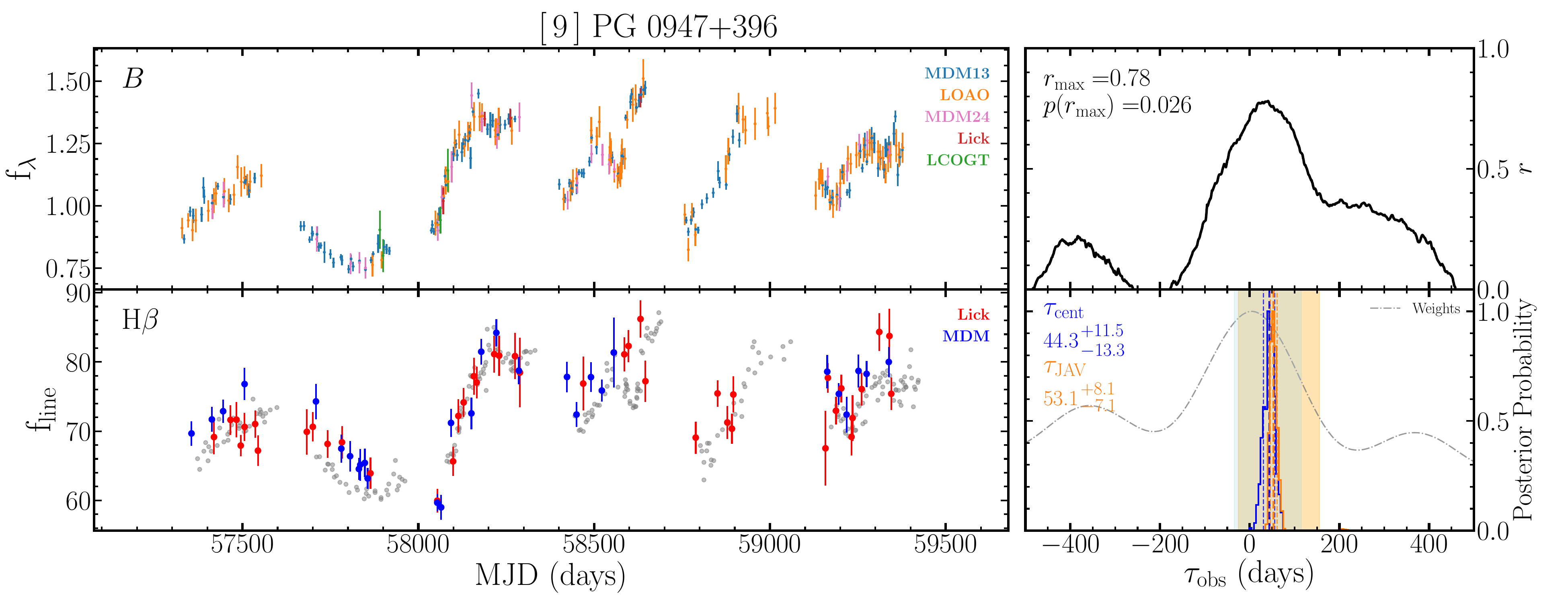}
\caption{Same as Figure \ref{fig:LC_1} but for J0801+512, J0939+375, and PG0947+396.}
\label{fig:LC_4}
\end{figure*}

\begin{figure*}[htbp]
\centering
\includegraphics[width=0.95\textwidth]{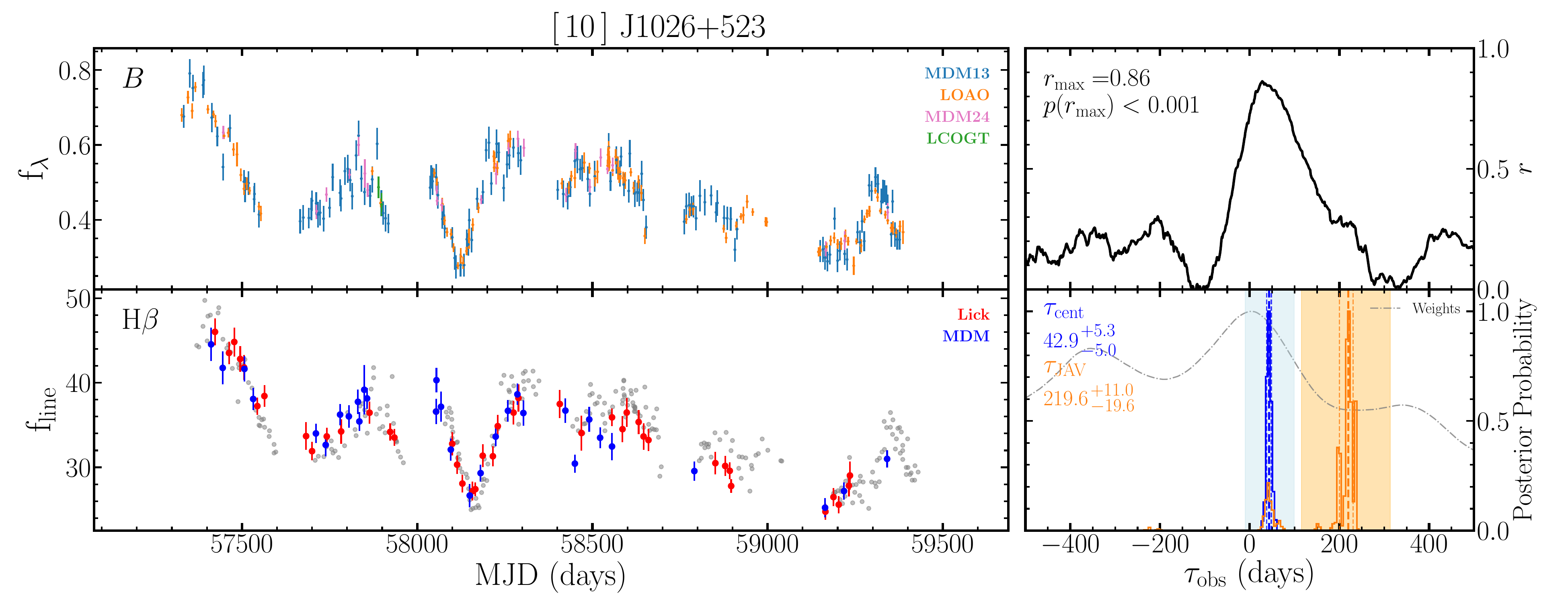}
\includegraphics[width=0.95\textwidth]{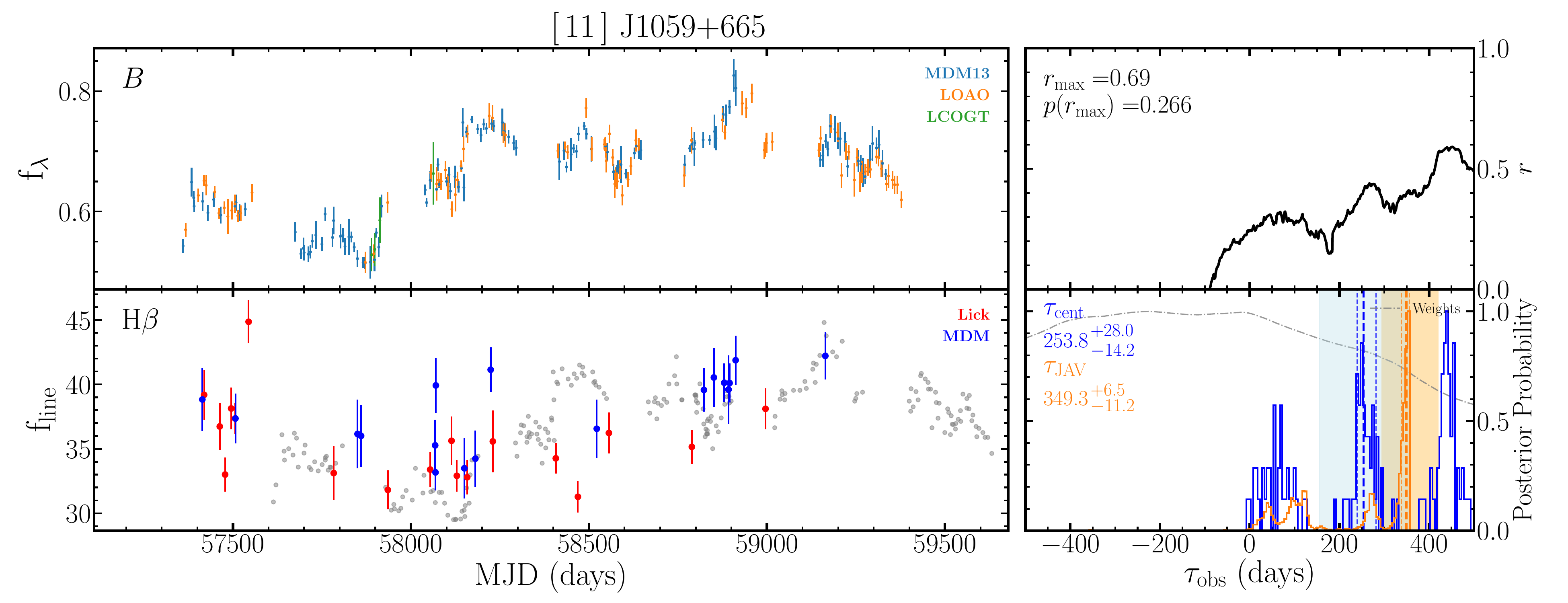}
\includegraphics[width=0.95\textwidth]{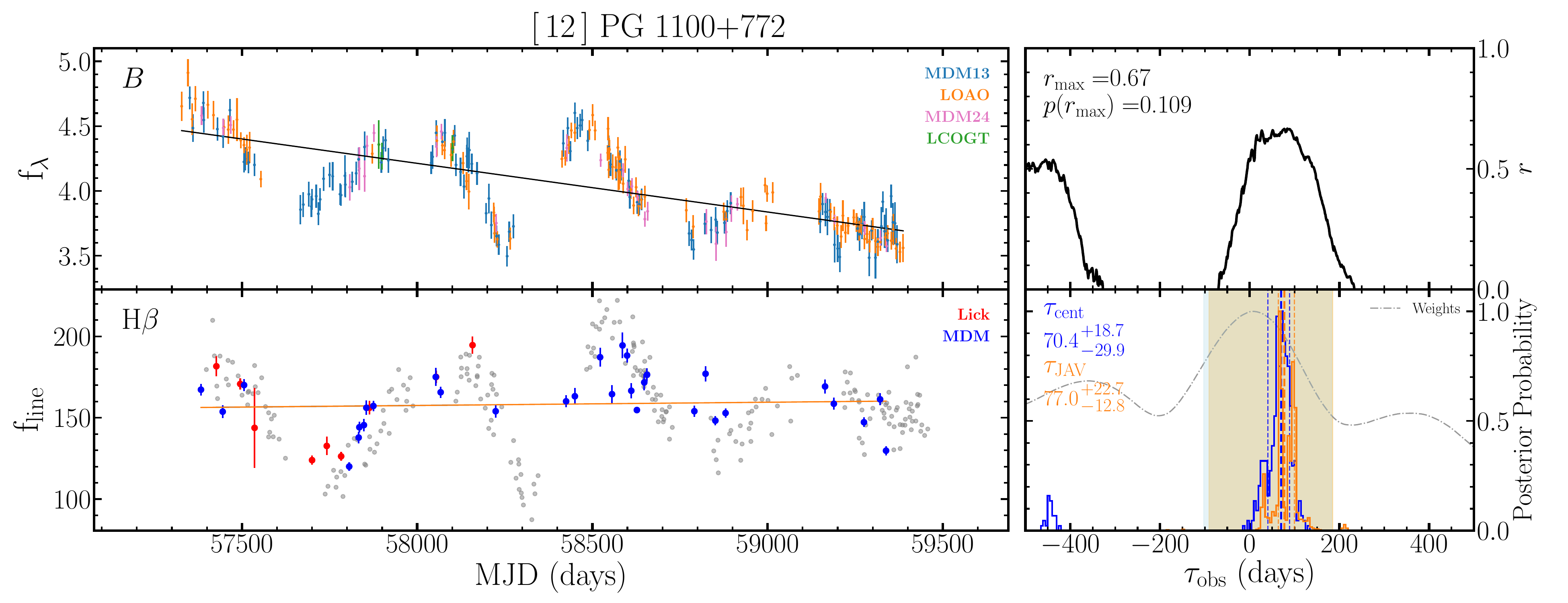}
\caption{Same as Figure \ref{fig:LC_1} but for J1026+523, J1059+665, and PG 1100+772.}
\label{fig:LC_4}
\end{figure*}

\begin{figure*}[htbp]
\centering
\includegraphics[width=0.95\textwidth]{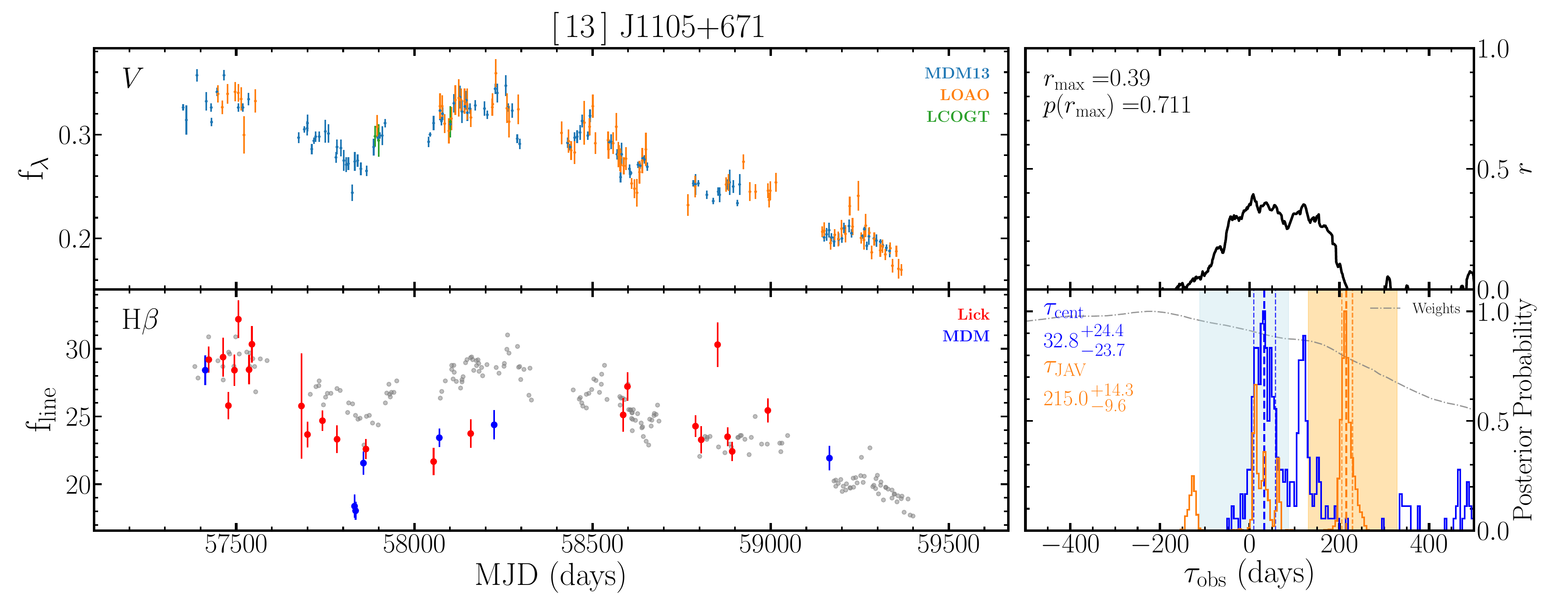}
\includegraphics[width=0.95\textwidth]{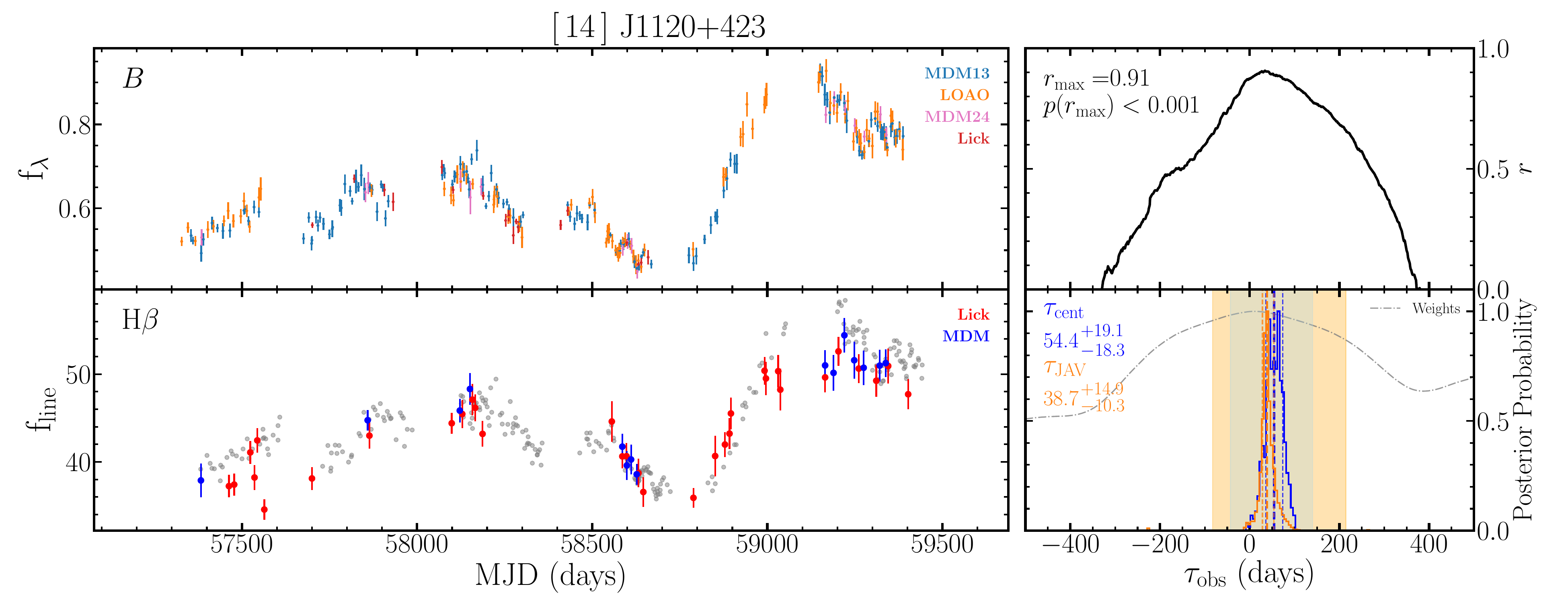}
\includegraphics[width=0.95\textwidth]{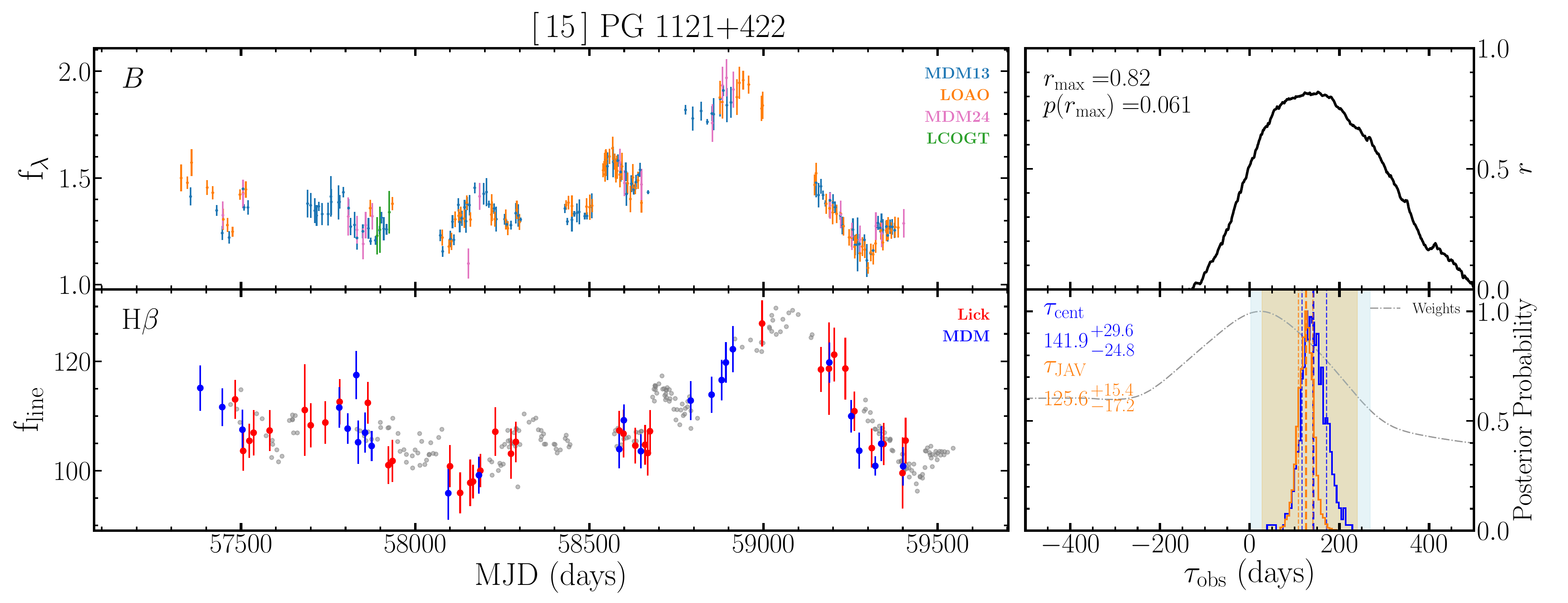}
\caption{Same as Figure \ref{fig:LC_1} but for J1105+671, J1120+423, and PG 1121+422.}
\label{fig:LC_5}
\end{figure*}

\begin{figure*}[htbp]
\centering

\includegraphics[width=0.95\textwidth]{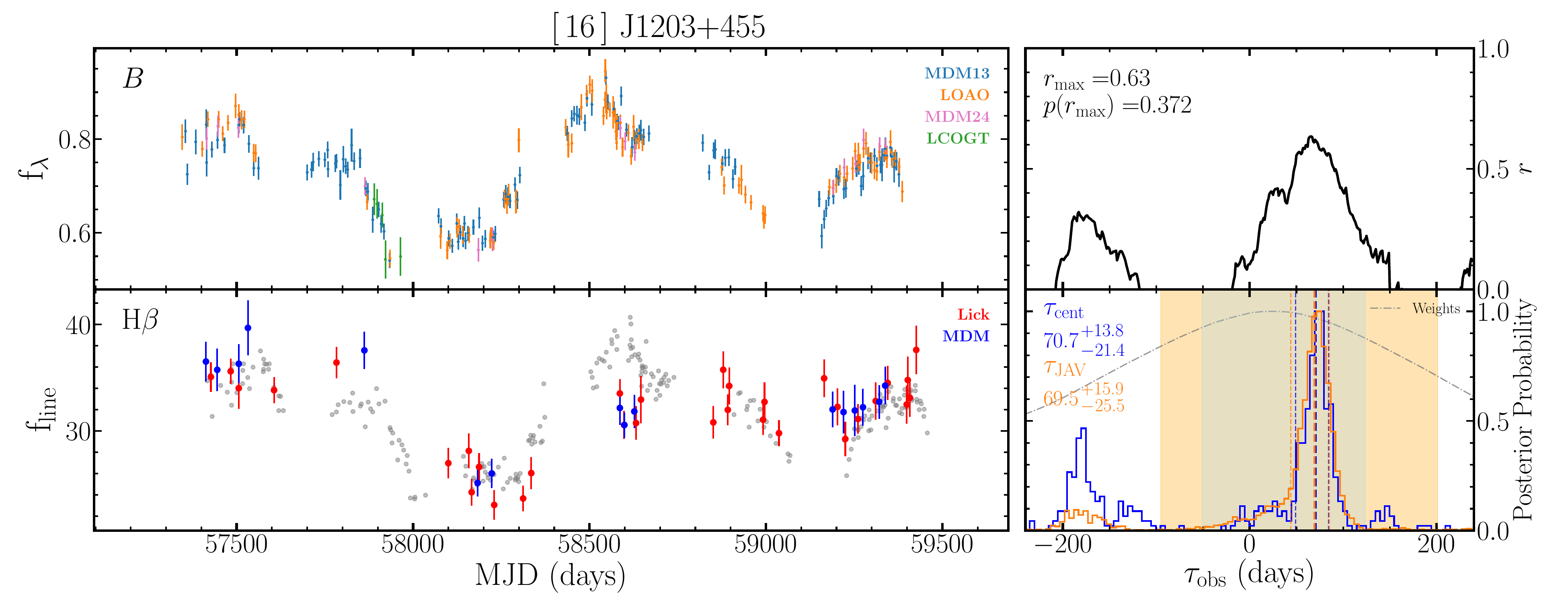}
\includegraphics[width=0.95\textwidth]{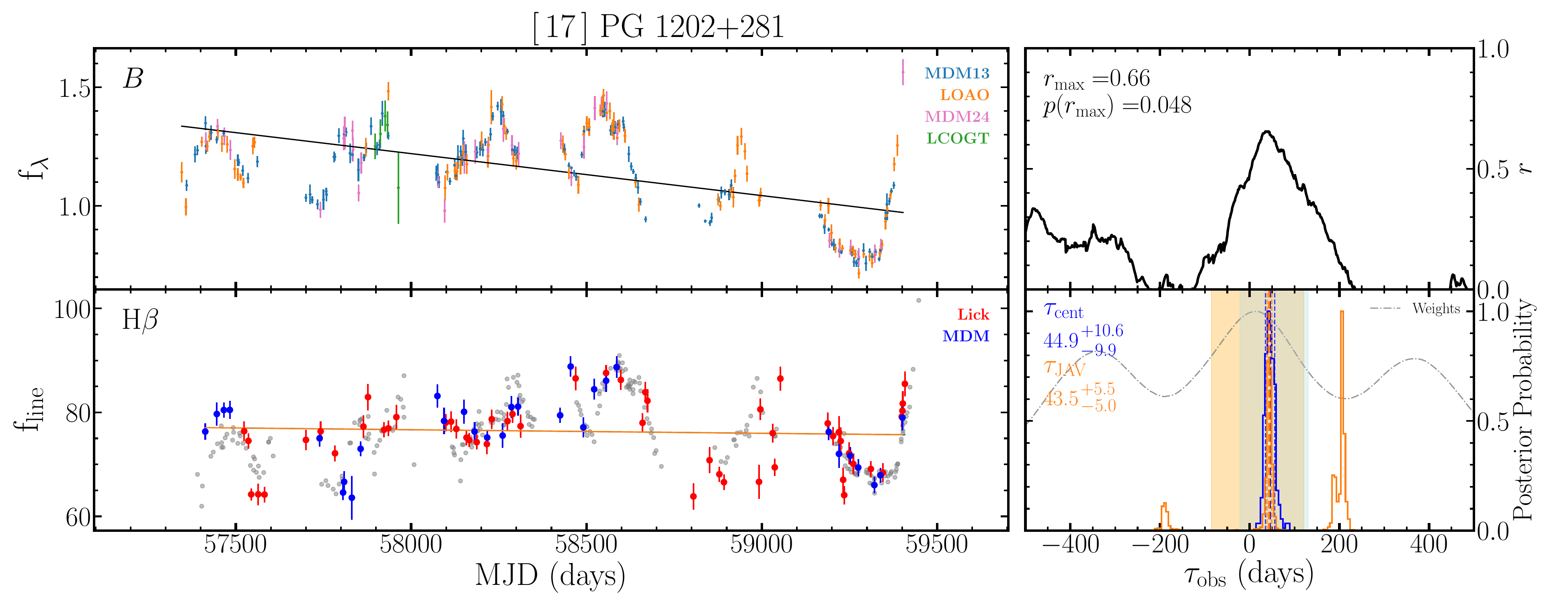}
\includegraphics[width=0.95\textwidth]{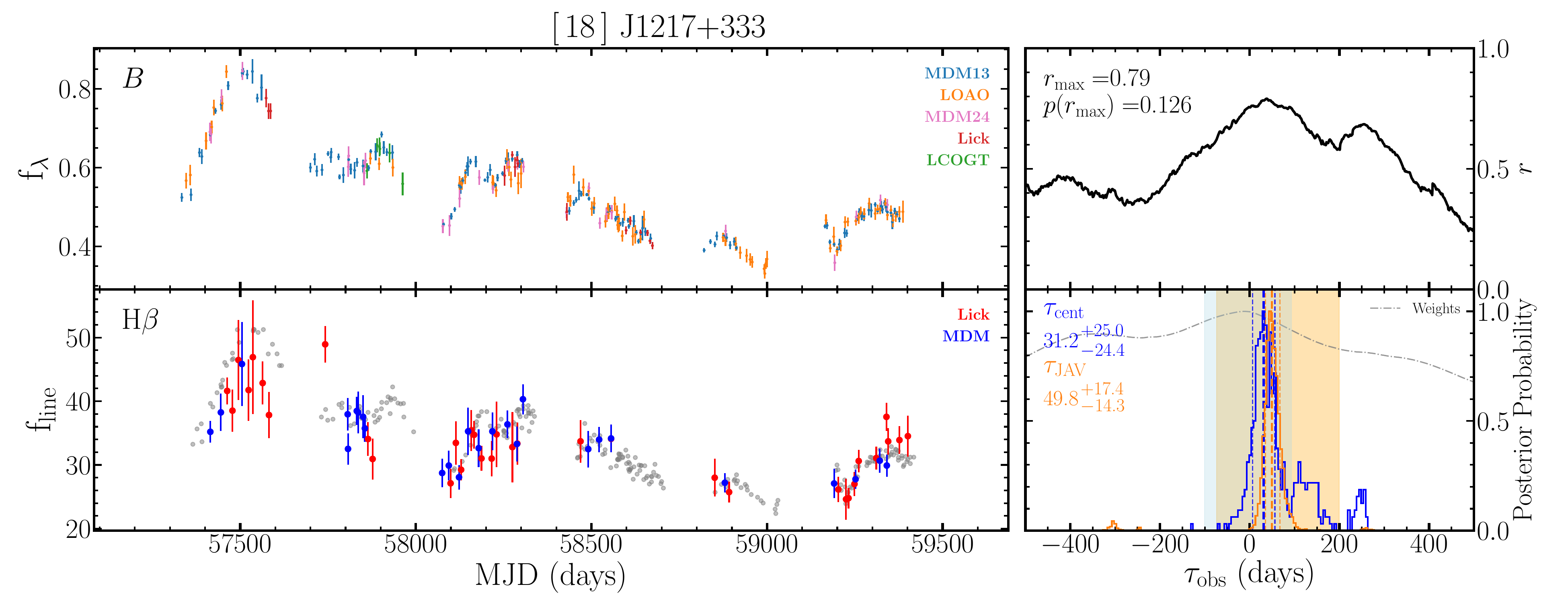}
\label{fig:LC_6}
\caption{Same as Figure \ref{fig:LC_1} but for J1203+455, PG 1202+281, and J1217+333.}
\end{figure*}

\begin{figure*}[htbp]
\centering

\includegraphics[width=0.95\textwidth]{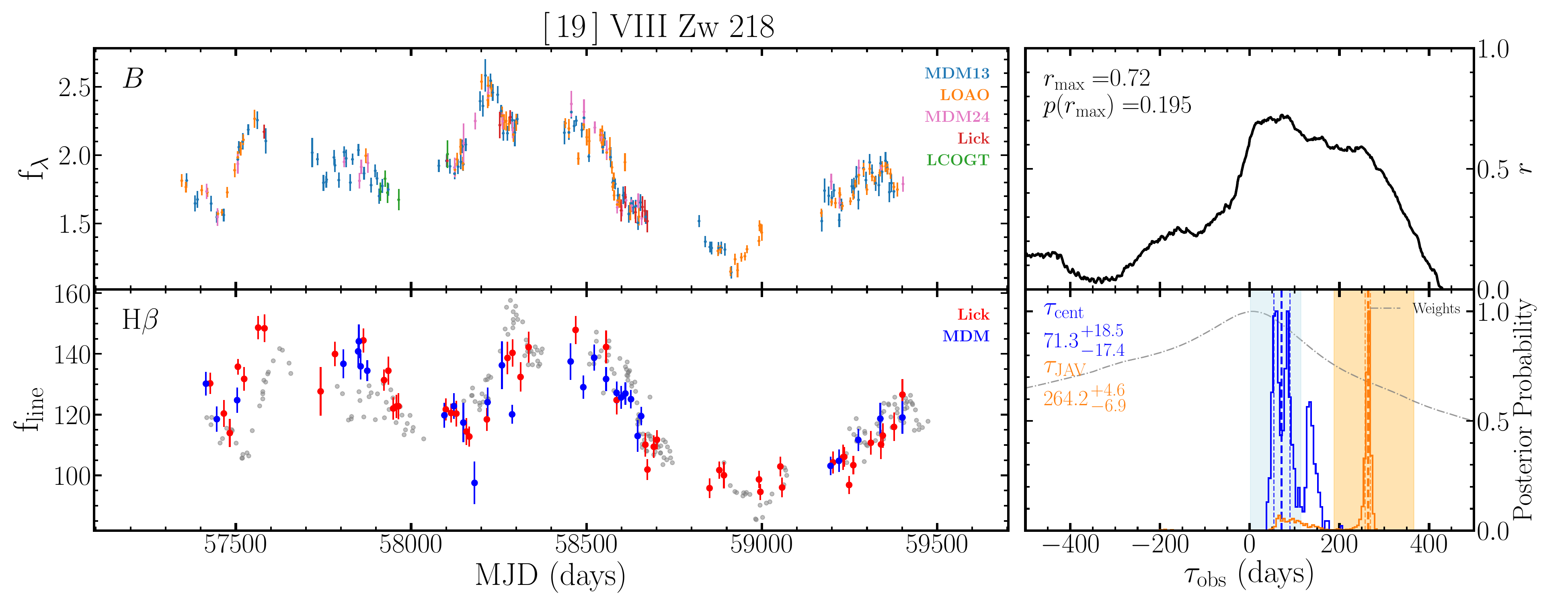}
\includegraphics[width=0.95\textwidth]{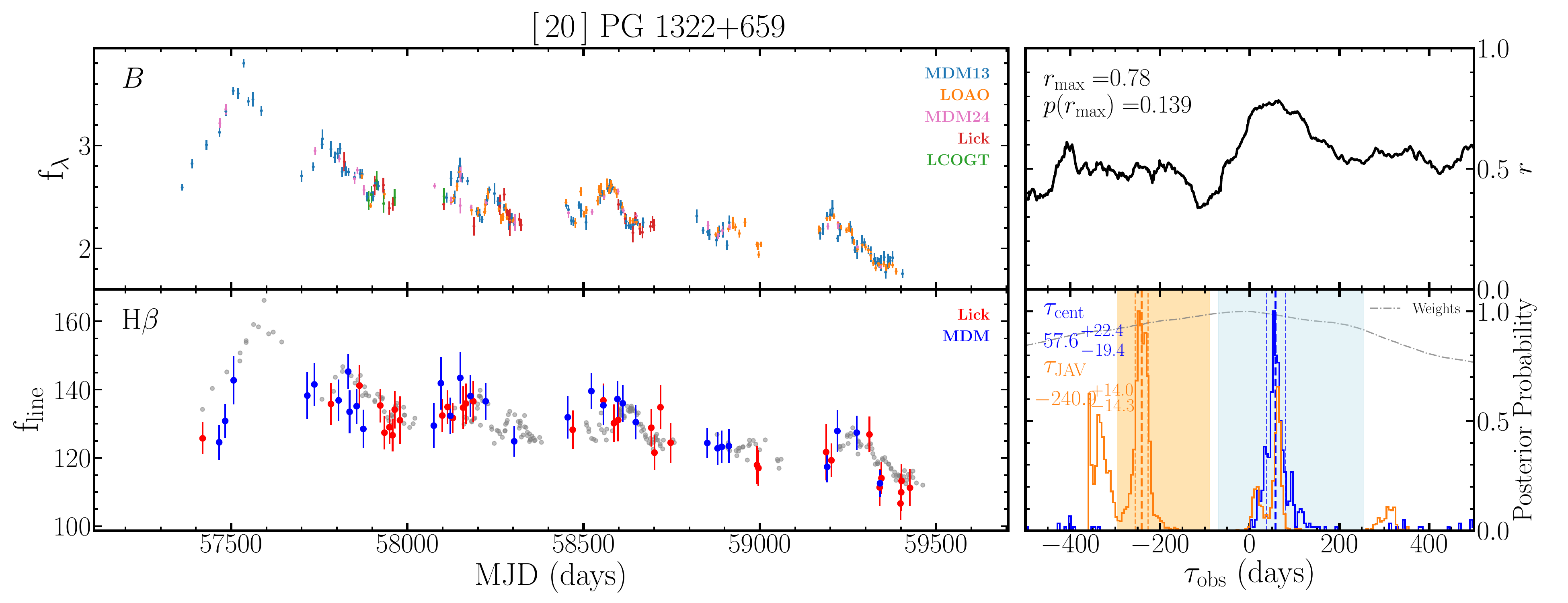}
\includegraphics[width=0.95\textwidth]{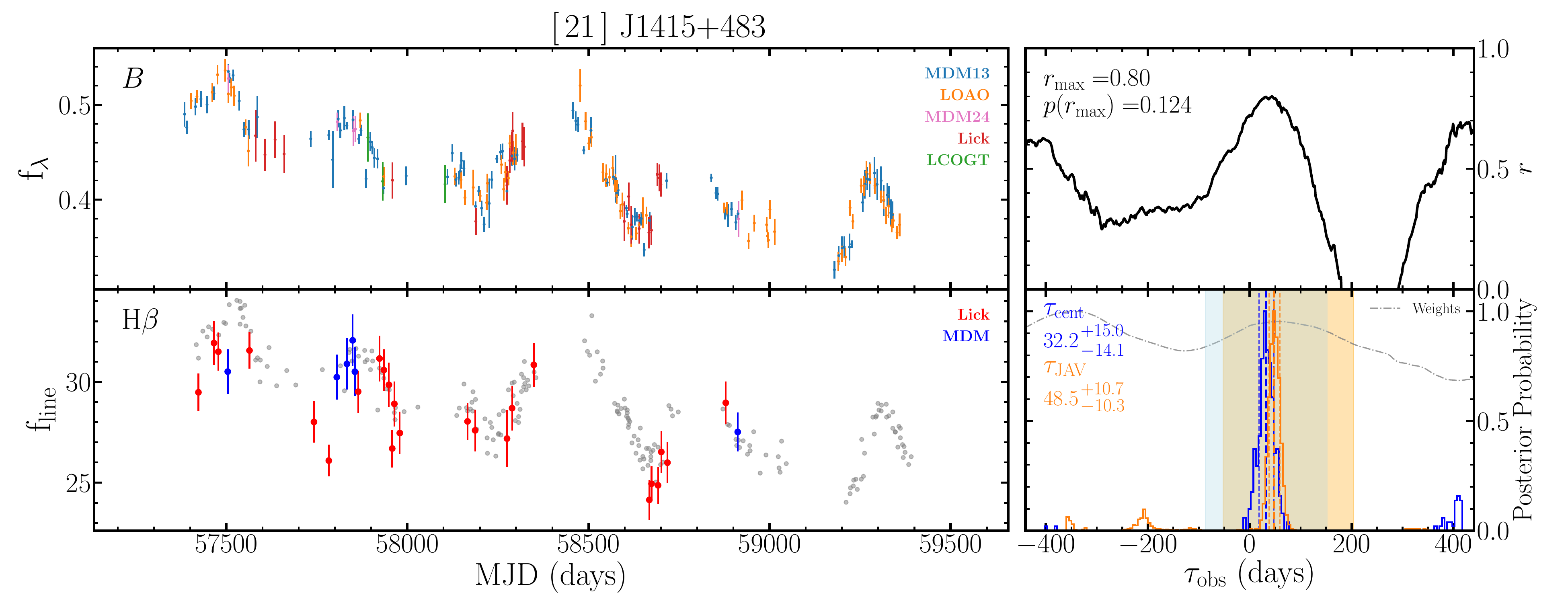}
\caption{Same as Figure \ref{fig:LC_1} but for VIII Zw218, PG 1322+659, and J1415+483.}
\label{fig:LC_7}
\end{figure*}

\begin{figure*}[htbp]
\centering

\includegraphics[width=0.95\textwidth]{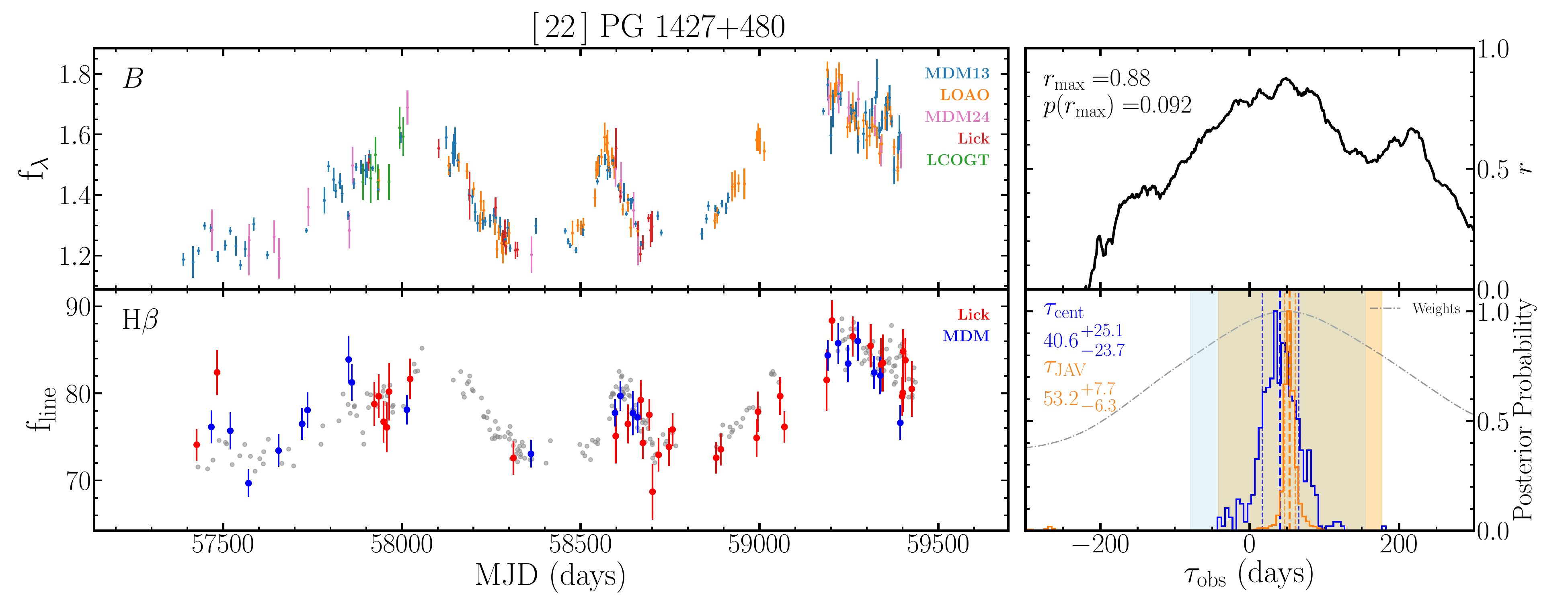}
\includegraphics[width=0.95\textwidth]{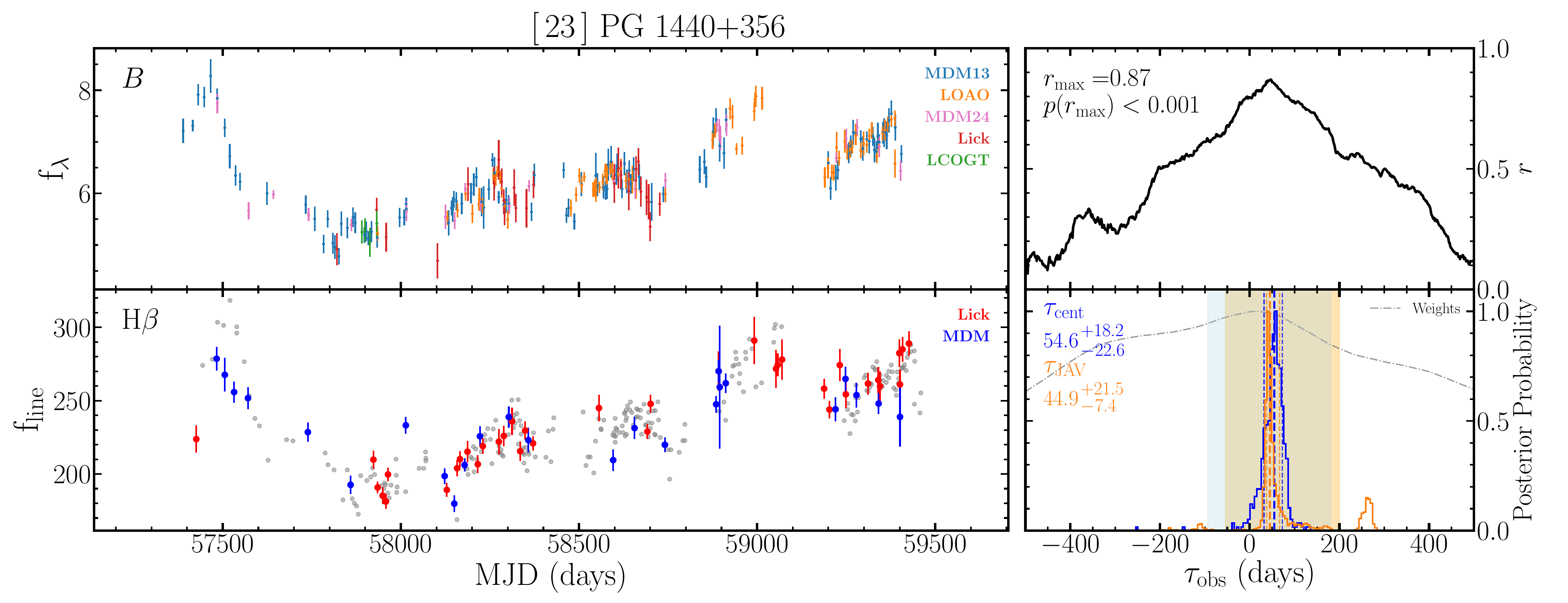}
\includegraphics[width=0.95\textwidth]{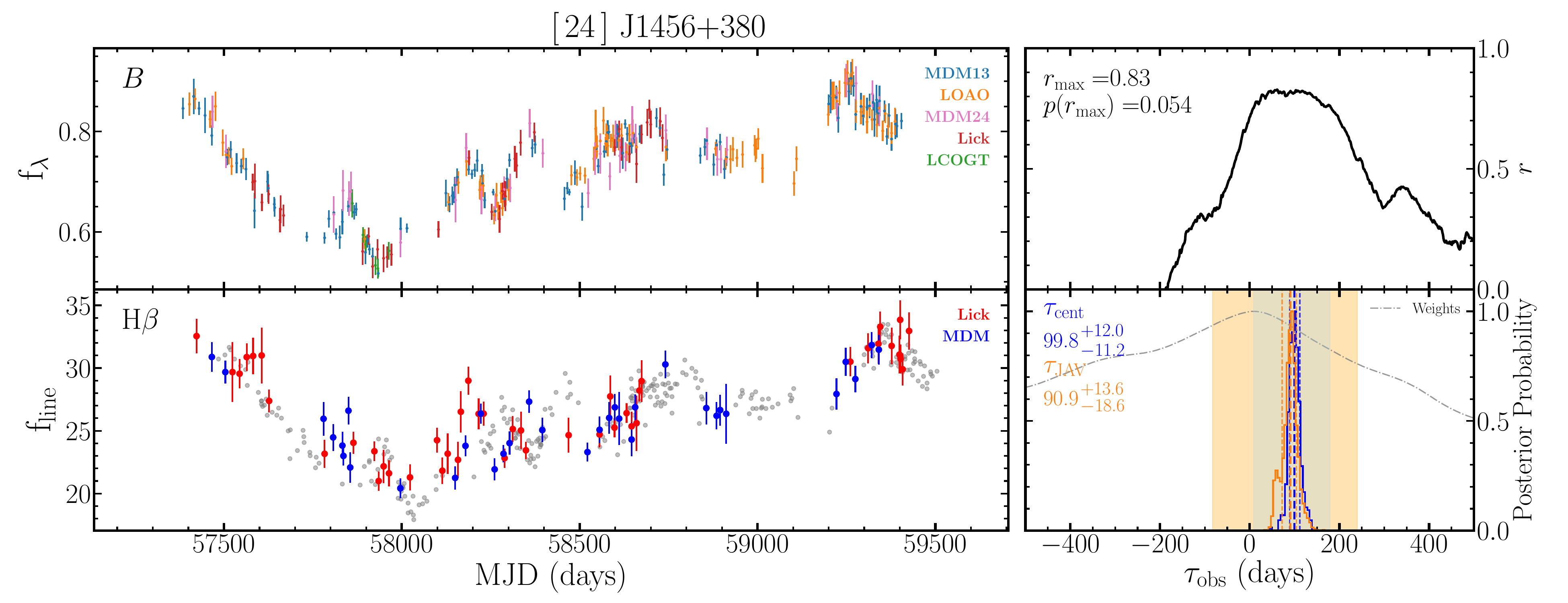}
\caption{Same as Figure \ref{fig:LC_1} but for PG 1427+480, PG 1440+356, and J1456+380.}
\label{fig:LC_8}
\end{figure*}

\begin{figure*}[htbp]
\centering

\includegraphics[width=0.95\textwidth]{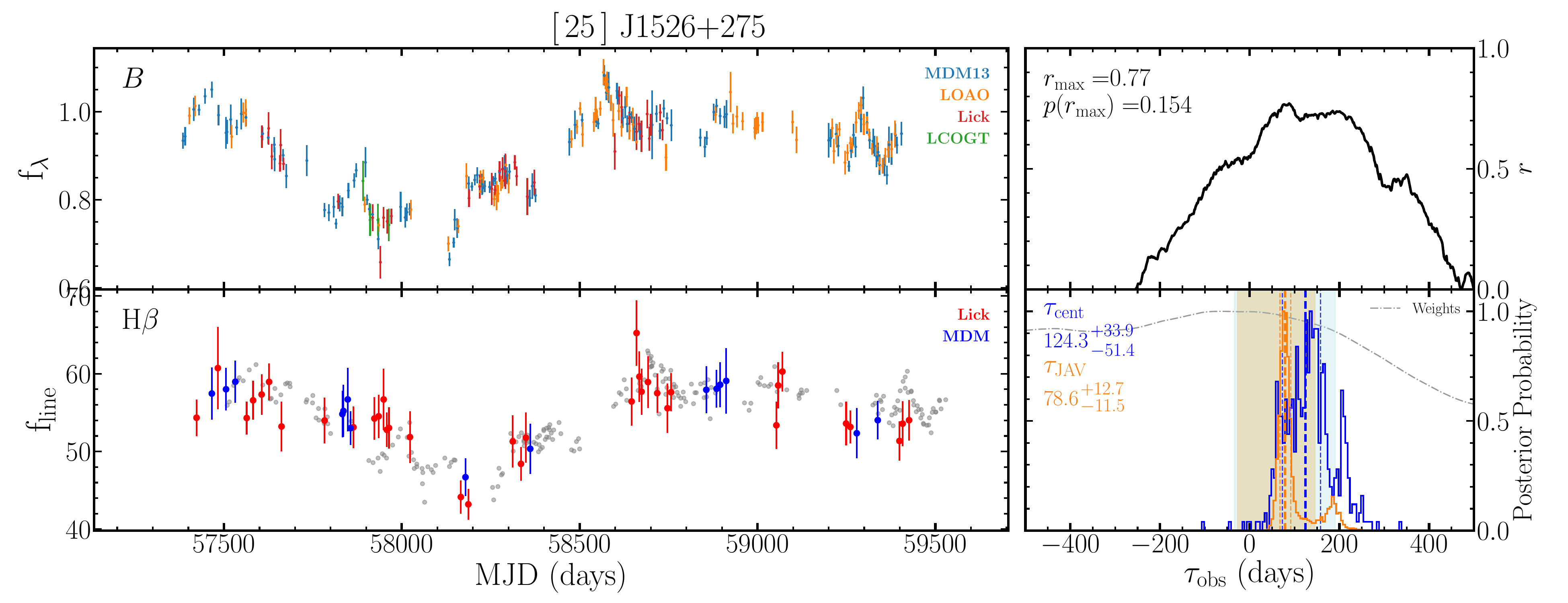}
\includegraphics[width=0.95\textwidth]{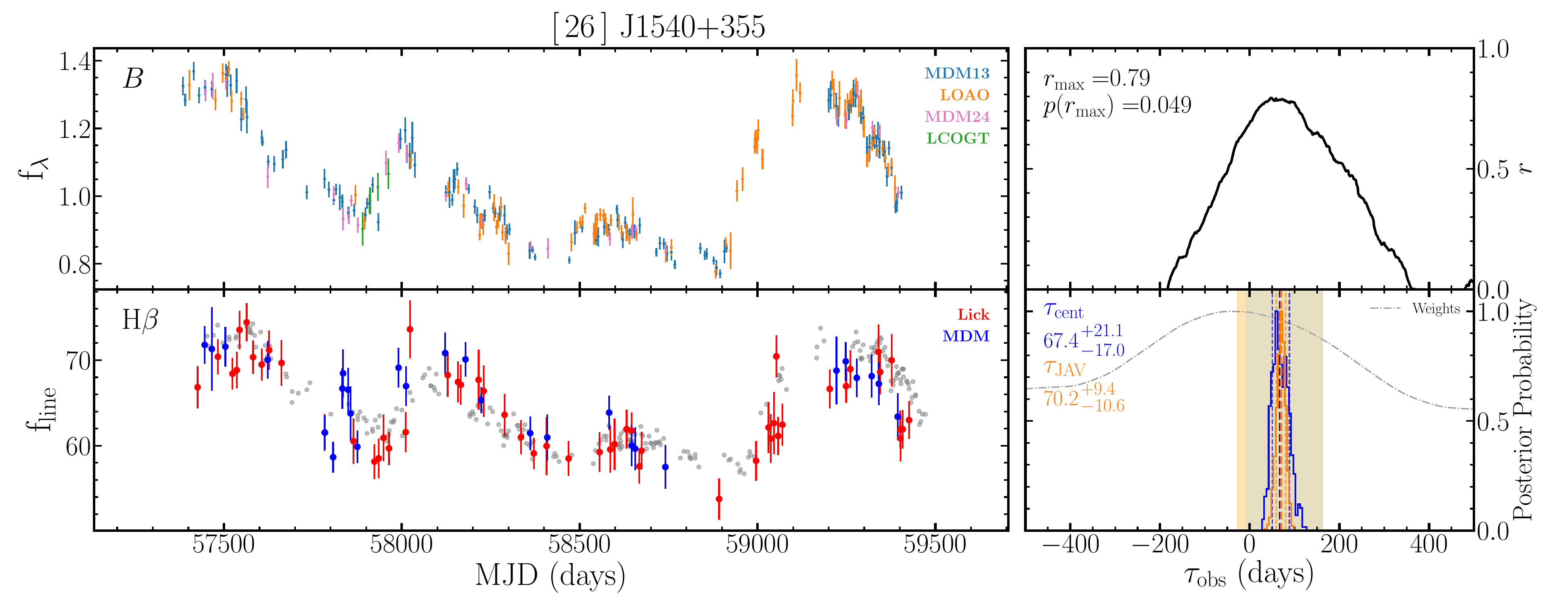}
\includegraphics[width=0.95\textwidth]{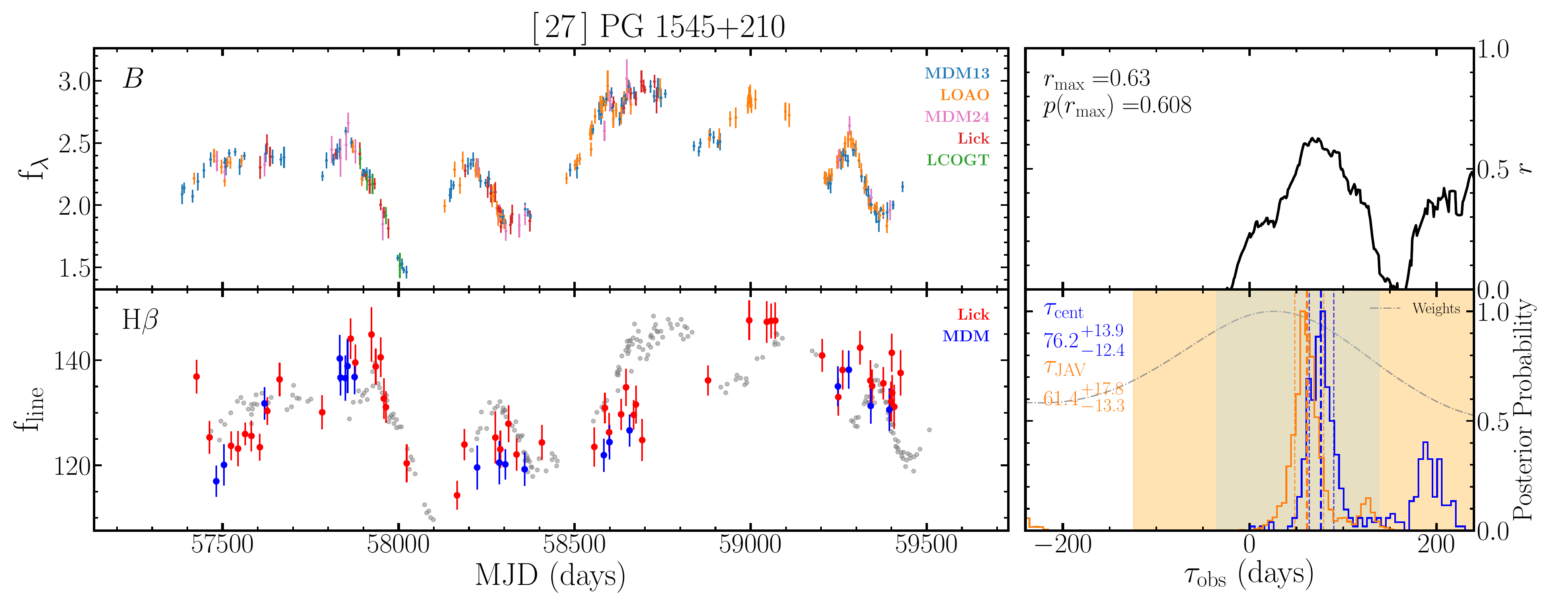}
\caption{Same as Figure \ref{fig:LC_1} but for J1526+275, J1540+355, and PG 1545+210.}
\label{fig:LC_9}
\end{figure*}

\begin{figure*}[htbp]
\centering

\includegraphics[width=0.95\textwidth]{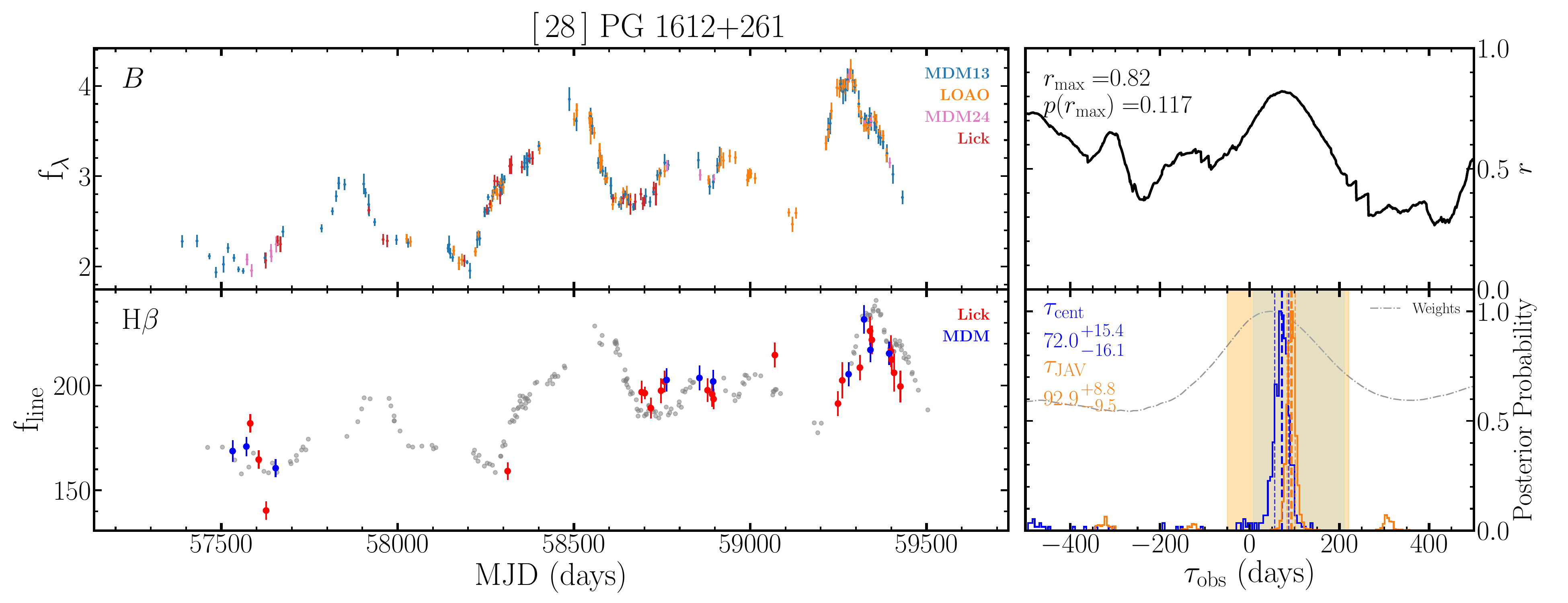}
\includegraphics[width=0.95\textwidth]{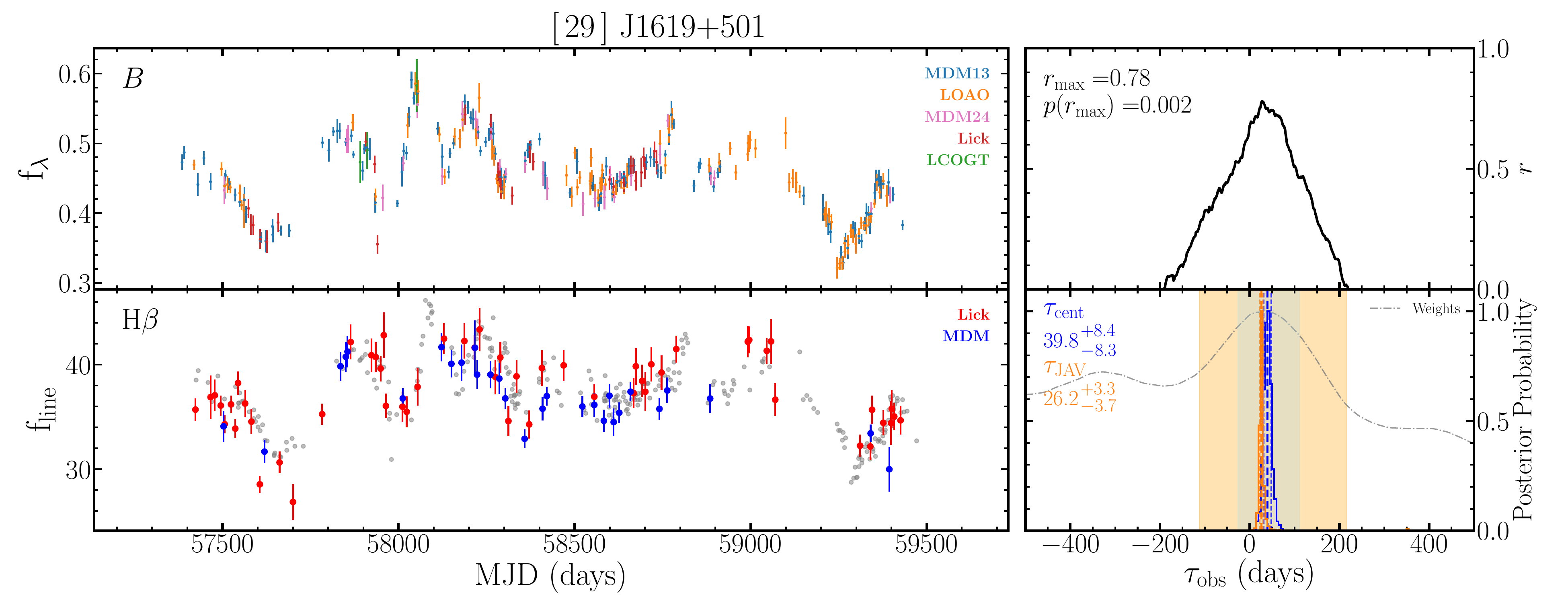}
\includegraphics[width=0.95\textwidth]{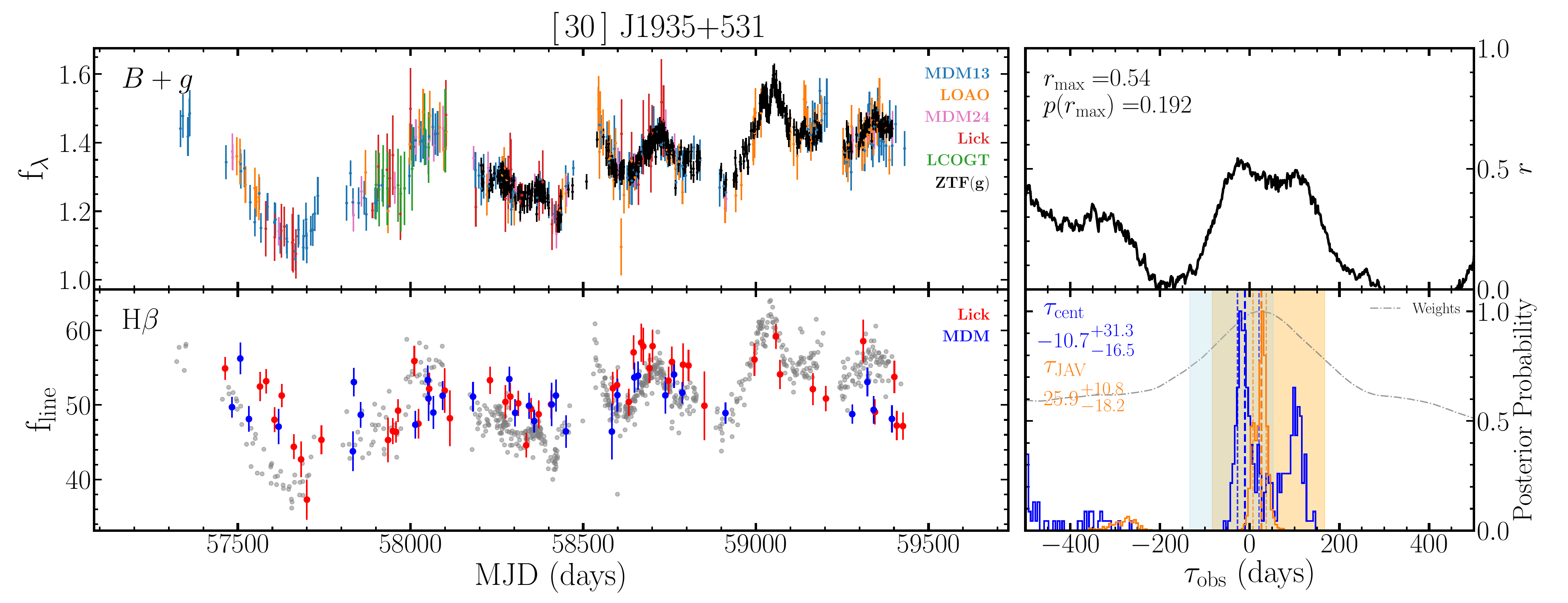}
\caption{Same as Figure \ref{fig:LC_1} but for PG 1612+261, J1619+501, and J1935+531.}
\label{fig:LC_10}
\end{figure*}

\begin{figure*}[htbp]
\centering
\includegraphics[width=0.95\textwidth]{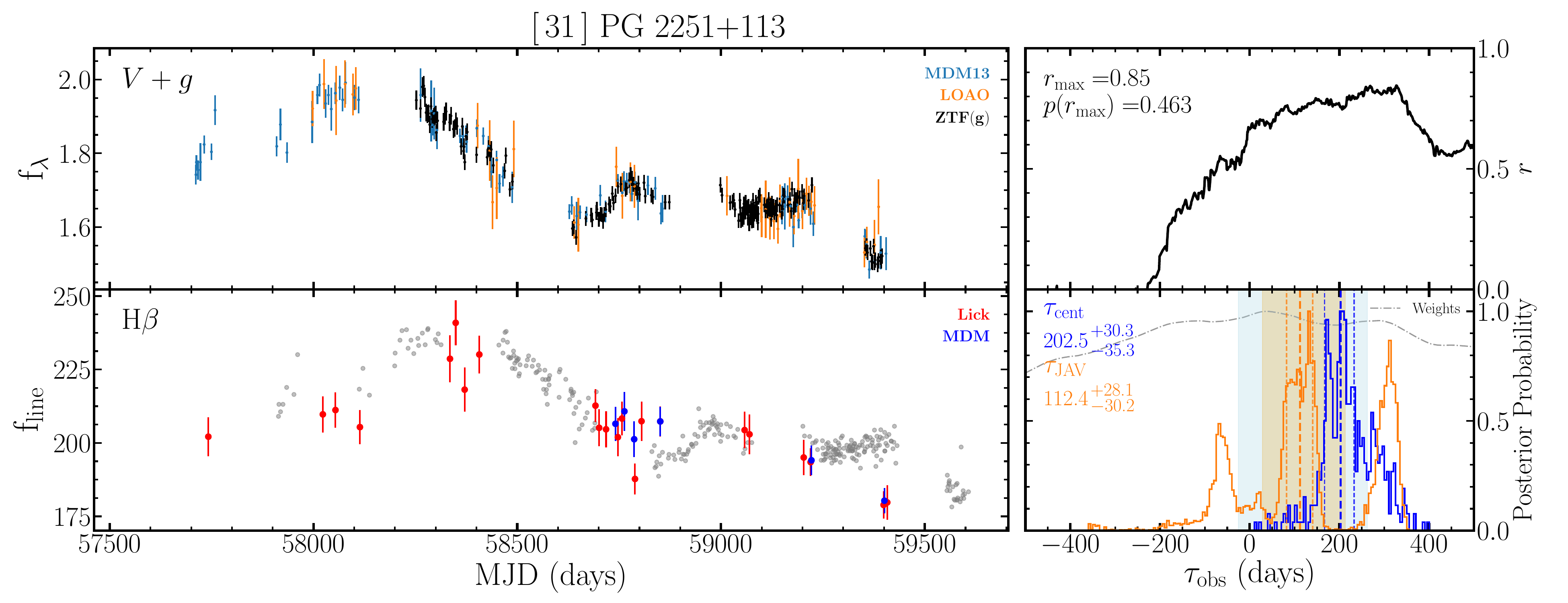}
\includegraphics[width=0.95\textwidth]{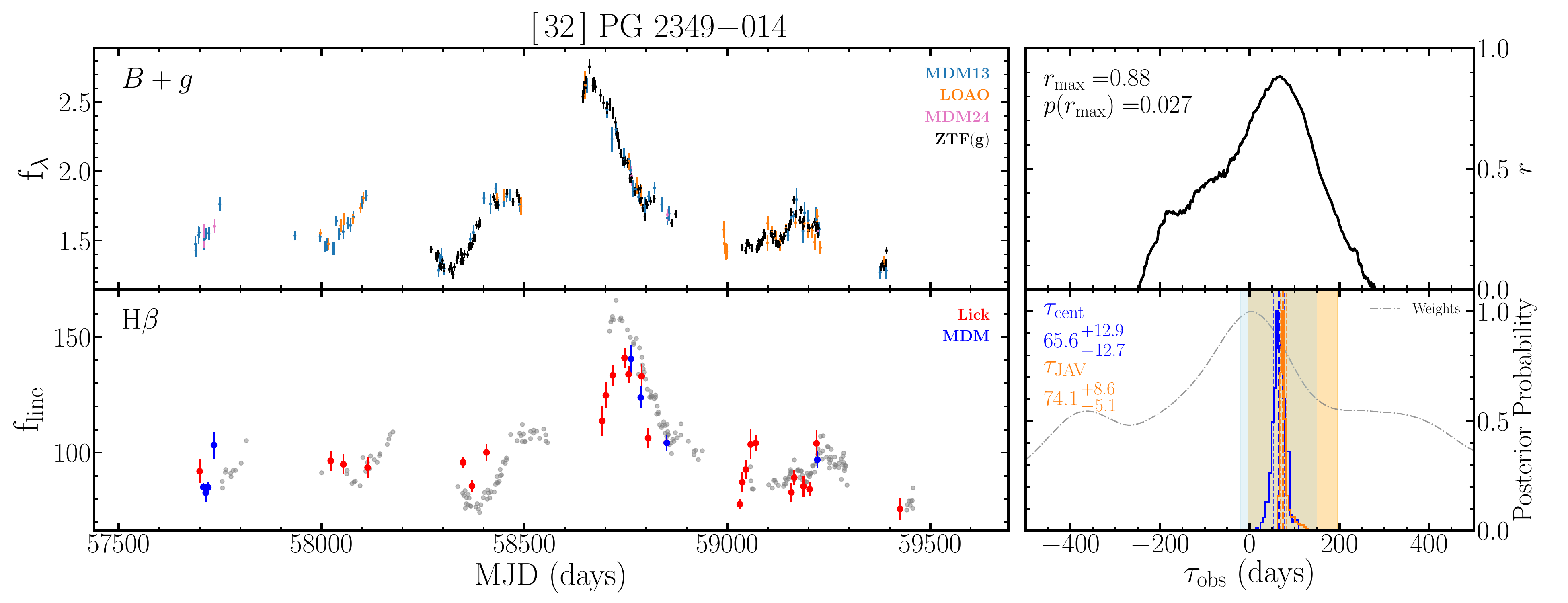}
\caption{Same as Figure \ref{fig:LC_1} but for PG 2251+113 and PG 2349$-$014.}\label{fig:LC_11}
\end{figure*}

\begin{table}[htbp]
    \centering
    
    \caption{$B$ band and H$\beta$ light curves}
    {\footnotesize
    \begin{tabular}{l c c c c c }
     \hline \hline

    Name & Band & Telescope & MJD & $F$ & $F_{\rm err}$  \\ \hline
    
Mrk~1501 & $B$ & MDM13 & 57689.6641 & 1.694 & 0.049 \\ 
Mrk~1501 & $B$ & MDM13 & 57690.6992 & 1.612 & 0.049 \\ 
Mrk~1501 & $B$ & MDM13 & 57697.6797 & 1.675 & 0.051 \\ 
Mrk~1501 & $B$ & MDM13 & 57698.6914 & 1.772 & 0.056 \\ 
Mrk~1501 & $B$ & MDM13 & 57699.6133 & 1.660 & 0.059 \\ 
Mrk~1501 & $B$ & MDM24 & 57710.7148 & 1.734 & 0.069 \\ 
Mrk~1501 & $B$ & MDM13 & 57711.6953 & 1.802 & 0.049 \\ 
Mrk~1501 & $B$ & MDM13 & 57712.7617 & 1.839 & 0.051 \\  \\    
    \hline

    \multicolumn{6}{p{0.46\textwidth}}{{\bf Notes.} For $B$ band light curves the quantity F represent the flux density in the unit of $10^{-15}$ erg s$^{-1}$ cm$^{-2}$ \AA$^{-1}$, while for H$\beta$ (Band$=$\hbeta) the quantity F represent the integrated H$\beta$ flux in the unit of $10^{-15}$ erg s$^{-1}$ cm$^{-2}$. This table is available in its entirety online. }

    \end{tabular}
    }
    \label{tab:light_curves}
\end{table}

\section{Results}

\subsection{\hbeta\ lag measurement}\label{sec:lag-measurements}

The emission line lags have been measured in a number of different ways in the previous RM studies. These methods are categorized into two main classes: the traditional cross-correlation analysis and the methods with statistical approach for describing AGN variability, e.g., damped random walk (DRW) models  \citep[e.g.,][]{Kelly09,MacLeod10}. The former class includes the interpolated cross-correlation function \citep[ICCF, e.g.,][]{Gaskell_Peterson87, Peterson98}, the discrete correlation function \citep[DCF,][]{Edelson_Krolik_1988}, and the $z$-transformed discrete correlation function \citep[$z$DCF,][]{Alexander13arXiv}, while the second class includes {\tt JAVELIN} \citep{Zu11}, {\tt CREAM} \citep{Starkey16}, and  {\tt MICA} \citep{Li16}. It has been demonstrated that for sparsely sampling light curves, DCF and $z$DCF are less efficient in recovering the lag than ICCF and {\tt JAVELIN} \citep{White_Peterson94, Li-J19}. While comprehensive comparison among the second-class methods is not yet available, these methods generally share a similar basic algorithm and at least some of them provide consistent results \citep[e.g., ][]{Grier17b, Grier19}.

We adopt the most commonly used approach ICCF as the primary method for the lag measurements of SAMP targets. As the ICCF method has been adopted by the majority of previous studies, it is possible to directly compare with the previously reported measurements. We use the python package {\tt PyCCF} \citep{Peterson98, Sun18} to perform the lag measurements, by calculating the cross-correlation coefficient $r$ between the continuum and \hbeta\ emission line light curves after linearly interpolating one light curve to match the time grid of the other light curve. One light curve is shifted by a series of $\tau$ values in a searching window, and as a function of $\tau$, ICCF is calculated and the cross-correlation coefficient r is determined. In this process, either the continuum or the emission-line light curve is interpolated to calculate the ICCF and the final ICCF is obtained by averaging the two ICCFs. Then, the centroid ($\tau_{\rm cent}$) or the peak ($\tau_{\rm peak}$) is determined using the range in the averaged ICCF, where $r$ is larger than 80\% of its maximum, and adopted as the lag of the two light curves. We consider $\tau_{\rm cent}$ as the primary ICCF lag measurements as done by previous studies \citep[e.g.,][]{Peterson04}, while we also present $\tau_{\rm peak}$ of each AGN
for comparison and completeness in Table 5. The lag uncertainties are estimated with the flux randomization/random subset sampling (FR/RSS) method, which randomizes the flux of each epoch based on its uncertainty and randomly selects a subset of epochs in the light curve for each simulation \citep{Peterson98}. 

In the ICCF analysis, we use a searching window of [$-600$, $600$] days with a step of 1 day. Note that this window is roughly 30\% of our 6-year baseline ($\sim2000$ days on average) and at least a factor of 3 larger than the expected lag estimated based on the size-luminosity relation \citep{Bentz13}. For two targets, PG~1100$+$772 and PG~2251$+$113, the searching window is less than a factor of 2 of their expected lags. Thus, we test the effect of the size of the searching window by increasing the searching window as [$-1000$,1000] days, finding that the lags are essentially the same.
For J1415$+$483, which is the only target that was monitored for 5 years, we use a searching window of [$-445$, 445] days, corresponding to $\sim30$\% of its baseline. In the case of the {\tt JAVELIN} analysis, we initially use the same searching window. However, we finally use a smaller searching window, [$-360$, $360$] days, because we find {\tt JAVELIN} tends to provide strong aliases if we use a larger searching window, and sometimes unacceptably large lag values are reported (i.e., $>500$ days or $<-500$ days ).

In addition, we test the improvement of the lag measurements by using subsamples of the 6-year light curves, finding that cross-correlation results are more reliable with higher cross-correlation coefficients, because of the densely distributed observations in particular seasons. Note that the total 6-year light curve is composed of six seasons separated by seasonal gaps. Specifically, we find an improvement relative to the total light curves when we use only 2018 observations for J0939$+$375, 2020-2021 observations for J1203$+$455, 2016-2017  observations for PG~1545$+$210, and 2019-2021 observations for PG~1427$+$480, presumably owing to the much higher quality light curves during these specific seasons.  Under such circumstances, we adopt the measurements based on the light curves of these specific seasons.

Both the posterior distributions of ICCF and {\tt JAVELIN} can show multiple peaks as a feature of sparsely sampled multiple-year data with seasonal gaps  \citep[e.g.,][]{Grier19, Homayouni20, Yu22arXiv}, which is likely to be caused by quasi-periodic variations, mismatch of weakly variable features \citep{Homayouni20}, and seasonal gaps. Following \citet{Grier19}, we employ an alias identification procedure to remove these aliases and identify the primary peak for measuring lags. We apply a weight function to the posterior distribution of ICCF and {\tt JAVELIN}, which is a convolution of the following two components. The first component is a probability function based on the overlapped fraction defined as:
\begin{equation}
    P(\tau) = [N(\tau)/N(0)]^2 ,
\end{equation}
where $N(\tau)$ and $N(0)$ is the number of overlapped points between continuum and emission-line light curves with or without shifting one light curve by $\tau$, respectively. This component helps to reduce the weight of the lag values that are similar to seasonal gaps because no real data are overlapped between the light curves of the continuum and \hbeta\ if one light curve is shifted by these values. The second component is the auto-correlation function (ACF), which represents how fast the continuum variability is. If the continuum light curve varies slowly, then the distribution of ACF is wide, indicating that the seasonal gaps are less likely to affect the lag measurement. On the other hand, fast continuum variability leads to a narrow ACF distribution, and in this case, the important features of the variability pattern are more likely blocked by the seasonal gaps \citep{Homayouni20}. Thus, we combine the overlap probability function (i.e., Eq. 6) and the ACF to generate a weight function, which then is multiplied by the lag posterior distribution of ICCF and {\tt JAVELIN}.

To identify the primary peaks in the weighted posterior distribution, we first smooth the distribution by a Gaussian kernel with a width of 12 days, which was determined based on experiments and visual inspection as similarly adopted by previous studies \citep[e.g.,][]{Grier19, Homayouni20, Yu22arXiv}. By identifying the highest peak as the primary peak in the smoothed distribution, we define its range between the two adjacent local minimums. Then, we remove the posteriors outside this range and measure the lag from the truncated distribution. Note that we use the weighted posterior distribution for defining the range of truncation. Finally, we determine the lag and its uncertainty by measuring the median and 16-to-84th quadrature of the truncated \textit{unweighted} posterior distribution. 

Based on the aforementioned analysis, we measure the \hbeta\ lag of the sample (see Table \ref{tab:lags}).
As a consistency check, we compare the new measurements with the measurements presented in Paper \RNum{2}, which reported the initial \hbeta\ lag measurements of two AGNs, J0801$+$512 (2MASS~J10261389$+$5237510) and J1619$+$501 based on the first-three-year data. The lags measured in Paper \RNum{2} are $41.8^{+4.9}_{-6.0}$ days and $60.1^{+33.1}_{-19.0}$ days in the observed frame for J0801$+$512 and J1619$+$501, respectively. Our new lag measurement of J0801$+$512 ($43.2^{+4.7}_{-4.8}$ days) is consistent with the measurement reported in Paper \RNum{2} within 1$\sigma$ uncertainty, while for J1619$+$501 our new measurement ($39.3^{+8.6}_{-8.7}$ days) is much smaller but consistent with Paper \RNum{2} considering the large uncertainty of the initial measurement. 

\subsection{Effect of detrending }

A long-term trend in the line light curves can bias the cross-correlation result because the long-timescale (low frequency) trend can overwhelm the relatively shorter intrinsic pattern due to the lag \citep[e.g.,][]{Welsh99, Denney10, Zhang19}. In such cases, detrending of the light curves may improve the lag analysis. 

We examine the effect of long-term trends by detrending the continuum and/or \hbeta\ light curves 
using a first-order polynomial,
which is usually enough to remove the various linear trends between the continuum and emission line light curves. Higher order polynomials are more dangerous to use since they are suspected to introduce artificial signals in the light curves and largely reduce the cross-correlation strength at the same time \citep{Peterson95}. We find that only for two objects, PG~1100$+$772  and PG~1202$+$281, $r_{\rm max}$ is significantly improved  
after detrending. Thus, we decide to use the results based on detrending for the two objects. For completeness, we present the lag analysis for these two objects without detrending in Appendix \ref{sec:appendixB}.

Based on the optical and radio variability of a radio-loud AGN, 3C 273, \citet{Li20} suggested that the underlying physics of the detrending process is to correct for the contamination of jet emission in the optical continuum light curves since jet emission has no corresponding effect on the emission-line light curves. We note that PG~1100$+$772 is a radio-loud AGN with a radio loudness $R=f_{\textrm{5GHz}}/f_{2500}\sim400$, suggesting that PG~1100$+$772 can be treated as an analogy of 3C 273 \citep{Zhang19}. However, it is difficult to disentangle the jet contribution from the observed continuum light curves for this target due to the lack of radio monitoring data during our reverberation campaign. It is possible that radio-loud AGNs may introduce scatter in the size-luminosity relation if this scenario is correct. A larger sample of radio-loud AGNs is required in the RM studies to verify this scenario.

\subsection{Lag reliability}\label{sec:lag-reliability}

\begin{figure}[htbp]
    \centering
    \includegraphics[width=0.48\textwidth]{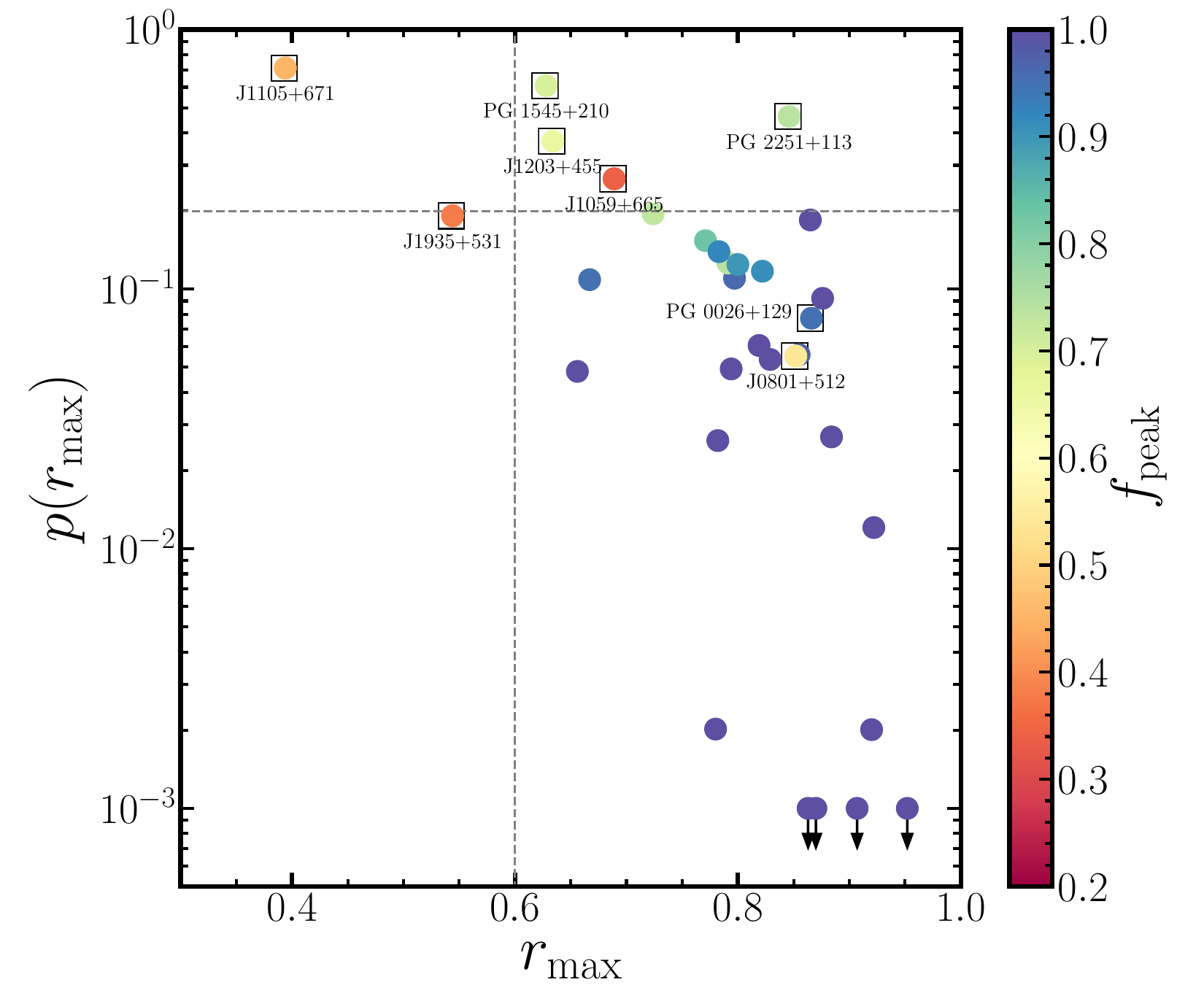}

    \caption{Assessment of the lag reliability using the three criteria, i.e., the maximum cross-correlation coefficient ($r_{\rm max}$), the probability that uncorrelated light curves would produce a cross-correlation coefficient of at least $r_{\rm max}$ ($p (r_{\rm max})$),   and the fraction of the primary peak in the posterior distribution ($f_{\rm peak}$). The best lags are defined with $r_{\rm max}$$\geq$0.6 (vertical dashed line), $p(r_{\rm max})$$<$0.2 (horizontal dashed line), and $f_{\rm peak}$ $>$0.6. Two objects show $r_{\rm max}< 0.6$. There are four objects with $r_{\rm max}\geq 0.6$, that do not satisfy $p(r_{\rm max})$<$ 0.2$, and there is one object with  $r_{\rm max}\geq 0.6$ and $p(r_{\rm max})$$<$0.2, but with $f_{\rm peak} \leq 0.6$. 
    }
    \label{fig:lag-reliability}
\end{figure}

In this section, we investigate the reliability of the lag measurements from the cross-correlation analysis. The two main criteria of lag reliability 
are the goodness of the cross-correlation between the continuum and emission line light curves, and the constraints against artificial signatures in the light curves, i.e., aliases. In general, the lag measurements are more reliable if the input light curves have a higher cadence, more regular sampling, longer time baseline, and stronger non-quasi-periodic variability patterns. However, well-defined quantitative criteria are yet to be available while a few new attempts have been reported in the literature. 

One of the criteria of reliability is the maximum cross-correlation coefficient $r_{\rm max}$, which is the maximum value in the ICCF, representing the strength of the correlation between two light curves. Previous studies often used a threshold of $r_{\rm max}$ $>$ 0.6 or 0.5, to adopt reliable lag measurements \citep[e.g.][]{Grier17b,U22}. However,  $r_{\rm max}$ alone is not sufficient to determine the lag reliability. For instance, $r_{\rm max}$ would be relatively high even though the light curves have very few data points but show strong linear variation, or even if the variability amplitude is small compared to the flux uncertainties, but some features in the light curves are well matched. In these cases, the measured lag can be biased or reflect the randomly uncorrelated light curves.
Not to mention that flux uncertainties are not taken into account in the $r_{\rm max}$ calculation.

Recently, a new method {\tt PyI$^2$CCF}\footnote{https://github.com/legolason/PyIICCF} has been proposed to assess the lag reliability \citep[][Guo et al., in prep]{Guo21AAS, U22}, using a null-hypothesis to evaluate the probability that the observed cross-correlation can be equally obtained by two uncorrelated red-noise light curves. This idea originates from X-ray reverberation studies and has been used to assess the correlation of multi-wavelength variability of AGNs \citep{Uttley03,Arevalo08,Chatterjee08}. 
Using the publicly available {\tt PyI$^2$CCF}, we investigate the reliability of our lag measurements. We generate 10$^3$ realizations of a pair of mock light curves for each object by keeping the same cadence and sampling as the observed light curves using DRW models. Then, by counting the number of positive lags ($\tau>0$) with the $r_{\rm max}$ value higher than the observed $r_{\rm max}$ among all simulations, we determine the probability $p(r_{\rm max})_{\tau>0}$ as a lag significant indicator. In this process, we keep the aforementioned searching window as defined in \S4.1. 

We further examine the fraction of the primary peak ($f_{\rm peak}$) in the posterior distribution that describes how much of the posteriors are within the range of the primary peak. For example,  $f_{\rm peak}$ less than 0.6 is often considered as that there are possible significant solutions other than the identified primary peak \citep{Grier19, Homayouni20, Yu22arXiv, Malik22ArXiv}. We listed $f_{\rm peak}$ of $\tau_{\rm cent}$ of all targets in Table \ref{tab:lags}.

We present the three criteria of the lag reliability: $r_{\rm max}$ based on the observed light curves, $p(r_{\rm max})$ based on simulations, and $f_{\rm peak}$ calculated from the posterior distribution, for individual AGNs in Table \ref{tab:lags}. Note that for the majority of our targets, the measured lag is reliable based on these three assessments. 
In Figure \ref{fig:lag-reliability} we directly compare the two indicators, i.e., $r_{\rm max}$ and $p(r_{\rm max})$ along with the color-coded ($f_{\rm peak}$).
We quantitatively define the best lag measurements, by requiring $r_{\rm max}\geq0.6$, 
$p(r_{\rm max})\leq0.2$ ,  
and $f_{\rm peak}\geq0.6$.  Note that there is no strict cutoff of $p(r_{\rm max})$ between best and less-reliable lag measurements.
 We adopt $p(r_{\rm max})_{\tau>0}\leq0.2$ following the previous study by \citet{U22}. 
In these assessments, we find that the measured lag of seven targets, namely, J0801$+$512, J1059$+$665, J1105$+$671, J1203$+$455, PG~1545$+$210, J1935$+$531, and PG~2251$+$113, do not satisfy the lag reliability criteria. By excluding these seven targets, 
we obtain the best \hbeta\ lag measurements for 25 AGNs. 
In the case of PG~0026+129, the obtained lag is consistent with zero within 1$\sigma$ uncertainty, suggesting that the lag is not resolved. 

Finally, we perform a visual inspection of all light curves to check how well the continuum light curve matches the emission-line light curve after shifting the continuum light curve by the measured lag. This method is qualitative, providing an additional check on the reliability of the measured lag. We present the two light curves, after shifting by the lag and scaling with the median and standard deviation of the fluxes,
in the lower panel of each target in Figure \ref{fig:LC_1}. Overall, we find a good match between the shifted continuum and the \hbeta\ light curves of the majority of targets in the sample, confirming the reliability of the lag measurements.

\begin{figure}[htbp]
    \centering
    \includegraphics[width=0.45\textwidth]{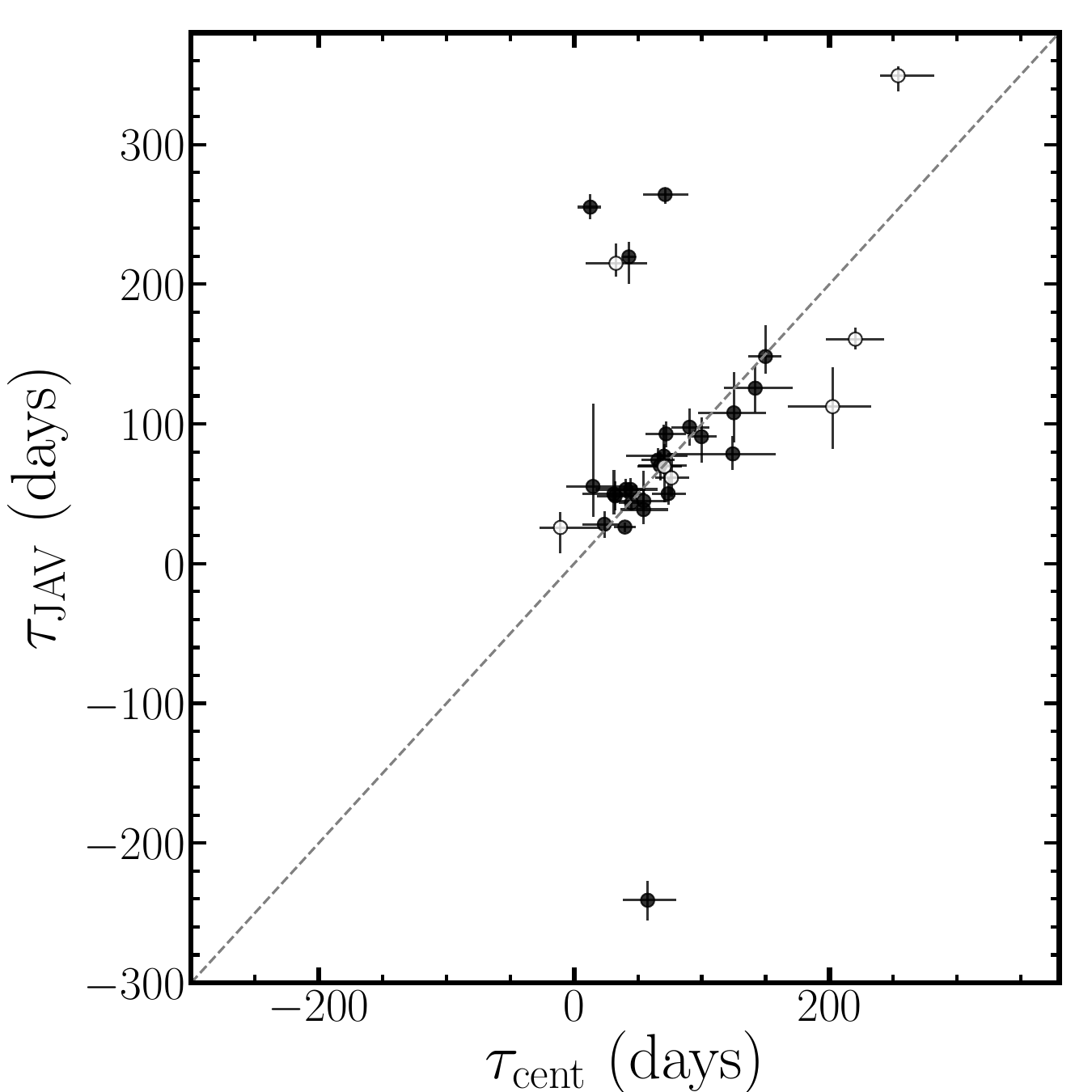}

    \caption{Comparison of the lag measurements in the observed-frame between $\tau_{\rm cent}$ (ICCF) and $\tau_{\rm JAV}$ ({\tt JAVELIN}). Best and less-reliable lag measurements are denoted with filled and open circles, respectively, using the threshold of  $r_{\rm max}=0.6$, $p(r_{\rm max})=0.2$ and $f_{\rm peak}=0.6$. 
    }
    \label{fig:jav_vs_iccf}
\end{figure}

As a consistency check, we compare the ICCF and {\tt JAVELIN} measurements in Figure \ref{fig:jav_vs_iccf}, finding that they are generally consistent. However, six objects, namely, Mrk~1501, J1026$+$523, J1059$+$665, J1105$+$671, PG~1322$+$659, and PG~2251$+$113, show a large discrepancy with {\tt JAVELIN} lag larger than ICCF lag by $150$ days or smaller than by $-150$ days while ICCF lags are within [0, 100] days. Three of them are already identified as less-reliable, for the other three AGNs the {\tt JAVELIN} lag seems overestimated or underestimated because of the combined effect of seasonal gaps and quasi-periodic variability based on visual inspection of their light curves. This result suggests that although the typical error of {\tt JAVELIN} lag is indeed smaller than ICCF, {\tt JAVELIN} measurements are sometimes easier to be affected by seasonal gaps than ICCF and generate spurious measurements, particularly for sparsely sampled light curves. Thus, we decided not to disqualify these ICCF lags as unreliable, albeit with the discrepancy with the  {\tt JAVELIN} results.

A simple statistical approach to quantify the lag reliability of a given sample is to examine the sample's false positive rate, i.e., what fraction of the lag measurements is false detection caused by inopportune features, seasonal gaps as well as inappropriately identified primary peaks and alias identification. The incidence of negative lags over the entire sample can be an indicator of the false positive rate \citep[e.g.,][]{Shen16a, Grier17b, Grier19}. In our SAMP sample, there is one object with a negative $\tau_{\rm cent}$ and one object with a negative $\tau_{\rm JAV}$. If we conservatively count these two measurements, the false positive rate is $\sim$6\% of the SAMP sample, indicating the high quality of our lag measurements.

{\begin{table*}[h!]
    \centering
    \caption{Observed-frame lags and lag reliability parameters} 
    \begin{tabular}{r l c c c c c c}
    \hline \hline
    
    & Name & $\tau_{\rm cent}$ & $\tau_{\rm peak}$  & $\tau_{\rm JAV}$ & $r_{\rm max}$ & $p(r_{\rm max})$  & $f_{\rm peak}$\\
    & & (days) & (days) & (days) & & & \\
    & (1) & (2) & (3) & (4) & (5) & (6) & (7) \\
         \hline

1 & Mrk~1501 & $12.7^{+8.3}_{-9.7}$ & $24.0^{+4.0}_{-9.0}$ & $255.2^{+9.3}_{-8.4}$ & 0.85 & 0.056 & 0.97 \\ 
2 & PG~0026$+$129 & $14.8^{+34.5}_{-21.1}$ & $25.0^{+19.0}_{-18.0}$ & $55.3^{+59.1}_{-21.7}$ & 0.87 & 0.077 & 0.95 \\ 
3 & PG~0052$+$251 & $73.7^{+13.9}_{-12.6}$ & $59.0^{+16.0}_{-14.0}$ & $50.1^{+11.4}_{-8.1}$ & 0.95 & $<0.001$ & 1.00 \\ 
4 & J0101$+$422 & $90.5^{+15.7}_{-14.0}$ & $91.0^{+30.0}_{-40.0}$ & $97.7^{+13.2}_{-13.2}$ & 0.80 & 0.110 & 0.96 \\ 
5 & J0140$+$234 & $149.9^{+12.5}_{-13.5}$ & $148.0^{+34.0}_{-22.0}$ & $148.3^{+22.6}_{-12.2}$ & 0.92 & 0.002 & 1.00 \\ 
6 & Mrk~1014 & $125.1^{+25.0}_{-28.1}$ & $145.0^{+22.0}_{-22.7}$ & $108.0^{+29.1}_{-21.4}$ & 0.92 & 0.012 & 1.00 \\ 
8 & J0939$+$375* & $23.7^{+11.2}_{-16.9}$ & $26.0^{+9.0}_{-23.0}$ & $28.2^{+9.6}_{-9.4}$ & 0.86 & 0.185 & 1.00 \\ 
9 & PG~0947$+$396 & $44.3^{+11.5}_{-13.3}$ & $41.0^{+17.0}_{-16.0}$ & $53.1^{+8.1}_{-7.1}$ & 0.78 & 0.026 & 1.00 \\ 
10 & J1026$+$523 & $42.9^{+5.3}_{-5.0}$ & $34.0^{+9.0}_{-7.0}$ & $219.6^{+11.0}_{-19.6}$ & 0.86 & $<0.001$ & 1.00 \\ 
12 & PG~1100$+$772 & $70.4^{+18.7}_{-29.9}$ & $79.0^{+16.7}_{-25.2}$ & $77.0^{+22.7}_{-12.8}$ & 0.67 & 0.109 & 0.95 \\ 
14 & J1120$+$423 & $54.4^{+19.1}_{-18.3}$ & $35.0^{+17.0}_{-12.0}$ & $38.7^{+14.9}_{-10.3}$ & 0.91 & $<0.001$ & 1.00 \\ 
15 & PG~1121$+$422 & $141.9^{+29.6}_{-24.8}$ & $151.0^{+29.0}_{-29.0}$ & $125.6^{+15.4}_{-17.2}$ & 0.82 & 0.061 & 1.00 \\ 
17 & PG~1202$+$281 & $44.9^{+10.6}_{-9.9}$ & $40.0^{+7.0}_{-9.0}$ & $43.5^{+5.5}_{-5.0}$ & 0.66 & 0.048 & 1.00 \\ 
18 & J1217$+$333 & $31.2^{+25.0}_{-24.4}$ & $38.0^{+33.0}_{-27.5}$ & $49.8^{+17.4}_{-14.3}$ & 0.79 & 0.126 & 0.74 \\ 
19 & VIII~Zw~218 & $71.3^{+18.5}_{-17.4}$ & $69.0^{+16.0}_{-30.0}$ & $264.2^{+4.6}_{-6.9}$ & 0.72 & 0.195 & 0.73 \\ 
20 & PG~1322$+$659 & $57.6^{+22.4}_{-19.4}$ & $52.0^{+15.0}_{-29.0}$ & $-240.9^{+14.0}_{-14.3}$ & 0.78 & 0.139 & 0.92 \\ 
21 & J1415$+$483 & $32.2^{+15.0}_{-14.1}$ & $36.0^{+14.0}_{-14.0}$ & $48.5^{+10.7}_{-10.3}$ & 0.80 & 0.124 & 0.90 \\ 
22 & PG~1427$+$480* & $40.6^{+25.1}_{-23.7}$ & $50.0^{+28.0}_{-36.0}$ & $53.2^{+7.7}_{-6.3}$ & 0.88 & 0.092 & 1.00 \\ 
23 & PG~1440$+$356 & $54.6^{+18.2}_{-22.6}$ & $44.0^{+6.0}_{-20.9}$ & $44.9^{+21.5}_{-7.4}$ & 0.87 & $<0.001$ & 1.00 \\ 
24 & J1456$+$380 & $99.8^{+12.0}_{-11.2}$ & $58.0^{+9.0}_{-16.5}$ & $90.9^{+13.6}_{-18.6}$ & 0.83 & 0.054 & 1.00 \\ 
25 & J1526$+$275 & $124.3^{+33.9}_{-51.4}$ & $80.0^{+11.0}_{-22.0}$ & $78.6^{+12.7}_{-11.5}$ & 0.77 & 0.154 & 0.83 \\ 
26 & J1540$+$355 & $67.4^{+21.1}_{-17.0}$ & $50.0^{+15.0}_{-12.6}$ & $70.2^{+9.4}_{-10.6}$ & 0.79 & 0.049 & 1.00 \\ 
28 & PG~1612$+$261 & $72.0^{+15.4}_{-16.1}$ & $76.0^{+16.0}_{-19.0}$ & $92.9^{+8.8}_{-9.5}$ & 0.82 & 0.117 & 0.91 \\ 
29 & J1619$+$501 & $39.8^{+8.4}_{-8.3}$ & $30.0^{+20.0}_{-6.0}$ & $26.2^{+3.3}_{-3.7}$ & 0.78 & 0.002 & 1.00 \\ 
32 & PG~2349$-$014 & $65.6^{+12.9}_{-12.7}$ & $66.0^{+12.0}_{-10.0}$ & $74.1^{+8.6}_{-5.1}$ & 0.88 & 0.027 & 1.00 \\ 
\hline
7 & J0801$+$512 & $220.3^{+22.4}_{-23.0}$ & $165.0^{+10.0}_{-12.0}$ & $160.7^{+8.5}_{-7.3}$ & 0.85 & 0.055 & 0.54 \\ 
11 & J1059$+$665 & $253.8^{+28.0}_{-14.2}$ & $259.5^{+21.9}_{-8.5}$ & $349.3^{+6.5}_{-11.2}$ & 0.69 & 0.266 & 0.34 \\ 
13 & J1105$+$671 & $32.8^{+24.4}_{-23.7}$ & $25.0^{+27.2}_{-17.2}$ & $215.0^{+14.3}_{-9.6}$ & 0.39 & 0.711 & 0.45 \\ 
16 & J1203$+$455* & $70.7^{+13.8}_{-21.4}$ & $68.0^{+18.0}_{-20.0}$ & $69.5^{+15.9}_{-25.5}$ & 0.63 & 0.372 & 0.66 \\ 
27 & PG~1545$+$210* & $76.2^{+13.9}_{-12.4}$ & $74.0^{+20.0}_{-13.0}$ & $61.4^{+17.8}_{-13.3}$ & 0.63 & 0.608 & 0.70 \\ 
30 & J1935$+$531 & $-10.7^{+31.3}_{-16.5}$ & $-19.0^{+16.0}_{-8.4}$ & $25.9^{+10.8}_{-18.2}$ & 0.54 & 0.192 & 0.38 \\ 
31 & PG~2251$+$113 & $202.5^{+30.3}_{-35.3}$ & $268.0^{+6.6}_{-12.0}$ & $112.4^{+28.1}_{-30.2}$ & 0.85 & 0.463 & 0.74 \\

\hline
\multicolumn{8}{p{0.67\textwidth}}{Notes. Notes. Column (1): object names; Column (2)--(4): the observed-frame lags  of ICCF centroid ($\tau_{\rm cent}$), peak ($\tau_{\rm peak}$), and {\tt JAVELIN}, respectively, through weighting and alias removal procedure. The lags and their two uncertainties are determined from the median, and 16th-50th/84-50th percentile interval from the unweighted lag posterior distribution covered by the primary peak.  For targets with * sign, lags are calculated based on the part of the light curves; Column (5)--(8): lag reliability parameters: $r_{\rm max}$ represents the maximum correlation coefficient; $p(r_{\rm max})$ is derived from {\tt PyI$^2$CCF} simulation which indicates how large the chance is to obtain the observed $r_{\rm max}$ from random light curves; $f_{\rm peak}$ is the fraction of the posterior distribution within the selected primary peak. Seven objects with less reliable lag measurements indicated by $r_{\rm max}\leq0.6$, $p(r_{\rm max})\geq0.2$, or $f_{\rm peak}\leq0.6$ are listed in the bottom. } \\

    \end{tabular}

\end{table*}
}\label{tab:lags}

\subsection{BH mass and Eddington ratio}

In this section, we present the \mbh\ determination of the sample, by measuring the width of the \hbeta\ emission line. We also determine Eddington ratio of individual AGNs using continuum luminosity. We compare the distribution of \mbh, luminosity, and Eddington ratio of the SAMP AGNs with that of the previous reverberation-mapped AGNs. 

\subsubsection{\hbeta\ line width measurements}

We measure the \hbeta\ line width from the decomposed broad \hbeta\ emission line profile using the mean spectra as well as the broad-line-only rms spectra. 
As a velocity measure, we adopt the full-width-at-half-maximum (FWHM) and the line dispersion $\sigma_{\rm line}$ (the 2nd moment of the line profile), following the definition by \citet{Peterson04}.
Note that we conservatively define two windows to represent the local continuum in the rms spectra for line width measurements as shown in Figure 2. 
Using the measured instrumental resolution of 481 km s$^{-1}$ from skylines, we correct for the instrumental broadening by subtracting the instrumental resolution from the measured widths in quadrature. Since the spectral resolution is measured from skylines, which uniformly fill the 4$^{\prime\prime}$ slit, the actual resolution for a point source is likely to be smaller. However, this effect is insignificant because \hbeta\ lines are intrinsically very broad. Since the change of the instrumental broadening during our observations is relatively small compared to the broad line width, we adopt a single representative spectral resolution, 481 km s$^{-1}$.

\subsubsection{\mbh\ measurements}

\begin{figure}[htbp]
    \centering
    \includegraphics[width=0.47\textwidth]{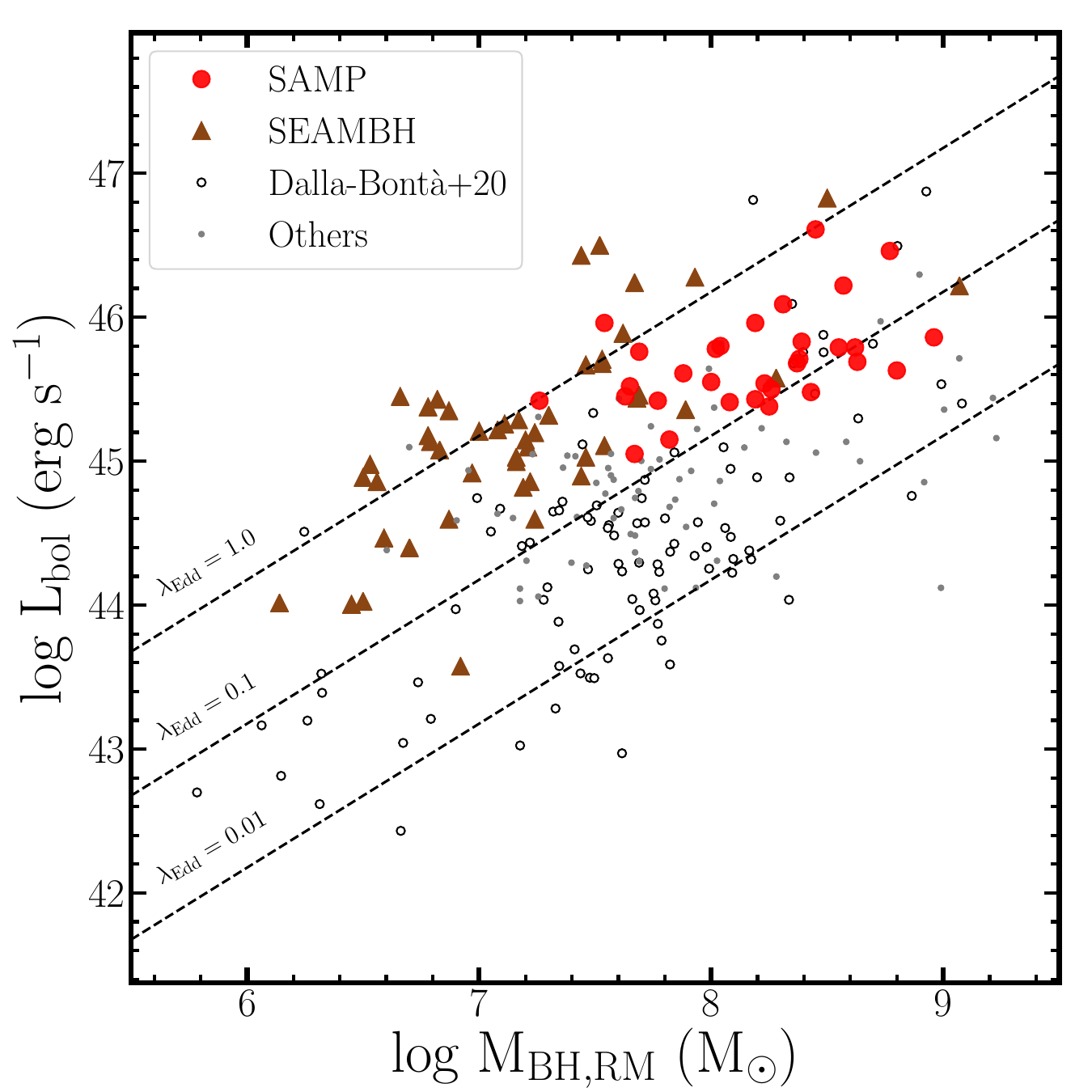}
    \includegraphics[width=0.47\textwidth]{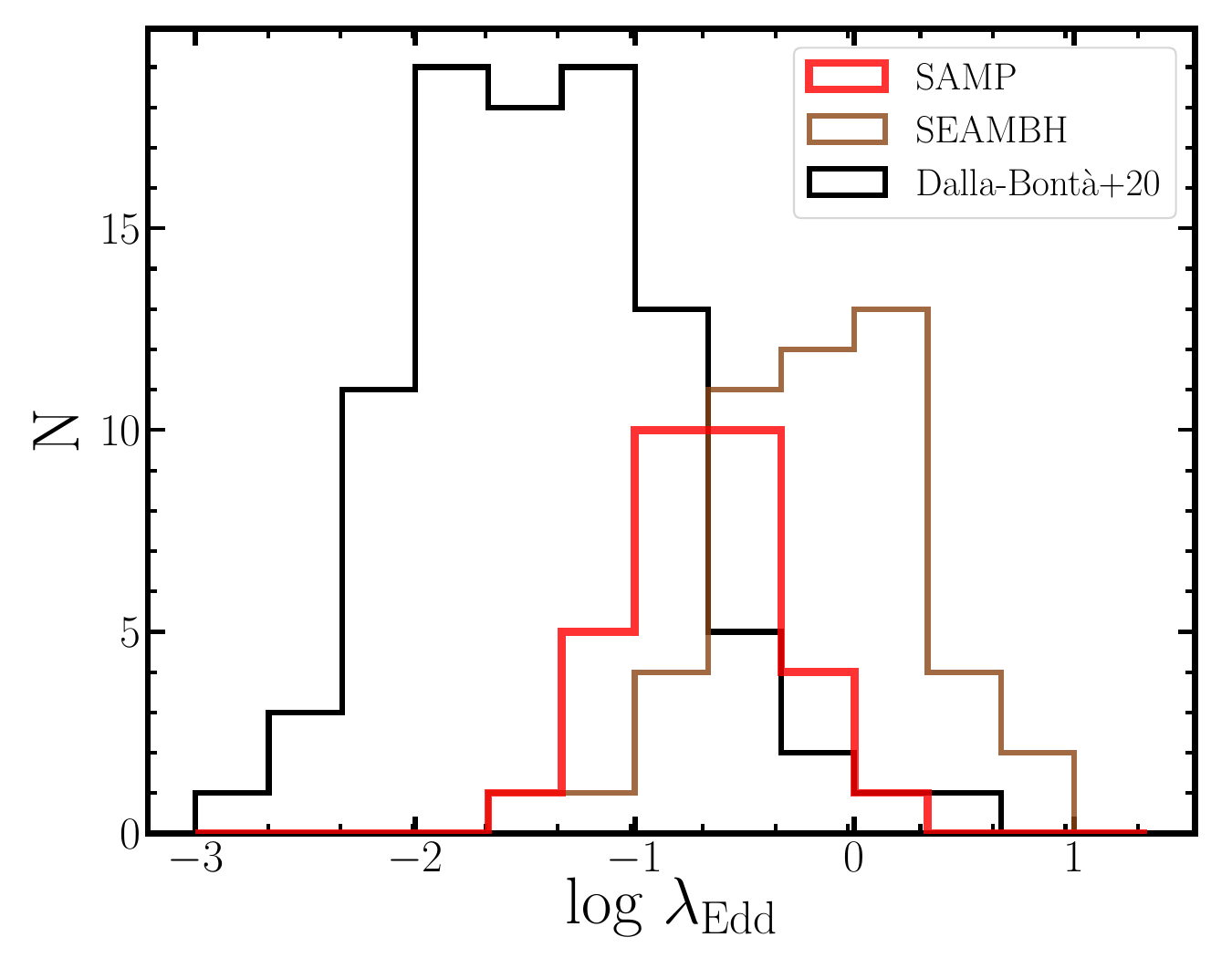}
    \caption{Upper: The distribution of SAMP final sample (red circle) on the BH mass--bolometric luminosity plane, overplotted with  \citet{DallaBonta20} collections (black dots) and SEAMBHs \citep[][brown triangles]{Du_Wang19, Hu21, Li-SS21}. Three dashed lines indicate the Eddington ratios from 0.01, 0.1, to 1.0 as labeled in the left end above the line. Our SAMP final sample has moderately high Eddington ratios. Lower: The distribution of Eddington ratios for SAMP, and other samples.}
    \label{fig:Lbol-Mbh}
\end{figure}

{\setlength{\tabcolsep}{1.0mm}
\begin{table*}[htbp]
    \centering
    \caption{Final Rest-Frame Lags, AGN luminosities, Line Widths, Virial Products, and Black Masses}
    \small
    \begin{tabular}{@{\extracolsep{4pt}} r l c c c c c c c c}
    
    \hline \hline
     & Name  & $\tau_{\rm final}$ & $L_{\rm 5100, AGN}$ & FWHM$_{\rm mean}$ & $\sigma_{\rm mean}$ & FWHM$_{\rm rms}$ & $\sigma_{\rm rms}$ & VP$_{\rm \sigma_{\rm rms}}$ & M$_{\rm BH}$\\
     
     & & (days) & ($10^{44}$ erg s$^{-1}$) & (km s$^{-1}$) & (km s$^{-1}$) & (km s$^{-1}$) & (km s$^{-1}$) & ($\times10^7{\rm M}_{\odot}$) & ($\times10^8{\rm M}_{\odot}$) \\

     & (1) & (2) & (3) & (4) & (5) & (6) & (7) & (8) & (9) \\
    \hline

1 & Mrk~1501 & $11.7^{+7.6}_{-8.9}$ & 1.22$\pm$0.06 & 4662$\pm$93 & 1935$\pm$30 & 4290$\pm$375 & 2154$\pm$64 & $1.1^{+0.7}_{-0.8}$ & $0.5^{+0.3}_{-0.4}$ \\ 
2 & PG~0026$+$129 & $13.0^{+30.2}_{-18.5}$ & 9.81$\pm$0.24 & 2608$\pm$62 & 2471$\pm$42 & 1334$\pm$216 & 1756$\pm$212 & $0.8^{+1.8}_{-1.1}$ & $0.3^{+0.8}_{-0.5}$ \\ 
3 & PG~0052$+$251 & $63.9^{+12.0}_{-10.9}$ & 5.58$\pm$0.16 & 4724$\pm$62 & 2196$\pm$19 & 4104$\pm$622 & 2068$\pm$267 & $5.3^{+1.0}_{-0.9}$ & $2.4^{+0.5}_{-0.5}$ \\ 
4 & J0101$+$422 & $76.1^{+13.2}_{-11.8}$ & 6.67$\pm$0.12 & 6171$\pm$31 & 2514$\pm$62 & 5406$\pm$278 & 2300$\pm$143 & $7.8^{+1.4}_{-1.2}$ & $3.5^{+0.7}_{-0.7}$ \\ 
5 & J0140$+$234 & $113.6^{+9.5}_{-10.2}$ & 13.35$\pm$0.14 & 2896$\pm$62 & 1601$\pm$12 & 2230$\pm$97 & 1438$\pm$138 & $4.6^{+0.4}_{-0.4}$ & $2.0^{+0.3}_{-0.3}$ \\ 
6 & Mrk~1014 & $107.6^{+21.5}_{-24.2}$ & 7.37$\pm$0.33 & 2890$\pm$97 & 1730$\pm$153 & 1786$\pm$67 & 1618$\pm$110 & $5.5^{+1.1}_{-1.2}$ & $2.5^{+0.6}_{-0.6}$ \\ 
8 & J0939$+$375 & $19.3^{+9.1}_{-13.7}$ & 2.85$\pm$0.07 & 2921$\pm$154 & 1561$\pm$39 & 1400$\pm$67 & 1038$\pm$113 & $0.4^{+0.2}_{-0.3}$ & $0.2^{+0.1}_{-0.1}$ \\ 
9 & PG~0947$+$396 & $36.7^{+9.5}_{-11.0}$ & 3.78$\pm$0.10 & 5258$\pm$36 & 2396$\pm$16 & 4910$\pm$648 & 2294$\pm$192 & $3.8^{+1.0}_{-1.1}$ & $1.7^{+0.5}_{-0.6}$ \\ 
10 & J1026$+$523 & $34.1^{+4.2}_{-4.0}$ & 3.04$\pm$0.07 & 3822$\pm$31 & 1938$\pm$7 & 2608$\pm$154 & 1198$\pm$66 & $0.9^{+0.1}_{-0.1}$ & $0.4^{+0.1}_{-0.1}$ \\ 
12 & PG~1100$+$772 & $53.7^{+14.3}_{-22.8}$ & 44.05$\pm$1.24 & 5993$\pm$62 & 2465$\pm$166 & 5591$\pm$154 & 2445$\pm$56 & $6.3^{+1.7}_{-2.7}$ & $2.8^{+0.8}_{-1.2}$ \\ 
14 & J1120$+$423 & $44.4^{+15.6}_{-14.9}$ & 3.38$\pm$0.10 & 6211$\pm$31 & 2466$\pm$11 & 5158$\pm$278 & 2180$\pm$45 & $4.1^{+1.4}_{-1.4}$ & $1.8^{+0.7}_{-0.7}$ \\ 
15 & PG~1121$+$422 & $115.8^{+24.2}_{-20.2}$ & 6.86$\pm$0.10 & 2613$\pm$31 & 1627$\pm$10 & 2230$\pm$62 & 1037$\pm$70 & $2.4^{+0.5}_{-0.4}$ & $1.1^{+0.3}_{-0.2}$ \\ 
17 & PG~1202$+$281 & $38.5^{+9.1}_{-8.5}$ & 2.75$\pm$0.05 & 5545$\pm$62 & 2503$\pm$25 & 3855$\pm$411 & 1893$\pm$111 & $2.7^{+0.6}_{-0.6}$ & $1.2^{+0.3}_{-0.3}$ \\ 
18 & J1217$+$333 & $26.5^{+21.2}_{-20.7}$ & 1.53$\pm$0.08 & 4476$\pm$31 & 1971$\pm$18 & 3233$\pm$154 & 1691$\pm$54 & $1.5^{+1.2}_{-1.2}$ & $0.7^{+0.5}_{-0.5}$ \\ 
19 & VIII~Zw~218 & $63.3^{+16.4}_{-15.4}$ & 2.90$\pm$0.05 & 5371$\pm$62 & 2523$\pm$21 & 3980$\pm$185 & 1679$\pm$113 & $3.5^{+0.9}_{-0.8}$ & $1.6^{+0.4}_{-0.4}$ \\ 
20 & PG~1322$+$659 & $49.3^{+19.2}_{-16.6}$ & 6.48$\pm$0.14 & 3004$\pm$31 & 1583$\pm$12 & 2104$\pm$97 & 1560$\pm$45 & $2.3^{+0.9}_{-0.8}$ & $1.1^{+0.4}_{-0.4}$ \\ 
21 & J1415$+$483 & $25.3^{+11.8}_{-11.1}$ & 4.41$\pm$0.11 & 3654$\pm$62 & 1742$\pm$16 & 4166$\pm$1172 & 1847$\pm$343 & $1.7^{+0.8}_{-0.7}$ & $0.8^{+0.4}_{-0.3}$ \\ 
22 & PG~1427$+$480 & $33.3^{+20.6}_{-19.4}$ & 6.26$\pm$0.12 & 2721$\pm$97 & 1582$\pm$38 & 1786$\pm$93 & 1292$\pm$155 & $1.1^{+0.7}_{-0.6}$ & $0.5^{+0.3}_{-0.3}$ \\ 
23 & PG~1440$+$356 & $50.6^{+16.9}_{-20.9}$ & 3.61$\pm$0.05 & 2545$\pm$62 & 1263$\pm$15 & 2167$\pm$97 & 1003$\pm$113 & $1.0^{+0.3}_{-0.4}$ & $0.4^{+0.2}_{-0.2}$ \\ 
24 & J1456$+$380 & $77.8^{+9.4}_{-8.7}$ & 5.32$\pm$0.11 & 7972$\pm$154 & 3002$\pm$72 & 5901$\pm$776 & 2504$\pm$167 & $9.5^{+1.1}_{-1.1}$ & $4.3^{+0.7}_{-0.7}$ \\ 
25 & J1526$+$275 & $63.9^{+10.3}_{-9.3}$ & 6.61$\pm$0.10 & 4538$\pm$31 & 1970$\pm$12 & 4476$\pm$216 & 2173$\pm$101 & $9.3^{+2.5}_{-3.9}$ & $4.2^{+1.2}_{-1.8}$ \\ 
26 & J1540$+$355 & $57.9^{+18.1}_{-14.6}$ & 2.83$\pm$0.06 & 2383$\pm$62 & 1493$\pm$14 & 2041$\pm$93 & 1077$\pm$79 & $1.3^{+0.4}_{-0.3}$ & $0.6^{+0.2}_{-0.2}$ \\ 
28 & PG~1612$+$261 & $63.7^{+13.6}_{-14.2}$ & 5.12$\pm$0.21 & 3165$\pm$93 & 1887$\pm$48 & 1977$\pm$195 & 2054$\pm$265 & $5.2^{+1.1}_{-1.2}$ & $2.3^{+0.6}_{-0.6}$ \\ 
29 & J1619$+$501 & $32.3^{+6.8}_{-6.7}$ & 2.62$\pm$0.04 & 5391$\pm$62 & 2388$\pm$17 & 6519$\pm$437 & 2511$\pm$367 & $4.0^{+0.8}_{-0.8}$ & $1.8^{+0.4}_{-0.4}$ \\ 
32 & PG~2349$-$014 & $55.9^{+11.0}_{-10.8}$ & 4.58$\pm$0.16 & 6924$\pm$123 & 3841$\pm$87 & 6890$\pm$555 & 3595$\pm$197 & $14.1^{+2.8}_{-2.7}$ & $6.3^{+1.4}_{-1.4}$ \\ 
\hline
7 & J0801$+$512 & $166.8^{+17.0}_{-17.4}$ & 3.24$\pm$0.08 & 2274$\pm$62 & 1469$\pm$23 & 1400$\pm$221 & 1359$\pm$157 & $6.0^{+0.6}_{-0.6}$ & $2.7^{+0.4}_{-0.4}$ \\ 
11 & J1059$+$665 & $189.4^{+20.9}_{-10.6}$ & 7.86$\pm$0.17 & 4457$\pm$62 & 2210$\pm$25 & 5034$\pm$1120 & 2343$\pm$150 & $20.3^{+2.2}_{-1.1}$ & $9.1^{+1.5}_{-1.2}$ \\ 
13 & J1105$+$671 & $24.8^{+18.5}_{-18.0}$ & 3.83$\pm$0.08 & 4724$\pm$62 & 2192$\pm$37 & 3793$\pm$807 & 2145$\pm$298 & $2.2^{+1.7}_{-1.6}$ & $1.0^{+0.8}_{-0.7}$ \\ 
16 & J1203$+$455 & $52.6^{+10.3}_{-15.9}$ & 9.79$\pm$0.17 & 5653$\pm$93 & 2854$\pm$42 & 4290$\pm$308 & 1840$\pm$154 & $3.5^{+0.7}_{-1.1}$ & $1.6^{+0.4}_{-0.5}$ \\ 
27 & PG~1545$+$210 & $60.3^{+11.0}_{-9.8}$ & 17.95$\pm$0.27 & 6557$\pm$93 & 2932$\pm$24 & 5282$\pm$1955 & 2655$\pm$193 & $8.3^{+1.5}_{-1.4}$ & $3.7^{+0.8}_{-0.8}$ \\ 
30 & J1935$+$531 & \nodata &13.77$\pm$0.20 & 5284$\pm$159 & 2705$\pm$45 & 2921$\pm$349 & 2361$\pm$157 & \nodata & \nodata \\ 
31 & PG~2251$+$113 & $152.7^{+22.9}_{-26.6}$ & 30.91$\pm$0.43 & 4811$\pm$31 & 2752$\pm$18 & 5034$\pm$252 & 2103$\pm$92 & $13.2^{+2.0}_{-2.3}$ & $5.9^{+1.1}_{-1.2}$ \\

\hline
\multicolumn{10}{p{0.99\textwidth}}{Notes. Column (1): object names; Column (2): the final rest-frame lags. We adopt the ICCF $\tau_{\rm cent}$ for all objects except J1526$+$275 for which we prefer $\tau_{\rm JAV}$ as the final lag measurement (see \S \ref{sec:lag-reliability}). Six unreliable measurements are listed at the bottom with one object showing negative lags not displayed. Column (3): extinction corrected and host contamination removed (if any) AGN luminosities in the unit of $10^{44}$ erg s$^{-1}$; Column (4)-(5): FWHM and $\sigma_{\rm line}$ measured from mean spectrum; Column (6)-(7): FWHM and $\sigma_{\rm line}$ measured from rms spectrum; Column (8): virial products calculated using the final lags in Column (2) and the rms spectrum $\sigma_{\rm line}$ in Column (7); Column (8): final BH masses calculated by multiplying the virial products in Column (7) with the virial factor $f=4.47$ \citep{Woo15}.}

    \end{tabular}
\end{table*}}\label{tab:BHmass}

We determine the \mbh\ based on Eq. 1, using the $\tau_{\rm cent}$ from the ICCF analysis as the lag and the line dispersion of \hbeta\ measured from the rms spectra ($\sigma_{\rm rms}$) as the velocity along with the average $f=4.47\pm0.43$ \citep{Woo15}.  Note that while we adopt a single $f$ value in this work, we will present the velocity-resolved lag measurements and dynamical modeling to derive the $f$ factor for each AGN in the future. 

For determining \mbh, the line dispersion ($\sigma_{\rm rms}$) is commonly used in RM studies because it best recovers the virial relation \citep{Peterson04, Park12a}. It is also suggested that the line dispersion is less sensitive to other factors, e.g., orientation, compared to the FWHM \citep{Collin06, Wang19}. On the other hand, the FWHM is frequently used in single-epoch \mbh\ estimation as it is easier to measure than the line dispersion based on low S/N spectra. We provide the \mbh\ of the sample based on the line dispersion in Table \ref{tab:BHmass}, but the \mbh\ with FWHM can be easily derived using Table \ref{tab:BHmass}. 

We also estimate the bolometric luminosity using the monochromatic luminosity $L_{5100}$ along with a constant bolometric correction factor $\kappa=9.26$ \citep[e.g.,][]{Richards06b, Shen11}. For calculating the Eddington luminosity $L_{\rm Edd}$, we assume $L_{\rm Edd} = 1.3\times10^{38}\,M_{\rm BH}/M_{\odot}$.

In Figure \ref{fig:Lbol-Mbh} we present \mbh\ and bolometric luminosity of the sample, in comparison with those of the previously reverberation-mapped AGNs by combining the collection of \citet{DallaBonta20} and the AGNs from the SEAMBHs \citep{Du_Wang19, Hu21, Li-SS21}. The \mbh\ of the SAMP AGNs range from 10$^{7}$ to 10$^{9}$ M$_{\odot}$ and the Eddington ratio covers from 0.05 to 1.82 with a median of 0.17, which is higher than that of the sample of \citet{DallaBonta20} but lower than the AGN in SEAMBHs.

\section{Discussion}

\subsection{\hbeta\ BLR size-luminosity relation}\label{sec:r-l_relation}

As the number of \hbeta\ lag measurements increased over the last decade, it has been demonstrated that the \hbeta\ size--luminosity relation 
has a considerably larger scatter and a shallower slope than previously accepted, indicating a more complex nature of the relation \citep[e.g.,][]{Du16, Grier17b, Du18b, Du_Wang19, Martinez-Aldama19,  Hu21, Li-SS21}. 
In particular, it is claimed that super-Eddington AGNs tend to be systematically offset from the relation of lower Eddington ratio AGNs \citep{Du16, Li-SS21}. In this section, we investigate the \hbeta\ BLR size-luminosity relation by combining our new measurements of high-luminosity AGNs with the previous measurements from the literature.

\begin{figure*}[htbp]
    \centering
    \includegraphics[width=0.48\textwidth]{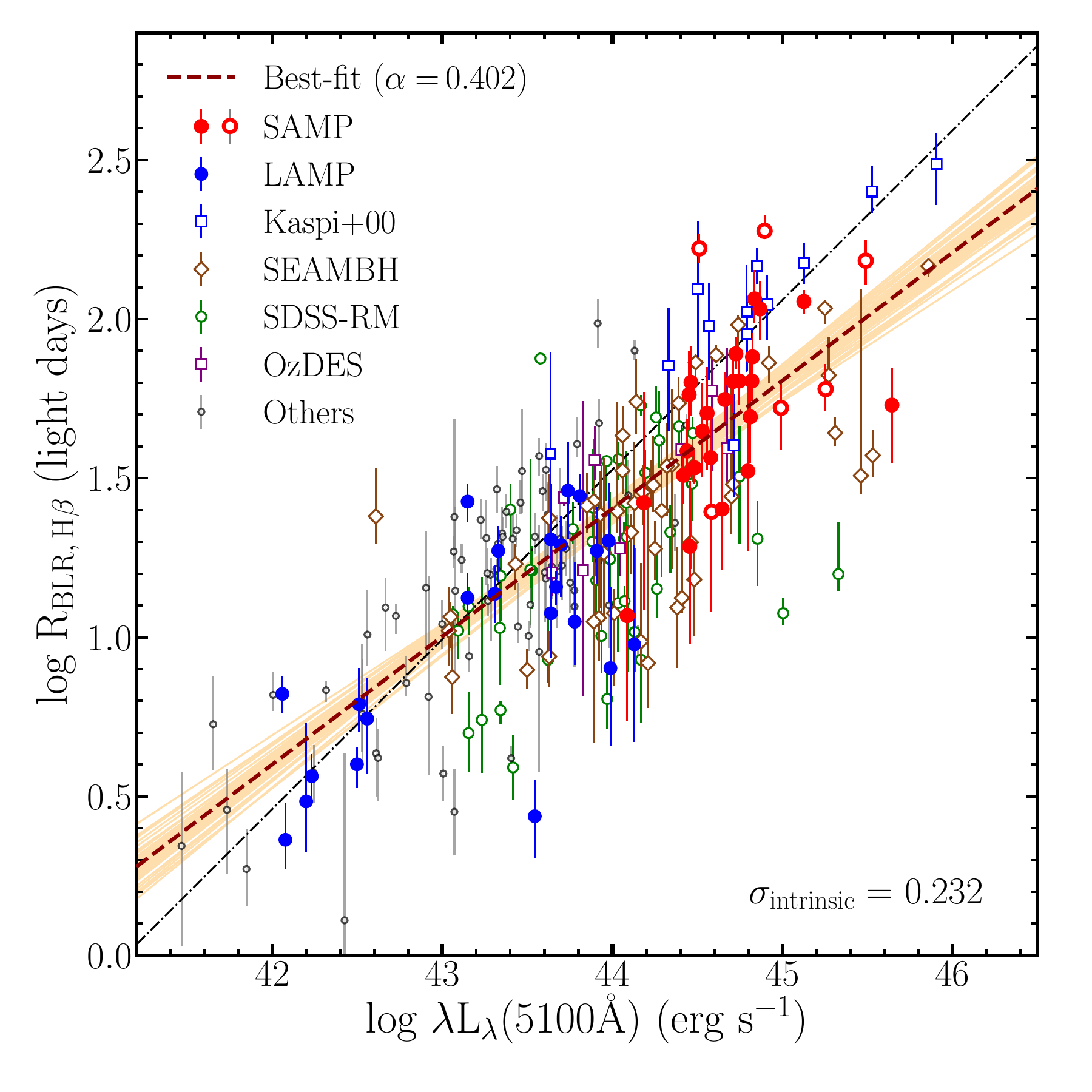}   
    \includegraphics[width=0.48\textwidth]{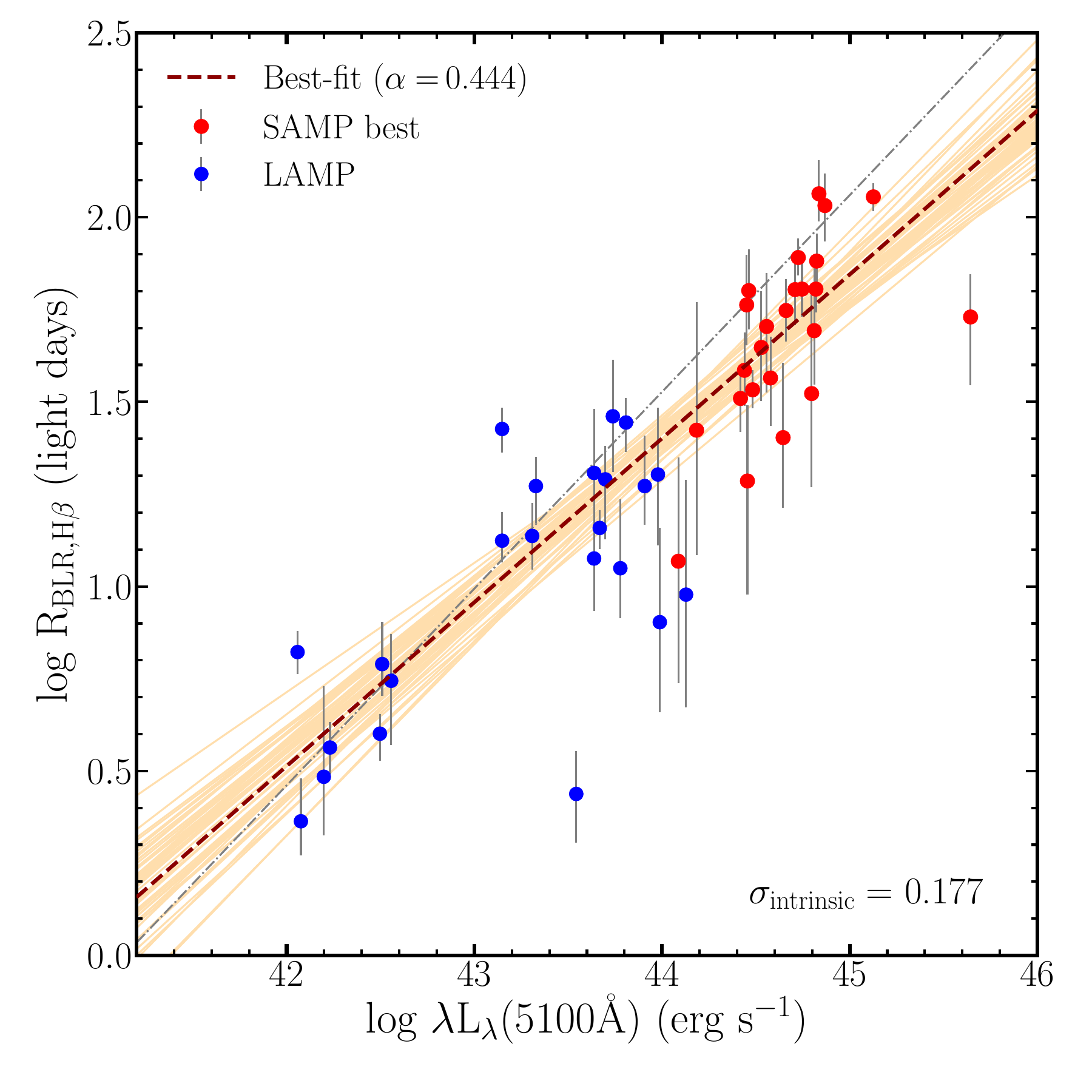} 
    \caption{Left: \hbeta\ BLR size -luminosity relation of the combined sample of SAMP (red filled and open circles) and the literature measurements, including LAMP 2008, and 2016 \citep[][blue circles]{Bentz09c, U22}, 12 PG quasars from \citet{Kaspi00} with the optical luminosity from \citet{Bentz13} (blue open squares), SEAMBHs \citep{Du16, Du18b, Zhang19, Hu21, Li-SS21} (brown open square diamonds), OzDES \citep{Malik22ArXiv} (purple open squares), SDSS-RM \citep[][green open circles]{Grier17b}, and others in the collection of \citet{DallaBonta20} (black open circles). The brown dashed line and the light orange lines represent the best-fit relation and 50 realizations randomly drawn from the MCMC chains. As a comparison, we denote the best-fit slope of 0.533 from \citet{Bentz13} (grey dotted-dashed line).
    Right: \hbeta\ BLR size -luminosity relation based on 24 best measurements from SAMP (red circles) and 23 measurements from LAMP (blue circles).
    }
     \label{fig:R-L_relation}
\end{figure*}

\begin{table*}[tbp]
    \centering
    \caption{Results of the best-fit \hbeta\ BLR size--luminosity relation}
    \begin{tabular}{l c c c c}
        \hline \hline
     & Sample   & $\alpha$  & $K$ & $\sigma$  \\ \hline

    Case 1 &  SAMP (30) + Literature (210)  &  $0.402^{+0.020}_{-0.022}$ & $1.405^{+0.018}_{-0.023}$ &  $0.232^{+0.013}_{-0.013}$ \\
   \hline 
    Case 2 &  SAMP best (24) + LAMP (23) &  $0.444^{+0.036}_{-0.035}$ & $1.401^{+0.034}_{-0.034}$ &  $0.177^{+0.028}_{-0.028}$ \\
    \hline
    \multicolumn{5}{p{0.9\textwidth}}{Note. The literature sample includes the objects from LAMP 2008 and 2016 \citep{Bentz09c}, SEAMBH project \citep{Du16, Du18b, Hu21, Li-SS21}, SDSS-RM \citep{Grier17b}, OzDES \citep{Malik22ArXiv}, and other measurements collected by \citet{DallaBonta20}.}
    \end{tabular}
    \label{tab:r-l-fitting}
\end{table*}

\subsubsection{Best-fit slope and scatter}\label{sec:r-l_relation_slope}
We utilize the collection by \citet{DallaBonta20}, who compiled a sample of AGNs with \hbeta\ lag and L$_{\rm 5100}$ after correcting for the host galaxy contamination.
As this collection does not include a large fraction of high-luminosity AGNs and AGNs without available HST images, we collect a number of AGNs from various other sources, i.e., the Super-Eddington-Accreting-Black-Holes (SEAMBHs) project \citep[e.g.,][]{Du16, Du18b, Hu21, Li-SS21}, the SDSS-RM project \citep{Grier17b}, the LAMP 2016 \citep{U22}, and other recent studies \citep{Li-SS21, Hu21, Malik22ArXiv}. 
The total sample consists of 242 \hbeta\ lag measurements (including multiple measurements of NGC 5548), among which 30 AGNs are based on the SAMP results. We exclude two SAMP AGNs, namely, PG~0026$+$129 and J1935$+$531, for which the obtained lag is negative or not larger than zero by 1$\sigma$ uncertainty (see Table \ref{tab:lags}). 
Note that by adding these two AGNs, we find an insignificant change of the slop and scatter. 

We perform a linear regression to derive the best-fit relation as:
{\small
\begin{equation}
    {\rm log} [R_{\rm BLR, H\beta}/({\rm lt \mbox{-}day})]= K + \alpha~ {\rm log}[\lambda L_{5100}/ (10^{44}{\rm erg}\,{\rm s}^{-1})],
\end{equation}
}
where the \hbeta\ BLR size is in the unit of light days and the monochromatic luminosity $\lambda L_{5100}$ is expressed in units of $10^{44}$ erg s$^{-1}$. The $\alpha$ and $K$ are the slope and intercept. To account for the asymmetric uncertainties of the lags, we performed MCMC regression fits using {\tt Python} package {\tt emcee}\footnote{\url{https://emcee.readthedocs.io/en/stable/}}, and adjusted the likelihood function so that it uses the upper error when the model value is larger than the data and uses the lower error in the opposite case. In brief, the regression minimizes the following quantity:
\begin{equation}
    \chi^2=\sum_{i=1}^{N}\frac{[y_i-(\alpha(x_i-x_0)+K)]^2}{(\alpha x)_{{\rm err},i}^2+y_{{\rm err},i}^2 + \sigma_{\rm int}^2},
\end{equation}
\noindent where $x_{{\rm err},i}$ and $y_{{\rm err},i}$ are the errors of variable $x_i$ and $y_i$, respectively, $x_0$ is the normalization x, and the $\sigma_{\rm int}$ is the intrinsic scatter.

We obtain a best-fit slope of 0.402$^{+0.020}_{-0.022}$ using the total sample of 240 AGNs with an intrinsic scatter of 0.232 dex (Figure \ref{fig:R-L_relation} left). As we find more similarity between the SAMP and the LAMP in terms of spectroscopic observations, spectral analysis, and lag measurements, we use the combined sample of 24 best SAMP measurements (after excluding PG~0026+129 due to its unresolved lag) and 23 LAMP measurements, finding that a best-fit slope of 0.444$^{+0.036}_{-0.035}$ with an intrinsic scatter of 0.177 dex (Figure \ref{fig:R-L_relation} right and Table 7).
Note that most of the SAMP objects lie below the previous relation (i.e., $\alpha$ = 0.533), which was defined based on $\sim$40 AGNs ($\sim$70 lag measurements) by \citet{Bentz13}.

It is unlikely that the deviation of the SAMP AGNs, with the observed-frame lag up to 254 days, is due to the underestimation of lags,
since the SAMP results are based on the 6-year ($\sim2000$ days) monitoring data with a $\pm$600-day lag search window. 
SAMP AGNs are generally luminous (i.e., L$_{5100}$ $>$ 10$^{44}$ erg s$^{-1}$) and the host galaxy contamination is not significant as demonstrated by the lack of stellar absorption lines in the spectra. Based on the spectral decomposition, we measure the host fraction at 5100\AA. The average host fraction of our sample is 0.07, ranging from 0 to 0.28. Thus, the continuum luminosity is unlikely to be overestimated because of the host contribution. 

Our results suggest that the slope is likely to be shallower than the popularly used 0.533 slope of \citet{Bentz13}, and that if the previous relation is utilized, single-epoch \mbh, particularly for high-luminosity AGNs at high-z, would be overestimated.

Note that these \hbeta\ lag measurements collected from the literature were not consistently determined since various studies adopted a number of different methods in their analysis. Thus, there could be various systematic uncertainties in deriving the size-luminosity relation based on this heterogeneous sample. 
Nevertheless, we note that the best-fit relation based on the total sample is relatively tight with a rms scatter of $\sim$0.3 dex and an intrinsic scatter of 0.234 dex, without requiring two different relations, respectively, for sub-Eddington and super-Eddington AGNs \citep[c.f.][]{Li-SS21}.
It is important to perform a consistent and uniform analysis of the cross-correlation and error measurement, in order to derive a better
size-luminosity relation. We will revisit the size-luminosity relation with uniformly measured lags and investigate the systematic effect of AGN parameters on the size-luminosity relation in our next study.

\subsubsection{Deviation from the size-luminosity relation}\label{sec:deviation}

\begin{figure*}[htbp]
    \centering
    \includegraphics[width=0.48\textwidth]{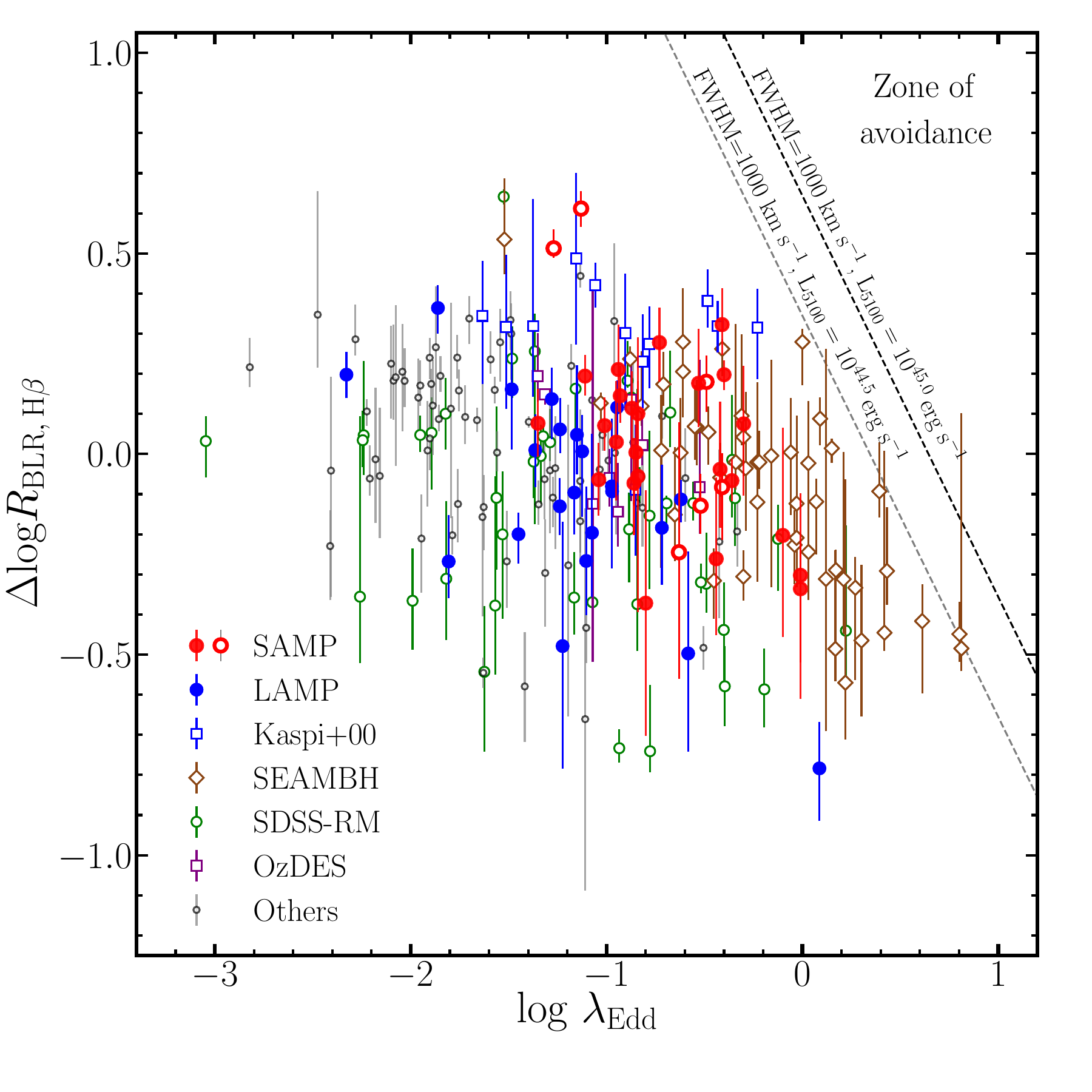}   
    \includegraphics[width=0.48\textwidth]{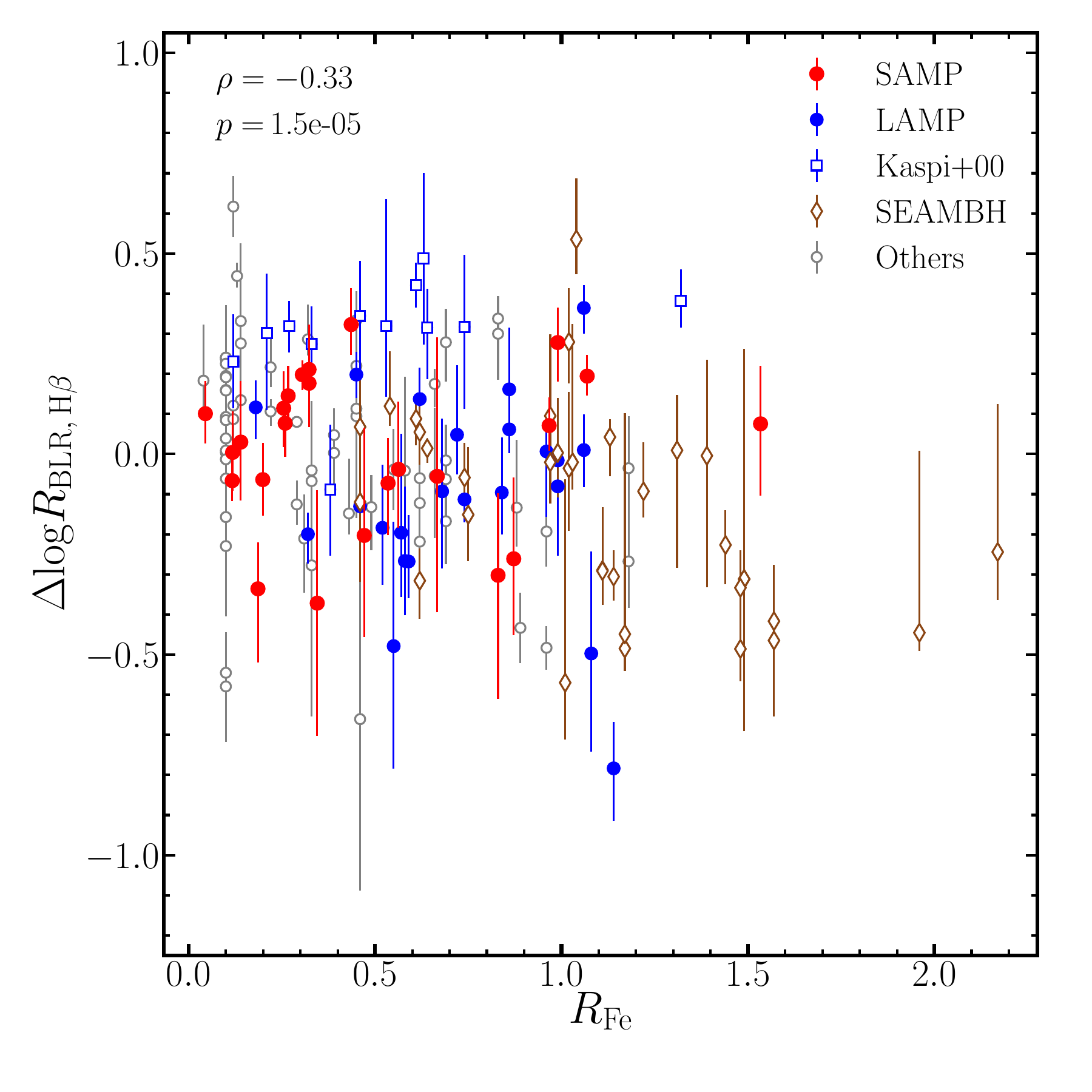} 
    \caption{\hbeta\ comparison of the deviation from the size-luminosity relation with Eddington ration (left) and strength of \FeII\ (right). Note that the decreasing trend with the Eddington ratio is partly due to the zone of avoidance (no type-1 AGNs with FWHM < 1000 km s$^{-1}$) and self-correlation.
    The dotted line indicates the upper envelope of the distribution defined by AGNs with FWHM$=$ 1000 km s$^{-1}$ for a fixed luminosity, L$_{5100}$$=$10$^{44.5}$ and 10$^{45}$, respectively.    
    }\label{fig:R-L_dev_relation}
\end{figure*}

We investigate whether the deviation from the size-luminosity relation correlates with any key parameters of AGN. Using the best-fit relation obtained based on all targets (Case 1), we calculate the deviation of each target and compare with Eddington ratio in Figure \ref{fig:R-L_dev_relation} (right). It was previously reported that supper-Eddington AGNs tend to have smaller \hb\ sizes for given AGN luminosity and that this deviation is stronger with higher Eddington ratios or high mass accretion rate. For example, \citet{Du15} claimed a negative correlation between the mass accretion rate with the deviation. 
However, this could be naturally caused by the zone of avoidance since there is typically no type-1 AGN with FWHM of the broad H$\beta$ line less than 1000 km s$^{-1}$ (see the dashed line in Figure \ref{fig:R-L_dev_relation}).
For example, for a fixed AGN luminosity (i.e., L$_{5100}$=10$^{45}$ erg s$^{-1}$) in the Eddington ratio between $\sim$0.3 and 1, there is an upper envelope, above which no AGN can be located because the widths of the \hbeta\ broad-line are not typically smaller than 1000 km s$^{-1}$. Note that this effect is not strong at lower Eddington ratios (i.e., $<$ $\sim$0.3), leading to a negative correlation only at the high Eddington regime. 

The decreasing trend, particularly at the high Eddington ratio is naturally expected (Figure \ref{fig:R-L_dev_relation} right) due to the self-correlation between the Eddington ratios and the deviation, since the former scales with $L/R$ and the latter scales with $R/L^{0.5}$ as pointed out by \citet{Fonseca-Alvarez20}. In other words, if the measured $R_{\rm BLR}$ is somewhat smaller than expected by the best fit, then the \mbh\ would be smaller. Consequently, the Eddington ratio of these targets would be systematically larger. Thus, the decreasing trend between the deviation and the Eddington ratio is expected due to the self-correlation.

To overcome these artificial trends due to the zone of avoidance and the self-correlation, we use the strength of \FeII\ emission ($R_{\rm Fe}$), which is the ratio between the \FeII\ emission flux in the range of 4434-4684\AA\ and the H$\beta$ emission flux, for comparing with the deviation (Figure \ref{fig:R-L_dev_relation} right). As $R_{\rm Fe}$ \ is closely related to Eigenvector 1 in PCA analyses of AGN properties, which is typically interpreted as an accretion rate parameter, we use $R_{\rm Fe}$ as an indicator of Eddington ratio. Note that for this practice, we only use the subsample including SAMP, LAMP, and other AGNs with available $R_{\rm Fe}$ in the literature. However, we find no strong trend between the deviation from the size-luminosity relation and R$_{Fe}$, suggesting no systematic effect of accretion rate on the deviation from the BLR size. In contrast, SEAMBH project reported that the deviation of their sample correlated with dimensionless accretion rate \citep{Du14, Du15} as well as the FWHM/$\sigma$ ratio of \hbeta\ line and $R_{\rm Fe}$ \citep{Du16, Du_Wang19}.
Note that we reproduce the same result as \citet{Du16, Du_Wang19} using their sample. However, by adding more luminous sub-Eddington AGNs (e.g., SAMP AGNs),  their correlation becomes weaker. We also note that by adopting a shallower slope of 0.4 in the size-luminosity relation, we obtain a systematically smaller correlation coefficient between the deviation and $either R_{\rm Fe}$ or Eddington ratio. 
A full analysis based on a large sample of reverberation-mapped AGNs with consistently measured \hbeta\ lag, line width, \mbh, and R$_{Fe}$  is required to clearly investigate the dependency of the size-luminosity relation on the Eddington ratio. We will present more detailed results based on our re-analysis of \hbeta\ lag measurements and the size-luminosity relation using a large archival sample (Wang et al. in preparation).

\subsubsection{What determines the slope of the size-luminosity relation?}\label{sec:deviation}

In this section we discuss several scenarios to explain the observed slope of the H$\beta$ BLR size - optical luminosity relation. First, we expect that the BLR size is proportional to the 0.5 power of the ionizing luminosity if the ionization parameter (U) and hydrogen gas density (n$_{\rm H}$) are similar for all AGNs, since U is proportional to L$_{\rm ion}$/(4$\pi$R$^{2}$n$_{\rm H}$), where L$_{\rm ion}$ is the photoionizing luminosity. However, this assumption may not be true as the gas clouds in the BLR could have a range of U and n$_{\rm H}$ \citep[e.g.,][]{Baldwin+95}. 
If higher luminosity AGNs have an average higher value of the product, U $\times$ n$_{\rm H}$, then the size of BLR gets smaller than expected, leading to a shallower slope. Currently, we do not have clear observational evidence for this trend. 

Second, the optical luminosity measured at 5100\AA\ may not properly represent the ionizing luminosity although the UV continuum luminosity at around the Lyman edge may show a better correlation with a 0.5 slope \citep[see discussion by][]{Czerny+19, Fonseca-Alvarez20}. 
In this scenario, the UV-to-optical flux ratio has to be systematically lower for higher luminosity AGNs, in order to be consistent with the observation that L$_{5100}$ is larger than expected from the 0.5 slope for a given BLR size. Some of the thin disk models showed that the UV-to-optical flux ratio decreases with increasing bolometric luminosity, decreasing Eddington ratio, and decreasing black hole spin \citep[e.g.,][]{Davis+11, Castello-Mor+16, Czerny+19}. However, model predictions show various trends of the UV-to-optical flux ratio, depending on the model assumptions, i.e., radiative efficiency and wind \citep[e.g.,][]{Laor+14}. In general, the UV-optical SED in the thin disk models is somewhat inconsistent with observations \citep[e.g.,][]{Davis+07, Davis+11}. Clearly, more detailed studies are required to understand the effect on the spectral slope. 

As an empirical test, we compare available H$\beta$ line luminosity with L$_{5100}$ for a subsample of our collected H$\beta$ reverberation-mapped AGNs, finding a slightly sublinear relation, L$_{\rm H\beta}$ $\propto$ (L$_{5100})^{0.92\pm0.01}$ (Figure 20). If we assume L$_{\rm H\beta}$ is proportional to the ionizing luminosity, we expect a $\sim0.46$ slope in the BLR size - optical luminosity relation. These results suggest that the systematic change of the SED slope between UV and optical ranges may contribute to the deviation from the 0.5 slope by increasing L$_{5100}$ for given BLR size and ionizing luminosity.

\begin{figure}[htbp]
    \centering
    \includegraphics[height=0.55\textwidth, width=0.49\textwidth]{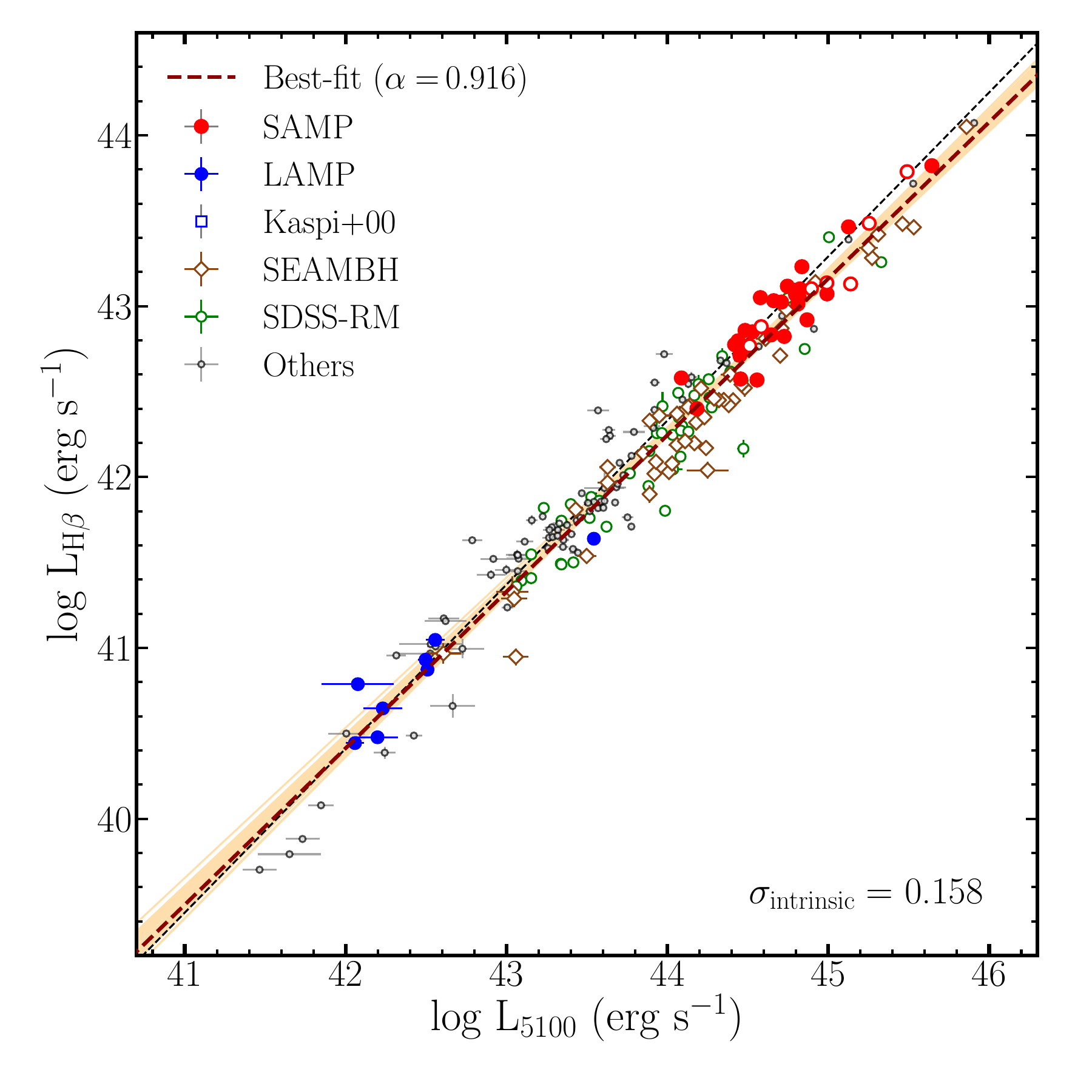}   
    \caption{Comparison of the optical luminosity at 5100\AA\ with the H$\beta$ line luminosity for a subsample of the H$\beta$ reverberation AGNs.  }
    \label{fig:L5100_LHb}
\end{figure}

Third, super-Eddington AGNs may suffer a shortening of the BLR size owing to the self-shadowing effect of the slim disk as detailed by \citet{Wang14}. Gas clouds in the BLR with a high inclination angle (with respect to the rotation axis) receive less ionizing photons than gas clouds with a lower inclination angle due to the shadowing of the funnel in the inner disk, and this effect in super-Eddington AGN leads to a shortening of the BLR size. Thus, super-Eddington AGNs have systematically smaller BLR sizes compared to sub-Eddington AGNs. 
Fourth, the BLR size-luminosity relation is driven by the outer boundary of the BLR, which is defined by the dust sublimation radius at the inner edge of the torus \citep{Suganuma06}. Interestingly, the dust size - optical luminosity relation also shows a shallower slope than 0.5. For example, \citet[e.g.,][]{Minezaki+19} reported a 0.424 slope between the K-band torus size and V-band continuum,
and Amit et al. (2023 submitted) found a $\sim$0.4 slope between the torus size based on the WISE W1-band and optical luminosity. Perhaps, this is also due to the self-shadowing effect of the slim disk model as pointed out by \citet{Chen+23}. 

We consider a sample selection effect of our collected AGNs. While we try to increase the number of high-luminosity AGNs in this study, there is also a lack of high Eddington ratio AGNs in the low-luminosity range. To demonstrate the difference in Eddington ratio distribution as a function of luminosity, we divide the sample into 4 luminosity bins and present the Eddington ratio distribution in Figure 21. We clearly notice that the median Eddington ratio is increasing for higher luminosity bins. This trend can naturally cause the shallower slope if higher Eddington ratio AGNs tend to have smaller BLR sizes for given luminosity. In other words, a higher fraction of high Eddington ratio AGNs in higher luminosity bins can generate a shallower slope.

We discussed various effects, which could be responsible for the shallower slope of the correlation between the BLR size and optical luminosity. 
It is puzzling to observe a shortening of the BLR size for a given object. For example, the best-studied low-Eddington AGN, NGC 5548 showed a factor of 5-10 smaller BLR size for a given (or similar) optical luminosity \citep[see Figure 13 in][]{Pei+17}. This suggests that the BLR size can change significantly without changing luminosity or Eddington ratio. More detailed studies are required to better understand the scatter and slope of the BLR size - luminosity relation.

\begin{figure}[htbp]
    \centering
    \includegraphics[height=0.55\textwidth, width=0.49\textwidth]{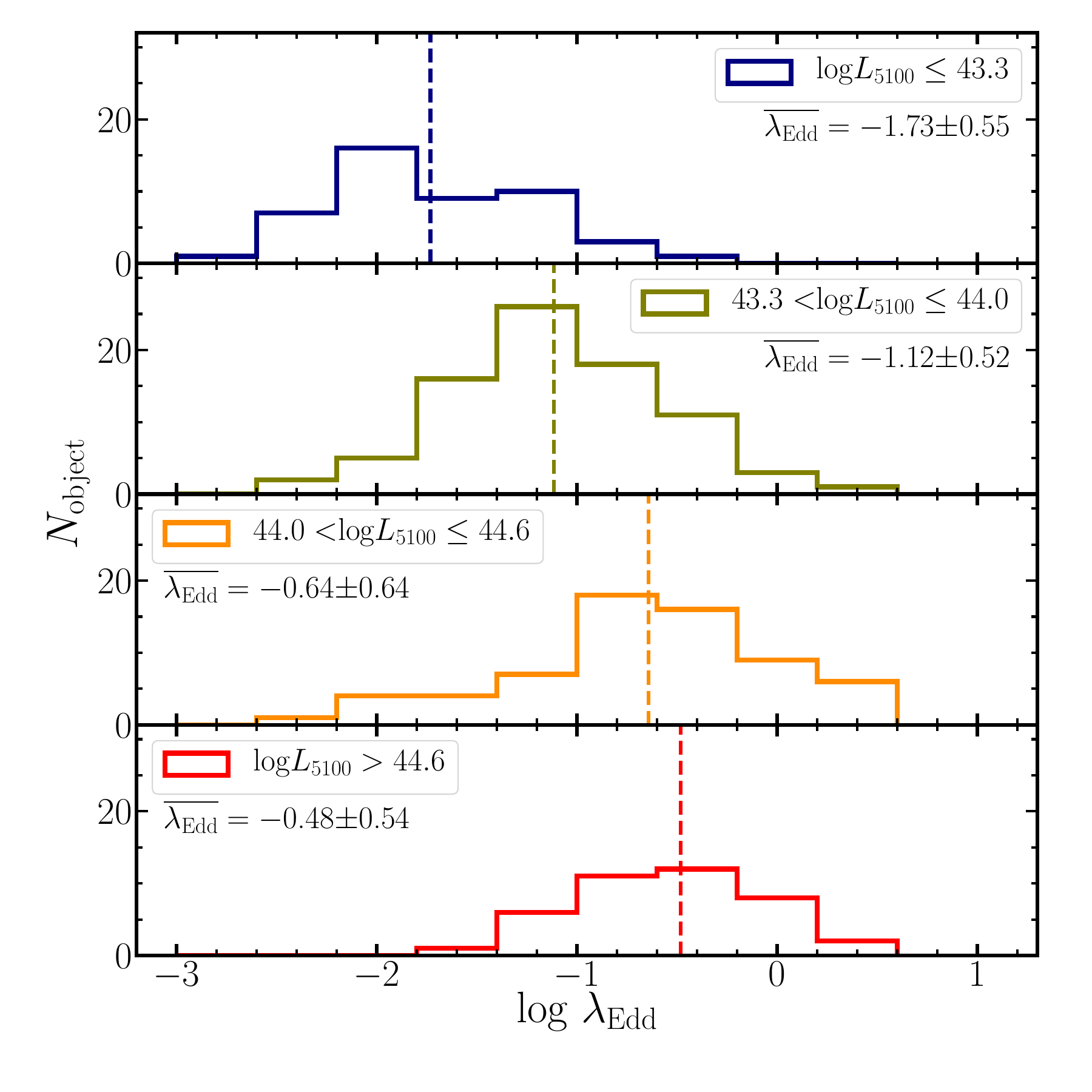}   
    \caption{Eddington ratio distribution in each luminosity bin. It is clearly shown that higher luminosity bins contain on average higher Eddington ratio AGNs.   }
    \label{fig:ER_dist}
\end{figure}

\section{Summary}
We present \hbeta\ reverberation mapping results based on the six-year (2015-2021) data from the SNU AGN Monitoring Project. For a sample of 32 high luminosity AGNs ($L_{\rm 5100, AGN} > 10^{44.1\sim45.6}$ erg s$^{-1}$) at z $<$$\sim$0.4, we measure the lags between the \hbeta\ and continuum light curves, using both the ICCF and {\tt JAVELIN} methods. By applying three reliability parameters ($r_{\rm max}$, $p(r_{\rm max})$, and $f_{\rm peak}$), we
quantitatively access the lag measurements, and use the accepted measurements to investigate the BLR size-luminosity relation. Our main conclusions are:

\begin{enumerate}
    \item Among the lag measurements of 32 targets, we report 25 best \hbeta\ lag measurements, which satisfy $r_{\rm max}$ $>$ 0.6,  $p(r_{\rm max})$$<$ 0.2, and $f_{\rm peak}$ $<$0.6. These new measurements significantly increase the current RM sample at high luminosity end. 
    \item We compare the {\tt JAVELIN} lag $\tau_{\rm JAV}$ with ICCF lag $\tau_{\rm cent}$, finding that they are generally consistent with each other but $\tau_{\rm JAV}$ tends to systematically offset due to more alias.  Thus, we adopt the ICCF $\tau_{\rm cent}$ results as our final lag measurements.

    \item By comparing the \hbeta\ BLR size and AGN continuum luminosity, we find that most of the SAMP AGNs are located below the previous relation measured by \citet{Bentz13}, suggesting that the slope is shallower than that expected from a simple photoionization model. We find the best slope of 0.39-0.46 by combining with previous \hbeta\ lag measurements in the literature. This result indicates that the single-epoch \mbh\ estimates based on the previous size-luminosity relation can be overestimated. 
    
    \item It is possible that the deviation from the size-luminosity relation correlates with Eddington ratio. However, we do not clearly confirm the correlation except for that caused by the self-correlation between AGN luminosity and Eddington ratio. Nevertheless, we detect a hint of correlation,
using the \ion{Fe}{2} relative strength. A consistent analysis of \hbeta\ lag and error measurements based on a uniform method for the large RM sample is necessary to unveil the nature of the size-luminosity relation.

\end{enumerate}

While the sample size of the reverberation-mapped AGNs significantly increased over the last decade by various studies including the SAMP, there is a still scarcity of very high luminosity AGNs at $L_{5100} \sim10^{46}$, and a uniform analysis including quantitative assessment of the measured lag is required to properly investigate the \hbeta\ BLR size-luminosity relation. These issues are beyond the scope of this paper and we will revisit them in the future.

\acknowledgments{J.-H.W. would like to thank all members of the SAMP for their efforts since 2015. We thank the anonymous referee for her/his useful comments, which improved the presentation and clarity of the manuscript. J.-H.W. thanks Shane Davis and Omer Blaes for the discussion on the think disk models. 
This work has been supported by the Basic Science Research Program through the National Research Foundation (NRF) of Korean Government (2021R1A2C3008486) and the Samsung Science \& Technology Foundation under Project Number SSTF-BA1501-05.
S.W. acknowledges the support from the NRF grant funded by the Korean government (MEST) (No. 2019R1A6A1A10073437). 
M.K. was supported by the NRF grant (No. 2022R1A4A3031306) funded by the Korean government.
Research at UCLA was supported by the NSF grant NSF-AST-1907208. 
Research at UC Irvine was supported by NSF grant AST-1907290.
V.N.B. gratefully acknowledges assistance from NSF Research at Undergraduate Institutions
(RUI) grant AST-1909297. Note that findings and conclusions do not necessarily represent views of
the NSF.
V.U acknowledges funding support from NASA Astrophysics Data Analysis Program (ADAP) grants \#80NSSC20K0450 and \#80NSSC23K0750, and Space Telescope Science Institute grants \#HST-AR-17063.005-A, \#HST-GO-17285.001, and \#JWST-GO-01717.001. 
S.R. acknowledges the partial support of SRG-SERB, DST, New Delhi through grant no. SRG/2021/001334. 
H.A.N. Le acknowledges the support of the National Natural Science Foundation of China (NSFC-12003031) and the "Fundamental Research Funds for the Central Universities"
We thank M. Fausnaugh for the advice in running {\tt mapspec}, and Yanrong Li and Yu-yang Songsheng for installing {\tt PyCALI}.

}

\clearpage

\appendix
\renewcommand\thefigure{\thesection.\arabic{figure}} 
\setcounter{figure}{0}

\section{Result of non-detrending for Pr1\_ID19 and Pr1\_ID27}\label{sec:appendixB}

\begin{figure*}[h]
\centering
\includegraphics[width=0.95\textwidth]{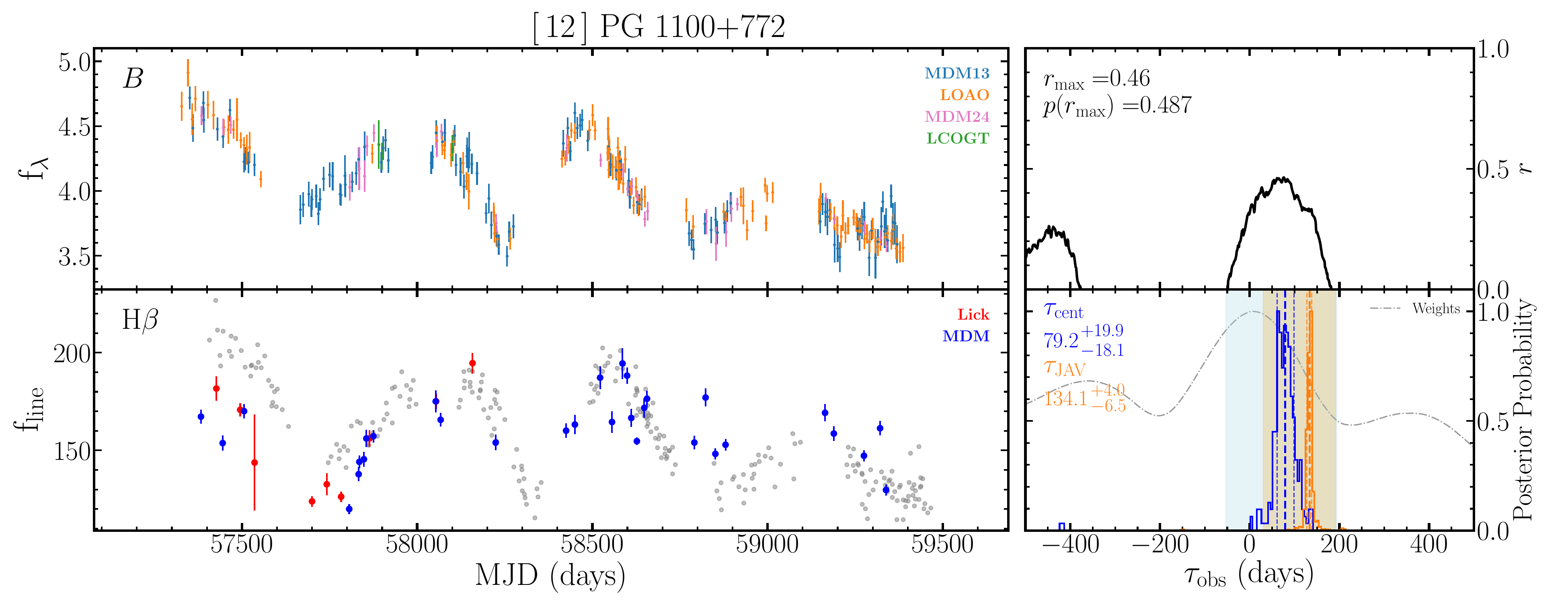}
\includegraphics[width=0.95\textwidth]{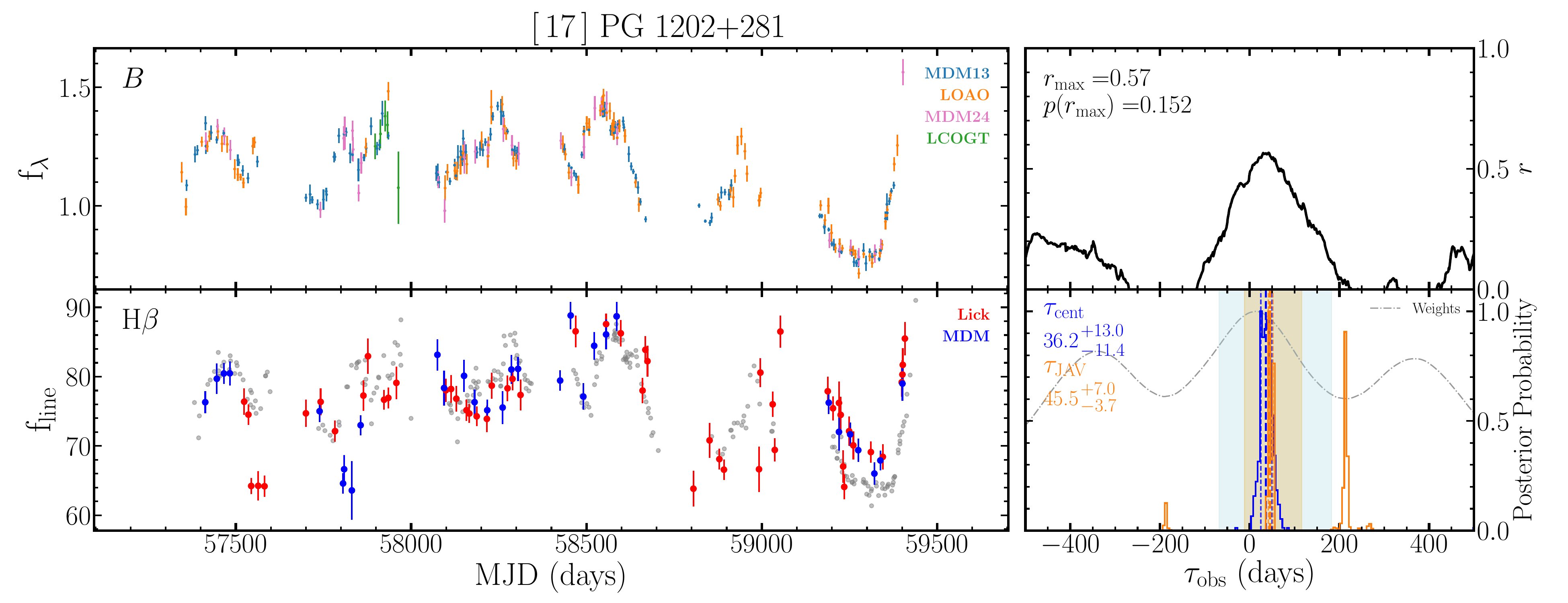}
\caption{The non-detrending lag estimation for PG~1100$+$772 and PG~1202$+$281. JAVELIN and p-value need to be updated.}
\label{fig:LC_AppendixB}
\end{figure*}

\section{Comparison with the previous measurements of individual objects}

\textit{Mrk 1501}: There were two previous monitoring campaigns of Mrk 1501 \citep{Grier12,Bao22Arxiv}. \citet{Grier12} observed this object from 2010 August to 2011 January and reported H$\beta$ $\tau_{\rm cent}=12.6\pm3.9$ days in the rest-frame. Combining with the \hbeta\ velocity dispersion measured from the rms specturm, $\sigma_{\rm rms}=3321\pm107$ km/s, the BH mass was determined as $1.84\pm0.27\,{\rm M_{\odot}}$.  In the case of the continuum luminosity, \citet{DallaBonta20} reported ${\rm log\,L}_{\rm 5100,AGN}$ = $43.980\pm0.053$ based on the decomposition analysis with HST image. Our new measurements H$\beta$ $\tau_{\rm cent}=11.7^{+7.6}_{-8.9}$ days and ${\rm log\,L}_{\rm 5100,AGN}=44.09$ based on the SAMP data are generally consistent with \citet{Grier12}. 
 
In addition, the target was observed from 2017 Oct. to 2021 Jan. by the Monitoring AGNs with \hbeta\ Asymmetry project \citep[MAHA,][]{Du18, Brotherton20, Bao22Arxiv}.  Note that their monitoring baseline is overlapped with the SAMP baseline between 2018 and 2021. 
They reported H$\beta$ $\tau_{\rm cent}=30.9^{+2.5}_{-2.4}$ days 
($\tau_{\rm MICA}=25.8^{+1.2}_{-1.1}$ days) using all 
seasons, while using the 2020 data they measured H$\beta$ $\tau_{\rm cent}=15.6^{+15.4}_{-11.1}$ days, which is in agreement with our measurement. 
Mrk 1501 is located below the best-fit slope of the size-luminosity relation by 0.41 dex. Since this object is a radio-loud AGN  with strong radio variability \citep[e.g.,][]{Unger87}, a possible scenarios is that the jet contribution to the optical luminosity may cause an overestimation of the continuum luminosity and underestimation of the \hbeta\ lag. Further study is required to investigate this effect.

\textit{PG~0026+129}: There were two previous monitoring campaigns of PG~0026+129 \citep{Kaspi00, Hu20}. \citet{Kaspi00} observed the target during a $\sim$7 year campaign, and \citet{Peterson04} reanalyzed the data, reporting H$\beta$ $\tau_{\rm cent}=111.0^{+24.1}_{-28.3}$ days in the observed frame. In contrast, \citet{Hu20} found a much smaller H$\beta$ lag based on a much higher cadence monitoring campaign as H$\beta$ $\tau_{\rm cent}$=$11.7^{+7.4}_{-7.8}$ days and H$\beta$ $\tau_{\rm cent}$=$27.7^{+5.0}_{-6.0}$ days using 2017 and 2019 data, respectively. While the baseline of their campaign overlapped with that of the SAMP, our data with lower cadence and smaller number of epochs could not provide a good temporal coverage. Thus, we removed this target from our measurements (i.e., \hbeta\ lag was comparable to 0 within the uncertainty). 

\textit{PG~0052+251}: \citet{Kaspi00} observed PG~0052+251 from 1991 July to 1998 September and \citet{Peterson04} reanalyzed the data, reporting the rest-frame H$\beta$ $\tau_{\rm cent}=89.8^{+24.5}_{-24.1}$ days. Our new measurement $\tau_{\rm cent}=63.9^{+12.0}_{-10.9}$ days is smaller but consistent with their lag within uncertainties. 
In the case of the continuum luminosity, our estimate ${\rm log\,L}_{\rm 5100,AGN}=44.75\pm0.01$ is comparable with 
${\rm log\,L}_{\rm 5100,AGN}=44.791\pm0.020$ based on the HST imaging analysis by \citep{DallaBonta20}. 
 
\textit{PG~0947$+$397}: PG~0947$+$397 was monitored by the MAHA from 2017 October to 2021 May, which overlapped with the SAMP's baseline from 2018 to 2021. \citet{Bao22Arxiv} reported H$\beta$ lag $\tau_{\rm cent}=34.4^{+4.5}_{-4.9}$ days in the rest-frame, consistent with ours within 1$\,\sigma$. 

\textit{PG~1100$+$772}: PG~1100$+$772 was monitored by MAHA from 2018 November to 2021 April which overlapped with SAMP's baseline from 2018 to 2021. They reported \hbeta\ lag  $\tau_{\rm cent}=44.9^{+30.5}_{-30.8}$ days in the rest-frame.  Our new lag measurement is well consistent with theirs within 1$\sigma$ uncertainty with slightly smaller uncertainties due to longer baseline. We find that the rms FWHM and $\sigma_{\rm line}$ in \citet{Bao22Arxiv} were reported to be $11229^{+29}_{-23}$ and $4002^{+87}_{-110}$ km/s, respectively, which is much higher than our measurements. As a consequence, their derived $M_{\rm BH}$ ($78.13^{+5.44}_{-4.72}\times10^7M_{\odot}$) is much larger than ours ($2.8^{+0.8}_{-1.2}\times10^8M_{\odot}$). We suggest that the difference is a result of the line variability between the different time baseline. As the line variability in 2016 and 2017 are larger than that of the following three years, the width of the rms line profile based on the six-year data becomes smaller. Note that the line widths measured from mean spectra show more consistency between the two studies. This object is very interesting as it is a strong outlier of $R$--$L$ relation but its low $R_{\rm Fe}$ and strong \OIII/\hbeta\ ratio indicates that it is not a typical SEAMBHs.

\textit{PG~1202+281}: PG~1202+281 was monitored by MAHA from 2016 December to 2021 April which overlapped with the SAMP's baseline from 2018 to 2021. They reported H$\beta$ lag $\tau_{\rm cent}=98.5^{+28.2}_{-30.1}$ days in the rest-frame together with a total $L_{5100}=3.42\pm0.32 \,\times10^{44}$ erg/s \citep{Bao22Arxiv}.  Our new measurement ($38.5^{+9.1}_{-8.5}$) is much smaller but still consistent with 3$\sigma$ uncertainties. Note that their lag measurement of individual season in 2018 and 2020, i.e., $50.0^{+6.6}_{-4.6}$ and $53.3^{+10.9}_{-8.5}$ days,  respectively, is much closer to our values.  Our derived luminosity is slightly lower than theirs that can also partly explain the difference. Combining with the measured rms FWHM and $\sigma_{\rm line}$ ($4255^{+23}_{-17}$ and $1301^{+18}_{-24}$ km/s, respectively), which is much smaller than our values ($5545\pm62$ and $2186\pm25$ km/s respectively), they derived a $M_{\rm BH}=9.80^{+0.44}_{-0.46}\times10^7 M_{\odot}$, close to our measurements ($1.2^{+0.3}_{-0.3}\times10^8M_{\odot}$).

\bibliography{ref}
\end{document}